\documentclass[structabstract]{aa} 
\usepackage{booktabs}
\usepackage{graphicx}
\usepackage{txfonts}
\usepackage{natbib}
\usepackage{longtable}
\usepackage{lscape}
\usepackage{multirow}
\usepackage[colorlinks=true, allcolors=blue]{hyperref}
\usepackage{xcolor}

\bibpunct{(}{)}{;}{a}{}{,}

\newcommand{\kms}{km\,s$^{-1}$}

\newcommand{\degree}{$^{\circ}$}

\newcommand{\Lsun}{L$_{\odot}$}

\newcommand{\nodata}{...}

\begin{document}

\title{Unveiling the enigma of the Nakashima-Deguchi object (IRAS 19312+1950): a candidate Orion KL analog}
\author{Y.~Gong\inst{1,2}, K.~M.~Menten\inst{2}\fnmsep\thanks{Prof. Dr. Karl M. Menten, who sadly passed away during the final stages of preparing this manuscript, initiated this project nearly twenty years ago and steadily accumulated data on this source ever since. He maintained a profound interest in it, always referring to it as the Nakashima–Deguchi object, a designation that we follow throughout this paper. The present paper essentially builds upon his original idea concerning the origin of this particularly intriguing source. We dedicate this paper to his enduring scientific curiosity and unwavering inspiration.}, T.~Kami{\'n}ski\inst{3}, F.~Schuller\inst{4}, F.~Navarete\inst{5}, A.~Belloche\inst{2}, P.~Schilke\inst{6}, F.~Wyrowski\inst{2}}

\institute{
  Purple Mountain Observatory and Key Laboratory of Radio Astronomy, Chinese Academy of Sciences, Nanjing 210008, China \\
  e-mail: ygong@pmo.ac.cn
  \and
  Max-Planck-Institut f{\"u}r Radioastronomie, Auf dem H{\"u}gel 69, D-53121 Bonn, Germany 
  \and
  Nicolaus Copernicus Astronomical Center, Polish Academy of Sciences, Rabia{\'n}ska 8, 87-100 Toru{\'n}, Poland \\
   e-mail: tomkam@ncac.torun.pl
  \and 
  Leibniz-Institut f{\"u}r Astrophysik Potsdam (AIP), An der Sternwarte 16, 14482 Potsdam, Germany 
  \and 
  Laborat\'orio Nacional de Astrof\'isica (LNA/MCTI), Rua dos Estados Unidos, 154, 37504-364, Itajubá, MG, Brazil
  \and 
  I. Physikalisches Institut der Universit{\"a}t zu K{\"o}ln, Z{\"u}lpicher Str. 77, D-50937 K{\"o}ln, Germany
}

\date{Received date ; accepted date}

\abstract
{The Nakashima-Deguchi object (NDO; aka IRAS 19312+1950) has been recognized as an enigmatic object, exhibiting characteristics of both an evolved star and a star-forming region. However, the actual nature of this object remains a topic of ongoing debate.}
{We want to provide constraints on the nature of NDO by elucidating its molecular composition and dust content.}
{We conducted observations of this object in (sub)millimeter continuum and molecular line emission using the IRAM-30m and APEX telescopes as well as the Submillimeter Array.}
{Based on spectral energy distribution fitting of single-dish continuum observations, we determined that the cloud hosting NDO has a dust temperature of $25\pm2$~K, a source-averaged (24\arcsec) H$_2$ column density of $(4.4\pm 0.6) \times 10^{22}$ cm$^{-2}$, a dust emissivity index of $1.7\pm0.2$, and a total gas mass of $220\pm21 M_{\odot}$. SMA continuum observations resolve the cloud into at least four dust continuum cores. The centrally peaked core MM1, associated with an SiO maser, is the primary mass reservoir. Single-dish spectroscopic observations of NDO led to the detection of 59 spectral lines attributed to 12 species and their isotopologues. This includes the first detection of deuterated molecules, DCO$^{+}$, DCN, DNC, and HDCO, toward NDO, indicating substantial deuterium abundance. SMA spectroscopic observations detected 75 spectral lines toward MM1, assigned to 12 species and their isotopologues. These lines exhibit diverse morphologies on a scale of $\sim$0.1~pc, likely due to outflow feedback. The observed high-velocity outflow is inconsistent with a spherical morphology and rather resembles a wide-angle bipolar outflow.}
{Based on the detected molecular inventory, the presence of deuterated species, and the overall large mass of the complex, the surrounding material of NDO appears more naturally associated with a star-forming environment than with a purely evolved-star scenario. Our investigations led us to postulate that this enigmatic object may represent an Orion KL analog about ten times farther away.}
\keywords{ISM: molecules --- Stars: individual (IRAS~19312+1950) --- Stars: AGB and post-AGB --- Stars: protostars -- Stars: winds, outflows}

\titlerunning{Unveiling the enigma of NDO: an Orion KL analog revealed}

\authorrunning{Y. Gong et al.}

\maketitle


\section{Introduction}\label{sec.intro}
Attention to the Nakashima-Deguchi object (NDO; aka IRAS 19312+1950), a point source discovered by the Infrared Astronomical Satellite (IRAS), was first raised by \citet{2000PASJ...52L..43N} who discovered SiO and H$_{2}$O maser emission associated with this object during a SiO maser survey of very red and dusty IRAS objects \citep{2003PASJ...55..229N}. The SiO and H$_{2}$O maser emission showed pronounced temporal variability and their radial velocity distribution presented a puzzling picture: the H$_{2}$O maser emission had different velocities compared to the SiO maser emission and both had velocities that differed from the velocity of the thermal 
emission \citep{2004PASJ...56.1083D}. The authors also reported that the 1612 MHz OH maser was detected by B.~M. Lewis at the Arecibo observatory. These SiO, H$_{2}$O, and OH masers were subsequently confirmed by follow-up Very Long Baseline Interferometry (VLBI) observations, which demonstrated that the two velocity components of the H$_{2}$O maser were 
separated by 10.9 mas and that the SiO masers showed an elongated 
distribution \citep{2011ApJ...728...76N}. The VLBI Exploration of Radio Astrometry (VERA) measurements of H$_{2}$O masers not only gave a parallax distance of 3.80$^{+0.83}_{-0.58}$ kpc which we adopt here, but also indicated that the source motion largely deviated from the circular galactic rotation by $\sim$40~\kms\,\citep{2011PASJ...63...81I}. The time variability of OH, H$_{2}$O, and SiO masers was monitored by \citet{2016JKAS...49..261K} and \citet{2025ApJ...981...41F}. Additionally, two methanol lines at 44 and 95 GHz were proposed to be masers in this source \citep{2015PASJ...67...95N}.

Thermal line emission from several C-, O-, and N-bearing molecules has also been discovered toward the region. These molecules include CO, SO, SiO, CS, CN, HCN, HNC, H$_{2}$O, NH$_{3}$, N$_{2}$H$^{+}$, HCO$^{+}$, CH$_{3}$OH, H$_{2}$CS, and HC$_{3}$N \citep{2000PASJ...52L..43N,2004PASJ...56..193N,2004PASJ...56.1083D,2015PASJ...67...95N,2016ApJ...828...51C,2023A&A...669A.121Q}. The observed spectra of these molecules show two distinct components, a broad and a narrow one. The most abundant di- and triatomic molecules (CO, H$_{2}$O, HCN, and HCO$^{+}$) show emission in the broad component with a full width at zero power (FWZP) up to $\sim$185~\kms\,\citep{2016ApJ...828...51C}, which is much broader than the velocity spread of the SiO and H$_{2}$O masers. In contrast, CS, CN, HNC, SO$_{2}$, and the polyatomic molecules are dominated by the narrow component. High angular resolution (3$\rlap{.}^{\prime\prime}$7$\times$2$\rlap{.}^{\prime\prime}$5) observations of HCO$^{+}$ (1--0) indicate a wide-angle bipolar outflow in the low-velocity wing structure and an unresolved morphology in the higher-velocity wing structure \citep{2004ApJ...610L..41N}. Further BIMA observations suggested that the angular size of the narrow-component region is larger than that of the broad-component region, and the distribution of CS peaks off the IRAS source position \citep{2005ApJ...633..282N}.

The infrared morphology of NDO was first studied by \citet{2000PASJ...52L..43N} who found this source to be associated with an extended near-infrared (NIR) bipolar nebula. Subsequent Subaru NIR observations revealed that this nebula consists of a nearly spherical core (radius $<$3\arcsec) and an S-shaped arm centered at the core \citep{2007A&A...470..957M}. Their polarization map displayed a centro-symmetric polarization vector pattern in this nebula with a linear polarization degree of 30\%--60\% and a lower linear polarization degree of $<$20\% along the S-shaped arm. The mid-infrared spectral energy distribution (SED) demonstrates the presence of the amorphous silicate absorption feature around 10~$\mu$m \citep{2007A&A...470..957M,2016ApJ...828...51C}. Furthermore, Spitzer Infrared Spectrograph (IRS) observations displayed absorption features at 15~$\mu$m, which are due to the bending vibrational mode of CO$_{2}$ ice \citep{2016ApJ...828...51C}. 

NDO exhibits characteristics indicative of both evolved stars and star-forming regions, making its nature puzzling. Despite extensive searches, SiO masers have been detected in only six star-forming regions, that is, Orion KL, Sgr B2, W51N \citep{2009ApJ...691..332Z}, G0.38+0.04 \citep{2015A&A...584L...7G}, G19.61$-$0.23, and G75.78+0.34 \citep{2016ApJ...826..157C}. In contrast, over 2000 evolved stars have been found to exhibit SiO maser emission \citep[e.g.,][]{2012IAUS..287..265D,2018MNRAS.473.3325W,2024ApJ...961..190Y}. Interestingly, SiO masers are also persistent in a merger remnant \citep[V838 Mon;][]{2020A&A...638A..17O}. The detection of SiO masers might indicate that NDO is an evolved star. This inference is supported by the fact that the broad-component emission is centered on the star and its CO distribution appears consistent with a spherically symmetric circumstellar outflow \citep{2004PASJ...56..193N,2005ApJ...633..282N}. However, the presence of methanol argues against it being an evolved star, as methanol molecules are rarely detected in evolved stars \citep{2017A&A...603L...2O,2018A&A...618A.164S,2020A&A...644A..59K}. Furthermore, the absence of the 15~$\mu$m CO$_{2}$ absorption features in circumstellar envelopes of evolved stars suggests that this feature observed toward NDO arises from a young stellar object (YSO) or a foreground cloud \citep{2016ApJ...828...51C}. Based on the detection of the absorption features, \citet{2016ApJ...828...51C} proposed that NDO could represent an unusual massive star-forming region like DR21(OH) and Orion KL or a chance encounter between a massive evolved star and a dense molecular cloud. Recently, \citet{2023A&A...669A.121Q} argued that NDO might be a red nova remnant primarily based on the comparison of isotope ratios with CK Vul \citep{2017A&A...607A..78K}, whereas \citet{2025ApJ...981...41F} proposed that NDO could instead be a candidate water-fountain star. Despite these efforts, the nature of NDO remains unsettled. Therefore, we have conducted an observational campaign aiming at unveiling the nature of NDO by studying its surrounding environment. 

This paper has been structured as follows. We present our observations in Sect.~\ref{Sec:obs}, followed by the observational results in Sect.~\ref{Sec:res}. In Sect.~\ref{Sec:dis}, we discuss these results, and Sect.~\ref{Sec:sum} provides the summary and conclusions.

\section{Observations}\label{Sec:obs}
To probe the nature of NDO, we observed this object in (sub)millimeter continuum emission and transitions of numerous molecular species with the Institut de Radioastronomie Millimétrique (IRAM) 30-m and Atacama Pathfinder Experiment (APEX) 12-m telescopes as well as the Submillimeter Array (SMA). All observations were centered on the IRAS position ($\alpha_{\rm J2000}$, $\delta_{\rm J2000}$) = (19$^{\rm h}$33$^{\rm m}$24$\rlap{.}^{\rm s}$40, 19\degr56\arcmin54$\rlap{.}^{\prime\prime}$8).

\subsection{Single-dish (sub)millimeter continuum observations}
Observations of the 1.2~mm (i.e., 250~GHz) and 870~$\mu$m (i.e., 345~GHz) continuum emissions were carried out using the IRAM 30-m and APEX 12-m telescopes equipped with the Max-Planck Millimeter Bolometer array \citep[MAMBO,][]{1999InPhT..40..191K} and the Large Apex BOlometer CAmera \citep[LABOCA,][]{2007Msngr.129....2S}, respectively. Pointing and focus checks were conducted every 2--3 hours. The flux calibration uncertainties were determined to be better than 20\%. Our observations achieved 1$\sigma$ sensitivities of 90 mJy~beam$^{-1}$ and 40 mJy~beam$^{-1}$, and angular resolutions of 10\rlap{.}\arcsec5 and 18\rlap{.}\arcsec2 at 1.2~mm and 870~$\mu$m, respectively.

\subsection{Single-dish molecular line observations}

Our single-dish molecular line observations were designed to characterize both the molecular gas and the molecular outflow associated with NDO. The observations consist of three parts. First, we used the IRAM~30~m telescope to observe a set of molecular transitions below 300~GHz, including CO, HCO$^+$, HCN, HNC, CS, SiO, SO, SO$_2$, N$_{2}$H$^{+}$, H$_2$CO, CH$_3$OH, HC$_3$N, C$_{4}$H, and several isotopologues and deuterated species. Second, we used APEX to observe six transitions in the 345~GHz atmospheric window, providing higher-frequency constraints on the excitation of selected molecules. Third, we obtained complementary APEX maps of CO (2--1), CO (3--2), and CO (4--3) over a $2\arcmin\times2\arcmin$ region to trace the morphology and energetics of the associated molecular outflow. The full list of observed transitions, their rest frequencies, angular resolutions, and spectral resolutions is given in Tables~\href{https://zenodo.org/records/21234367}{A.2--A.10}.

Molecular transitions with rest frequencies below 300 GHz were observed with the IRAM 30m telescope on 2005 July 22--28. The observations were arranged in several receiver setups with the available IRAM 30~m receiver combinations. These setups sampled the 3, 2, and 1~mm atmospheric windows and targeted the molecular species listed above. Most observations were done in the wobbler switching mode, with a 2\arcmin\, wobbler throw in azimuth. To properly determine the molecular layers along the line of sight, CO and $^{13}$CO lines were also observed in frequency-switching mode, and in position-switching mode using two off-positions at 1\degr\, from the source, on both sides of the Galactic plane. In each observational setup, four transitions were observed simultaneously, using the available receiver combinations at the IRAM 30m telescope. The Versatile Spectrometric and Polarimetric Array (VESPA) autocorrelator was used as the backend. It provided bandwidths of about 105 or 210~MHz, with channel widths of 0.1--0.4~\kms, depending on the observed frequency. Pointing was verified every 1--2 hours and was found to be stable within $\sim$3\arcsec. Observations taken during sunset or with high precipitable water vapor (pwv) exhibited increased pointing uncertainty and were excluded from the final spectra. To constrain the extent of molecular emission around NDO, we performed raster maps of HCO$^{+}$ (3--2), H$_{2}$CO (3$_{0,3}$--2$_{0,2}$), SO (2$_{3}$--1$_{2}$) and CH$_{3}$OH (3$_{0}$--2$_{0}$ A) with a grid spacing of 12\arcsec. The half-power beam width (HPBW), $\theta$, of the IRAM 30m telescope is $\theta = 2460^{\prime\prime}/\nu$, where $\nu$ is the observed frequency in units of GHz. 

On 2005 August 8--10, six transitions in the 345 GHz atmospheric window were observed with the APEX telescope\footnote{This publication is based on data acquired with the Atacama Pathfinder EXperiment (APEX). APEX is a collaboration between the Max-Planck-Institut f{\:u}r Radioastronomie, the European Southern Observatory and the Onsala Space Observatory.}. 
These observations were intended to complement the IRAM 30~m data by sampling higher-excitation transitions of selected species near 345~GHz. Observations were conducted in the position-switch mode using the APEX-2A facility receiver \citep{2006A&A...454L..13G}. The MPIfR Fast Fourier Transform spectrometer (FFTS) was used as the backend, providing a 1~GHz bandwidth. The FFTS consisted of 2048 channels, resulting in a channel spacing of 0.488~MHz (i.e., $\sim$0.4~\kms). Pointing was checked every 1--2 hours and was found to be accurate within $\sim$3\arcsec. 

To further study the associated CO outflow, we conducted complementary mapping observations of a 2\arcmin$\times$2\arcmin\,region centered on NDO in CO (2--1), CO (3--2), and CO (4--3) with APEX on 2023 June 16. The CO (2--1) and CO (4--3) observations were performed simultaneously with the nFLASH230 and nFLASH460 receivers\footnote{\url{https://www.apex-telescope.org/ns/observing-run/observing/the-telescope/instruments/nflash/}}, while the CO (3--2) observations utilized the Large APEX sub-Millimeter Array (LAsMA) receiver\footnote{\url{https://www.apex-telescope.org/ns/observing-run/observing/the-telescope/instruments/lasma-large-apex-sub-millimetre-array/}} \citep{2008SPIE.7020E..10G}. These receivers are equipped with a new generation of advanced FFTSs \citep{2012A&A...542L...3K}, providing instantaneous intermediate-frequency bandwidths of 16, 8, and 8~GHz at 230, 345, and 460~GHz, respectively. The corresponding channel width is 61~kHz in all three bands, equivalent to velocity spacings of 0.08, 0.05, and 0.04~\kms\, at 230, 345, and 460~GHz, respectively.

All APEX observations were carried out using the telescope control software APECS \citep{2006A&A...454L..25M}. The HPBW, $\theta$, of the APEX telescope is $\theta = 7.8^{\prime\prime}\times (800/\nu)$, where $\nu$ is the observed frequency in GHz. 

The observed spectra were calibrated every $\sim$10 minutes using the standard chopper-wheel method \citep{1976ApJS...30..247U}. Throughout this paper, antenna temperatures, $T_{\rm A}^{*}$, have been converted to a scale of main beam brightness temperatures, $T_{\rm mb}$, by applying the forward efficiency, $\eta_{\rm f}$, and the main beam efficiency, $\eta_{\rm mb}$\footnote{The $\eta_{\rm f}$ and $\eta_{\rm mb}$ values can be found in \url{https://www.iram.fr/GENERAL/calls/w08/w08/node20.html} for IRAM-30 m and in \citet{2006A&A...454L..13G} for APEX.}. Absolute flux calibration uncertainties are assumed to be 20\% for IRAM 30m and APEX. Velocities are presented with respect to the local standard of rest (LSR) frame. 

Data reduction was carried out with the Grenoble Image and Line Data Analysis Software (GILDAS\footnote{\url{http://www.iram.fr/IRAMFR/GILDAS}}, \citealt{2005sf2a.conf..721P}). Zeroth- or first-order baselines were subtracted from the observed spectra.

\subsection{SMA observations}
We carried out a broad-band line imaging survey toward NDO (project code: 2018A-S003) using the SMA \citep{2004ApJ...616L...1H} in 2018 June and July. The eight antennas of the SMA were in the compact array configuration with baselines ranging from 9~m to 78~m. 3C279 and 3C454.3 were adopted as bandpass calibrators. The flux density scale was determined using Titan. 2015+371 and 2025+337 were used as gain calibrators. The primary beam of the SMA is about 50\arcsec\,at 230~GHz, so single-pointing observations were sufficient to fully cover NDO. 
The absolute uncertainty in the flux scale is estimated to be within 20\%.

Using in parallel the 230 and 240 GHz receivers and taking advantage of the broad bandwidth of the SWARM correlator \citep{2016JAI.....541006P}, we covered the frequency range of 189--283 GHz with only three setups. The correlator provides a native channel width of 1.1~kHz. To improve the signal-to-noise ratios for our analysis, we smoothed the data set to a channel width of $\sim$1.1~MHz, which corresponds to a velocity spacing of $\sim$1.5~\kms\,at 230~GHz. 

The data were calibrated using the MIR\footnote{\url{https://lweb.cfa.harvard.edu/~cqi/mircook.html}} software and exported to CASA. Subsequently, the calibrated data from all setups were jointly imaged with the ``tclean" task in the CASA software \citep{2007ASPC..376..127M}. Briggs weighting with a value of 0.5 was adopted to achieve an optimal balance between sensitivity and angular resolution. Deconvolution of the spectral line cubes was performed using the Hogbom algorithm \citep{1974A&AS...15..417H}. A common restored beam of 3\rlap{.}\arcsec60$\times$3\rlap{.}\arcsec29 with a position angle of $-71$\degr\,was achieved for the spectral line data cube from the three frequency setups, enabling investigation of physical and chemical properties on linear scales of $\sim$0.06~pc toward NDO. To study the compact outflow, the CO (2--1) data were imaged individually, resulting in a smaller beam of 3\rlap{.}\arcsec19$\times$2\rlap{.}\arcsec42 with a position angle of $80.4$\degr. Based on the residual data cubes, the typical rms noise levels are estimated to be 19--368 mJy~beam$^{-1}$ with a median value of 62~mJy~beam$^{-1}$ at a channel width of $\sim$1.1~MHz across the observed frequency range. A broad-band continuum map centered at 236.351~GHz (i.e., 1.27~mm) was produced by using the line-free channels of the calibrated data from all frequency setups. The Multi-Term Multi-Frequency Synthesis algorithm \citep{2011A&A...532A..71R}, utilizing two Taylor coefficients, was adopted for deconvolution. The restored beam is 2\rlap{.}\arcsec88$\times$2\rlap{.}\arcsec47 with a position angle of $-$81.8\degree, and the rms noise level is $\sim$0.4~mJy~beam$^{-1}$.

\subsection{Image combination}
To compensate for the lack of short-spacing information in our SMA data, we incorporated our single-dish data to recover the extended emission. Specifically, for the dust continuum data, we extrapolated our IRAM 30~m continuum emission at 250~GHz to 236.351~GHz with a fixed dust spectral index of 1.7 (see Sect.~\ref{Sec:dust}). The extrapolated continuum data and APEX CO (2--1) data were subsequently integrated with our SMA data using the CASA ``feather" task. 

\subsection{Archival data}
Herschel Heterodyne Instrument for the Far-Infrared (HIFI) spectra of CO, HCN, and SiO were obtained from the Herschel science archive\footnote{\url{http://archives.esac.esa.int/hsa/whsa/}} as  supplementary data for our analysis. Details regarding these observations were presented by \citet{2016ApJ...828...51C}. 

\section{Results}\label{Sec:res}
\subsection{Dust emission}\label{Sec:dust}
\subsubsection{Single-dish results}\label{Sec:dust-single}
\begin{figure}[!htbp]
\centering
\includegraphics[width = 0.45 \textwidth]{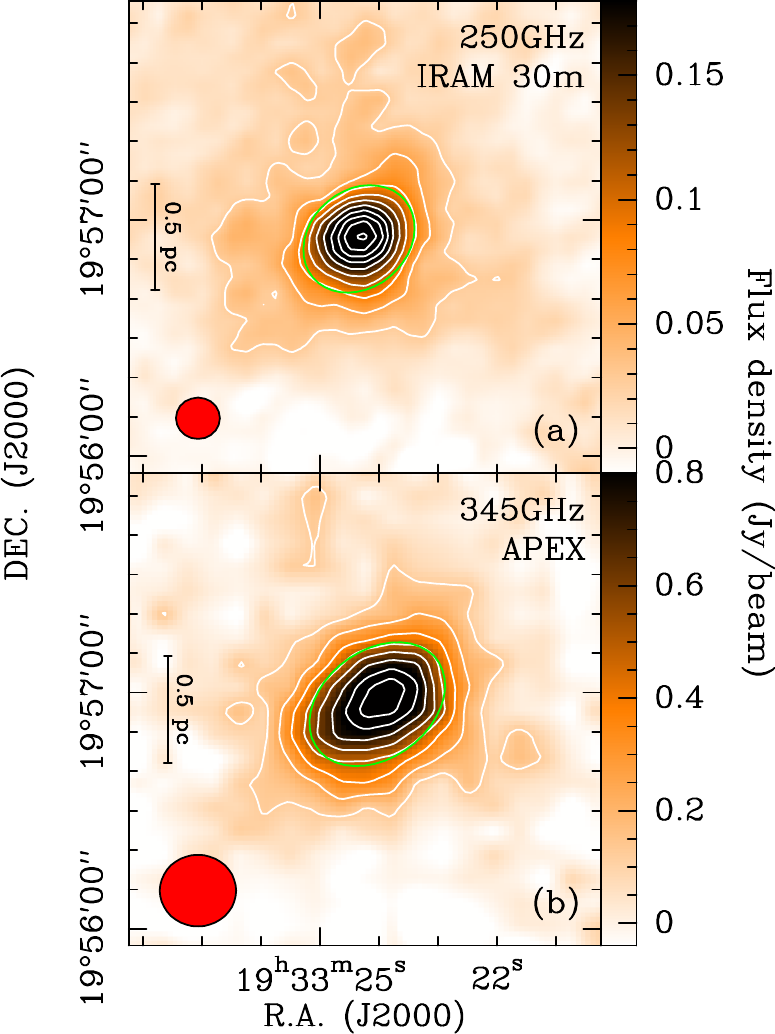}
\caption{{Continuum emission at 250~GHz (panel a) and 345~GHz (panel b) of NDO. Both are fitted with a single 2D Gaussian indicated by the green ellipses. Contour levels are from 3$\sigma$ in step of 3$\sigma$ (3$\sigma$ corresponds to 0.27~Jy~beam$^{-1}$ and 0.12~Jy~beam$^{-1}$ at 250 and 345 GHz, respectively). The beam size is shown in the lower-left corner of each panel.}\label{Fig:dust}}
\end{figure}

Figure~\ref{Fig:dust} shows the continuum emission of NDO at 1.2 mm and 870~$\mu$m. Both distributions reveal that this source is relatively isolated. Gaussian fitting was performed to determine the source sizes and flux densities, and the results are summarized in Table~\ref{Tab:size}. The obtained flux density (2.96$\pm$0.21~Jy) at 870~$\mu$m is in excellent agreement with the result (2.79~Jy) from the ATLASGAL survey \citep{2014A&A...565A..75C}. For both continuum emissions, the deconvolved full width at half maximum (FWHM) lengths of the major and minor axes agree with each other within 3$\sigma$. The effective FWHM size, defined as the square root of the product of the deconvolved major and minor FWHM axes, is determined to be in the range of 22\arcsec--25\arcsec\,with a value of 24\arcsec\,utilized for subsequent analysis. The fitted position angles (i.e., 139\degr$\pm$12\degr\,and 132.2\degr$\pm$6.4\degr\,at 250~GHz and 345~GHz, respectively) are also consistent within 3$\sigma$ at the two bands, demonstrating that this source exhibits elongation from the southeast to the northwest.

\begin{table*}[!hbt]
\caption{Results of Gaussian fits to the emission distribution in NDO.}\label{Tab:size}
\normalsize
\centering
\begin{tabular}{ccccccc}
\hline \hline
image                  & HPBW               & $\Delta \alpha_{\rm J2000}$  & $\Delta \delta_{\rm J2000}$ & $\theta_{\rm maj}$\tablefootmark{a} & $\theta_{\rm min}$\tablefootmark{a}  & P.A.\tablefootmark{a}   \\ 
                       & (\arcsec)           &  (\arcsec)               & (\arcsec)                     & (\arcsec)        & (\arcsec)         & (\degr, E of N)  \\ 
\hline                 
1.2 mm continuum      & 10.5           & $-$0.9$\pm$0.7           & 0.4$\pm$0.7                   & 27.7$\pm$2.0     & 21.6$\pm$1.6      & 139$\pm$12  \\ 
870~$\mu$m continuum  & 18.2           & $-$5.4$\pm$0.8           & 2.55$\pm$0.7                  & 30.0$\pm$2.5     & 16.6$\pm$1.8      & 128$\pm$6   \\
HCO$^{+}$ (3--2)        & 9.2            &  0.3$\pm$0.6             & 4.3$\pm$0.7                   & 27.6$\pm$1.7     & 26.0$\pm$1.6      & 147$\pm$42  \\
H$_{2}$CO (3$_{0,3}$--2$_{0,2}$) & 11.3    &  $-$0.5$\pm$1.0          & 1.8$\pm$0.7                   & 22.6$\pm$2.7     & 15.1$\pm$2.0      & 76$\pm$11   \\
SO (2$_{3}$--1$_{2}$)          & 24.8    &  $-$3.0$\pm$1.5           & 0.1$\pm$1.3                  & 26.3$\pm$5.2      & 19.5$\pm$5.0     &  86$\pm$27  \\
CH$_{3}$OH (3$_{0}$--2$_{0}$ A$^{+}$) & 17.0    &  $-$2.8$\pm$1.8           & 3.1$\pm$1.2                  & 25.9$\pm$5.2      & 15.3$\pm$4.3     &  90$\pm$15  \\
\hline
\end{tabular}
\tablefoot{\\
(a) These three columns give the deconvolved FWHM lengths of major and minor axes, and the position angles, respectively.}
\normalsize
\end{table*}

SED fitting can be used to study dust properties by assuming standard dust opacity laws typically used in ISM dust models. Although the SED fitting has been carried out by previous studies \citep[e.g.,][]{2016ApJ...828...51C,2016JKAS...49..261K}, their results did not provide substantial constraints on the cold dust component. In this study, we have revisited the SED to derive the dust temperature, $T_{\rm d}$, H$_{2}$ column density, $N_{\rm H2}$, and the dust emissivity index, $\beta$. Our analysis incorporates data from AKARI \citep{2007PASJ...59S.389K}, Herschel \citep{2016AA...591A.149M}, CSO \citep{2013ApJS..208...14G}, and our continuum observations at 250 and 345~GHz to constrain the SED at wavelengths longer than 60~$\mu$m. The flux densities from these observations are provided in Table~\href{https://zenodo.org/records/21234367}{A.1}. The emission at 160~$\mu$m was observed with both AKARI and Herschel. The comparison of both flux densities (see Table~\href{https://zenodo.org/records/21234367}{A.1}) suggests an uncertainty of 20\%\,in the absolute flux calibration which is also considered in the fitting. We assume a single-component modified blackbody emission model, which allows us to fit the SED with the following equations \citep[e.g.,][]{2008A&A...487..993K,2017ApJ...840...22L}:  
\begin{equation}\label{e.f1}
S_{\nu} = \Omega B_{\nu}(T_{\rm d})(1-e^{-\tau_{\nu}})\;,
\end{equation}
where $\Omega$ is the solid angle subtended by the source, $B_{\nu}(T_{\rm d})$ is the Planck function at a dust temperature of $T_{\rm d}$, and $\tau_{\nu}$ is the optical depth of the dust emission at the observed frequency; and  
\begin{equation}\label{e.f2}
N_{\rm H_{2}} = \frac{r_{\rm GDR}\tau_{\nu}}{\kappa_{\nu}\mu m_{\rm H}}\;,
\end{equation}
where the gas-to-dust ratio, $r_{\rm GDR}$, has been assumed to be 100 \citep[e.g.,][]{1974A&A....35..361K,1978ApJ...224..132B}, $N_{\rm H_{2}}$ is the H$_{2}$ column density, $\kappa_{\nu}$ is the dust opacity at the observed frequency, $\mu$ is the mean molecular weight which is taken to be 2.8, and $m_{\rm H}$ is the mass of the hydrogen atom. The dust opacity follows a relationship of $\kappa_{\rm \nu} = \kappa_{230}(\frac{\nu}{\rm 230})^{\beta}$ \citep{1983QJRAS..24..267H}, where $\kappa_{230}$ is 0.9 cm$^{2}$~g$^{-1}$ according to \citet{1994A&A...291..943O}, and $\nu$ is the observed frequency in GHz. 

\begin{figure}[!htbp]
\centering
\includegraphics[width = 0.45 \textwidth]{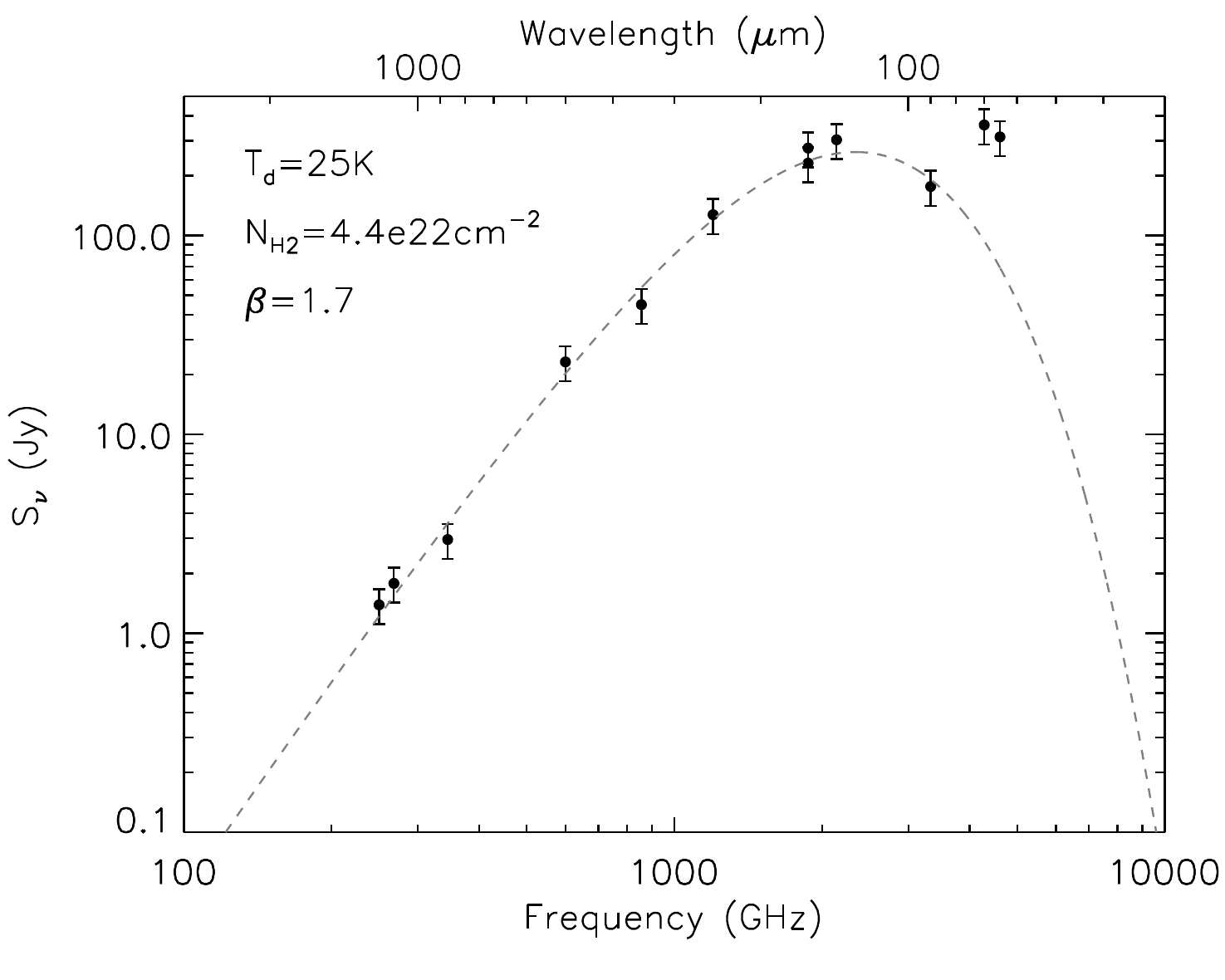}
\caption{{SED of NDO fitted with a single modified blackbody model shown by the dashed line. The fitted values are indicated in the upper-left corner of this panel, while the two data points at $\lambda<$80~$\mu$m are excluded from the fitting.}\label{Fig:sed}}
\end{figure}

The SED of NDO is shown in Fig.~\ref{Fig:sed}. Since the 65~$\mu$m and 70~$\mu$m emissions appear to deviate from the single-temperature modified-blackbody model, they are excluded from the fitting. This deviation may indicate the presence of an additional warmer dust component, indicating that the single-component fitting is a simplified approach. With this caveat, the fit to the data yields a dust temperature of $25\pm2$~K, a source-averaged (24\arcsec) column density of ($4.4\pm 0.6$)$\times 10^{22}$ cm$^{-2}$, and a dust emissivity index of 1.7$\pm$0.2. The dust emissivity index is broadly consistent with dust properties found in star-forming regions \citep[$\gtrsim$1.6; e.g.,][]{2013ApJ...767..126S}. Integrating over the solid angle subtended by the source, we obtain a total gas mass of 220$\pm$21~$M_{\odot}$ which is roughly consistent with the lower limit of the values inferred from wide-field $^{13}$CO observations \citep[225--478~$M_{\odot}$;][]{2016ApJ...825...16N}.

We note that these derived quantities are model dependent. In particular, the adopted dust opacity law is commonly used for ISM dust.  Alternative scenarios involving evolved objects may involve different dust compositions, grain-size distributions, and opacity laws \citep[e.g.,][]{1993ApJS...88..173K,2010A&A...513A..53L}. In such cases, the observed SED could in principle lead to different values of $\beta$, H$_{2}$ column density, and mass. Nevertheless, the detection of 
deuterated molecules in NDO supports the presence of cold interstellar material (see Sect.~\ref{sec:sd}). We therefore adopt the ISM dust-opacity prescription for the calculations in this work.

\subsubsection{SMA results}
\begin{figure*}[!htbp]
\centering
\includegraphics[width = 0.95 \textwidth]{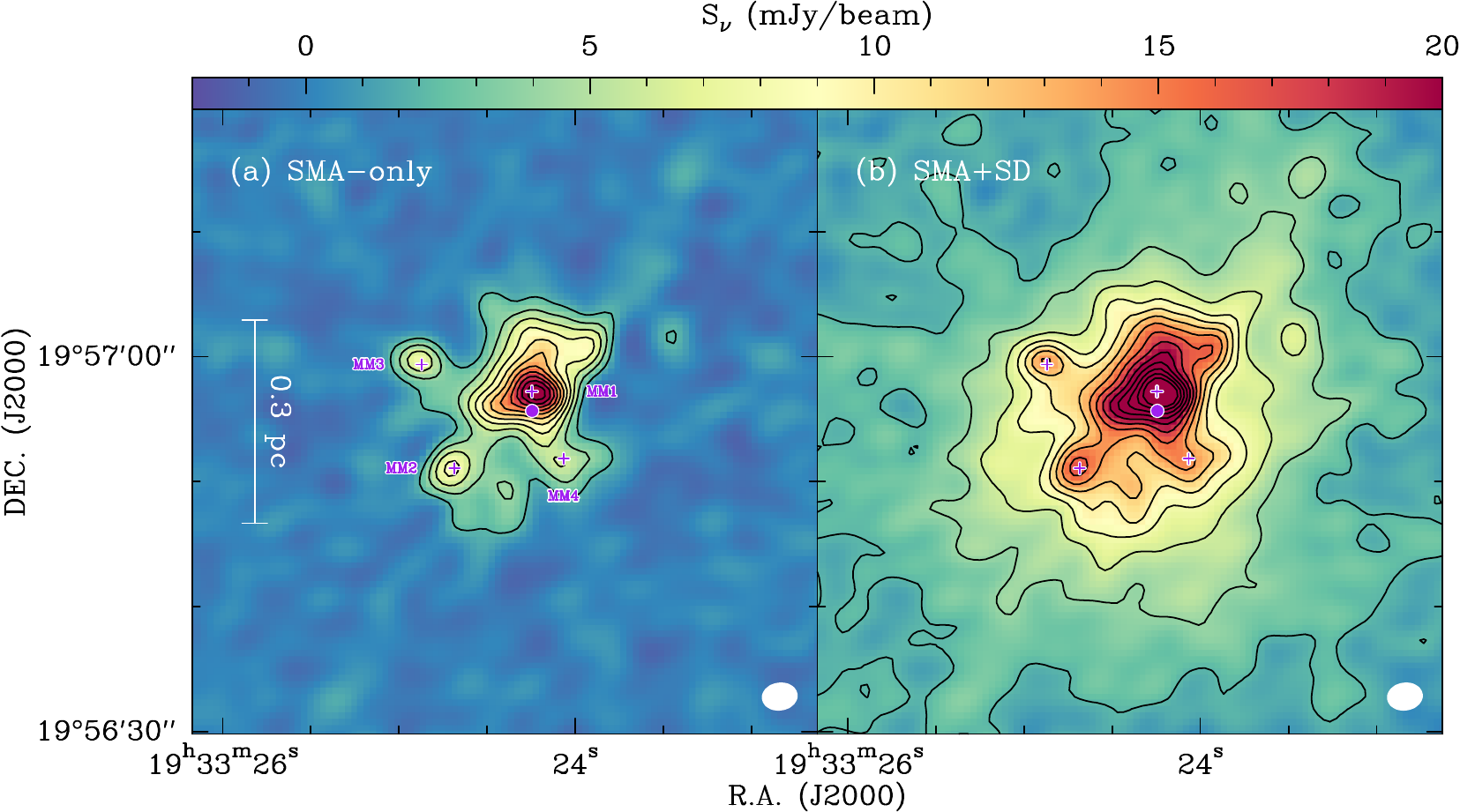}
\caption{{Comparison of the 236.351 GHz (i.e., 1.27~mm) dust continuum maps of NDO from the SMA-only (left panel) and from the SMA data combined with the IRAM~30m single-dish (SD) data, hereafter SMA+SD (right panel). In both panels, the contours start from 2~mJy~beam$^{-1}$ and increase by 2~mJy~beam$^{-1}$. The position of the SiO maser ($\alpha_{\rm J2000}$, $\delta_{\rm J2000}$)=(19$^{\rm h}$33$^{\rm m}$24\rlap{.}$^{\rm s}$245, 19$^{\circ}$56$^{\prime}$55\rlap{.}$^{\prime\prime}$689), taken from \citet{2011ApJ...728...76N}, is indicated by the purple circle, and the identified dust continuum cores are indicated by the purple plus symbols. The beam size is shown in the lower-right corner of each panel.}\label{Fig:sma-dust}}
\end{figure*}

Figure~\ref{Fig:sma-dust}a presents the SMA view of the 1.27~mm continuum emission of NDO. The high-angular-resolution observations have resolved the isolated structure in Fig.~\ref{Fig:dust} into at least four cores, namely MM1--MM4 in Fig.~\ref{Fig:sma-dust}a. The distribution is centrally peaked, with most of the emission originating in MM1. The other cores distributed around MM1 exhibit mm brightnesses that are at least three times lower than that of MM1. Such a morphology is similar to those observed in cluster-formation regions \citep[e.g.,][]{2013A&A...555A.112P,2023MNRAS.520.3259X,doi:10.1126/sciadv.ady6953}. The SiO maser appears to be associated with MM1, although the maser position is about 2\arcsec\, away from the 1.27~mm continuum peak of MM1. 

\begin{table*}[!hbt]
\caption{Observed and derived properties of dust continuum cores identified from the SMA 1.27~mm continuum emission.}\label{Tab:sma-dust}
\small
\centering
\begin{tabular}{cllccccccc}
\hline \hline
Source    & R.A. (J2000)\tablefootmark{a} & DEC. (J2000)\tablefootmark{a} & $\theta_{\rm maj}$\tablefootmark{b} &  $\theta_{\rm min}$\tablefootmark{b}  & P.A.\tablefootmark{b} & $S_{\nu}$ & $\int S_{\nu} {\rm d}\Omega$ & $N_{\rm H_{2}}$ & $M_{\rm c}$ \\ 
          & ($^{\rm h}$:$^{\rm m}$:$^{\rm s}$)         &(\degree:\arcmin:\arcsec)          &  (\arcsec) & (\arcsec)        & (\degree)  & (mJy~beam$^{-1}$) & (mJy)  & ($10^{22}$~cm$^{-2}$)     & (M$_{\odot}$)       \\ 
\hline                 
MM1      & 19:33:24.246(15)   & 19:56:57.20(17)   & 5.21$\pm$0.61 & 4.20$\pm$0.53  & 116$\pm$76 &  18.2$\pm$1.5 & 74.5$\pm$7.5&  6.4$\pm$0.5   & 13.3$\pm$1.3   \\
MM2      & 19:33:24.685(8)    & 19:56:51.06(12)   & 3.81$\pm$0.47 & 2.06$\pm$0.44 & 143$\pm$11 & 8.01$\pm$0.53 & 17.7$\pm$1.6 & 2.8$\pm$0.2  & 3.2$\pm$0.3   \\
MM3      & 19:33:24.871(11)   & 19:56:59.37(13)   & 3.26$\pm$0.59 & 1.88$\pm$0.66  & 48$\pm$21 & 6.60$\pm$0.55 & 12.8$\pm$1.5 & 2.3$\pm$0.2 & 2.3$\pm$0.3   \\
MM4      & 19:33:24.066(14)   & 19:56:51.83(14)   & 4.79$\pm$0.62 & 2.71$\pm$0.46 & 65$\pm$11 & 4.31$\pm$0.35 & 12.6$\pm$1.3 & 1.5$\pm$0.1   & 2.2$\pm$0.2 \\
\hline
\end{tabular}
\tablefoot{(a) Uncertainties in the last digits are given in parentheses. (b) These three columns give the fitted FWHM lengths of major and minor axes, and the position angles, respectively.}
\normalsize
\end{table*}

Figure~\ref{Fig:sma-dust} compares the SMA-only data and the SMA+IRAM30m combined data. The SMA+IRAM30m combined data exhibit higher flux densities and successfully recover extended emission. To further assess the difference in flux density, we compared cumulative flux densities at various radii. Remarkably, the SMA-only data demonstrate a significant deficiency in flux density, with a missing flux percentage ranging from 54\% to 86\% at an integration radius of $\leq$25\arcsec. This implies that the majority of the flux densities emanate from extended emission, with compact emission on smaller scales contributing to only a small portion of the total flux density.

We performed a Gaussian fitting of the SMA data to derive the positions, sizes, peak flux densities, and total flux densities of these dust cores. The results are listed in Table~\ref{Tab:sma-dust}. Adopting a dust spectral index of 1.7 and a dust temperature of 25~K (see Sect.~\ref{Sec:dust-single}), we derive the peak H$_{2}$ column density and core mass from the 1.27~mm dust continuum emission under the optically thin assumption with
\begin{equation}
    N_{\rm H_{2}} = \frac{r_{\rm GDR}S_{\nu}}{\Omega \kappa_{\nu}\mu m_{\rm H} B_{\nu}(T_{\rm d})} = 3.53\times 10^{24}\frac{S_{\nu}}{\rm Jy~beam^{-1}} {\rm cm}^{-2}  \;,
\end{equation}
and
\begin{equation}
    M_{\rm c} = \frac{d^{2}\int S_{\nu}{\rm d}\Omega}{\kappa_{\nu}B_{\nu}} = 178M_{\odot} \frac{\int S_{\nu}{\rm d}\Omega}{{\rm Jy}}\;,
\end{equation}
where $d$ is the distance to NDO. The derived properties are given in Table~\ref{Tab:sma-dust}. The peak H$_{2}$ column densities are in the range of (1.5--6.4)$\times 10^{22}$~cm$^{-2}$, and the core masses span from 2.2~M$_{\odot}$ to 13.3~M$_{\odot}$. Except for MM1, the other dust cores have masses of $<$4~M$_{\odot}$. This suggests that MM1 is the main mass reservoir in the SMA map. If the temperature in the outer cores is, however, lower than in the central one, then the masses of the outer cores would be higher.

Even ignoring the mass estimates, the distribution of continuum emission is atypical for low-mass evolved stars such as AGB stars. Although some massive evolved stars, such as red supergiants, are known to display clumpy circumstellar environments (e.g., VY CMa; \citealt{2019A&A...627A.114K}), the submillimeter luminosity of NDO far exceeds that of even the most luminous single red supergiants. In particular, when compared with VY~CMa \citep{2019A&A...627A.114K}, NDO is at least a factor of five more luminous at both 230 and 345~GHz. In this sense, the continuum data strongly support the view that a significant fraction of the surrounding material is interstellar rather than arising solely from a circumstellar envelope of a single evolved star.
However, the nature of the surrounding environment should be distinguished from the nature of the central source itself. The presence of nearby dust cores and dense gas does not exclude the possibility that the central object is related to a stellar merger or a red-nova-remnant-like event (with young progenitors), because such events may occur in close multiple systems and/or dense stellar environments. 

\subsection{Molecular emission}\label{sec:sd}
\subsubsection{Single-dish results}
We targeted 72 molecular transitions using the IRAM-30m and APEX telescopes (see Figs.~\href{https://zenodo.org/records/21234367}{B.1--B.10}), with 59 transitions exhibiting peak intensities exceeding 3$\sigma$. The detected transitions correspond to 12 species and their isotopologues: CO, HCN, HNC, CS, SiO, SO, HCO$^{+}$, c-C$_{3}$H$_{2}$, C$_{4}$H, H$_{2}$CO, HC$_{3}$N, CH$_{3}$OH. The transitions of SO$_{2}$, N$_{2}$H$^{+}$, and HDO were also observed, but none of them were detected by our single-dish observations. However, it is noted that N$_{2}$H$^{+}$ and SO$_{2}$ have already been reported by previous Nobeyama 45-m observations \citep{2004PASJ...56.1083D}.

Thanks to the high sensitivities achieved by our observations, we have successfully detected deuterated molecules (i.e., DCO$^{+}$, DCN, DNC, and HDCO) for the first time within this source. Previous attempts to identify these deuterated molecules were unsuccessful \citep[e.g.,][]{2023A&A...669A.121Q}, largely due to lower-sensitivity observations.

As mentioned in Sect.~\ref{sec.intro}, the observed line profiles exhibit a superposition of broad and narrow components. In order to derive the peak intensities, velocity centroids, and line widths of both components, we first masked out narrow components to perform a Gaussian fit to the broad component when it is present. Subsequently, Gaussian fits were applied to the residual spectra after subtracting the broad component to analyze the narrow components. For the $J$=1--0 and $J=$2--1 transitions of HCN, H$^{13}$CN and DCN which show nuclear quadrupole hyperfine structures (hfs), we used the hfs fitting method in CLASS to derive the observed properties of the narrow components. However, the hfs lines of higher $J$ HCN transitions are too close for a reliable fitting, so a single Gaussian fit was performed instead. The line widths derived from the Gaussian fit are broader than those obtained with the hfs fitting to low $J$ transitions, likely caused by opacity broadening and blending of the underlying hfs components in the higher-$J$ transitions. For undetected lines, $3\sigma$ values are used as upper limits for their peak intensities. The observed parameters for all spectra are given in Tables~\href{https://zenodo.org/records/21234367}{A.2--A.9}.

The broad component is observed in spectra of CO, $^{13}$CO, CS, SO, HCO$^{+}$, HCN, SiO, and H$_{2}$CO (see Figs.~\href{https://zenodo.org/records/21234367}{B.1--B.8}). 
 As shown in Tables~\href{https://zenodo.org/records/21234367}{A.2--A.9}, 
 the fitted velocity centroids vary from 30.9 to 36.5~\kms\,and the fitted line widths vary from 9.1 to 43.6~\kms. CO (1--0), CO (2--1), CO (3--2), and HCO$^{+}$ (1--0) exhibit FWZPs of 100--160~\kms\, (see Figs.~\href{https://zenodo.org/records/21234367}{B.1 and B.2}, 
consistent with observations of higher-$J$ CO lines \citep{2016ApJ...828...51C}. As demonstrated by previous interferometric studies \citep{2004ApJ...610L..41N,2005ApJ...633..282N}, the broad-component emitting region stems from the outflow emanating from NDO. Additionally, the broad components of HCN, SiO, and SO are more prominent than their narrow components (see Figs.~\href{https://zenodo.org/records/21234367}{B.3, B.5, and B.7}), suggesting that HCN, SiO, and SO are 
enhanced in the shocked outflowing gas. Such enhancements are commonly observed in molecular outflows \citep[e.g.,][]{1997ApJ...487L..93B}. In particular, SiO can be efficiently produced when shocks sputter grain 
mantles and/or erode grain cores, releasing Si-bearing material into the gas phase, where it is rapidly converted into SiO 
\citep[e.g.,][]{1997A&A...321..293S}. HCN may also be enhanced when grain-surface material is sputtered and released into the gas phase, causing a rapid increase in its gas-phase abundance \citep[e.g.,][]{2021MNRAS.507.1034L}. Similarly, SO can be enhanced when shocks release sulfur-bearing material from grains and drive warm gas-phase chemistry in the post-shock gas. \citep[e.g.,][]{1993MNRAS.262..915P,2001A&A...372..899B}.

In addition to the broad component, our spectra reveal multiple narrow components. Figure~\href{https://zenodo.org/records/21234367}{B.1}
highlights five CO narrow components indicated by the five blue dashed lines, with LSR velocities at about $-$4.4, 29.1, 33.1, 36.5, and 44.5~\kms, respectively. The $-$4.4~\kms\,component is seen in CO (1--0), but is absent in other higher-$J$ CO spectra (see Fig.~\href{https://zenodo.org/records/21234367}{B.1}),
implying that this velocity component comes from an offset position that is not covered by smaller beams of other CO transitions. Based on the parallax-based distance estimator \citep{2016ApJ...823...77R}, we find that this unrelated cloud would be located at a distance of $\sim$9.6~kpc. The 33.1~\kms\,component is seen in CO (1--0) and CO (2--1). This component may also contribute to $^{13}$CO (1--0) and $^{13}$CO (2--1), making them deviate from the Gaussian profile (see Fig.~\href{https://zenodo.org/records/21234367}{B.1}).
The 29.1, 36.5, and 44.5~\kms\,components, denoted as N1, N2, and N3, respectively, are more commonly detected in our CO observations. N2 exhibits the highest number of molecular lines, followed by N1. Self-absorption features near N2 and N3 have been reported in HIFI CO and H$_{2}$O spectra \citep{2016ApJ...828...51C}, similar to dips observed in our low-$J$ CO and HCO$^{+}$ spectra.

In the frequency range of 104660--104710~MHz, six C$_{4}$H hfs components form three spectral features which are indicated by red arrows in the top panel of Fig.~\href{https://zenodo.org/records/21234367}{B.9}.
Among these, the two lines at the rest frequencies of 104666~MHz and 104705~MHz have signal-to-noise ratios of $\sim$3. To improve the signal-to-noise ratio, we average the two lines in velocity space. The velocity-stacked C$_{4}$H spectrum is shown in the bottom panel of Fig.~\href{https://zenodo.org/records/21234367}{B.9} 
where its signal-to-noise ratio improves to $\sim$4. The fitted velocity and line width of the velocity-stacked C$_{4}$H spectrum are consistent with the N2 component derived from other lines (see Table~\href{https://zenodo.org/records/21234367}{A.7}),
confirming the presence of C$_{4}$H that was tentatively detected by \citet{2015PASJ...67...95N}.  

In this study, we have one unidentified line with the rest frequency determined to be 96744~MHz, assuming an LSR velocity of 36.5~\kms. This unidentified narrow line has a line width of 0.93$\pm$0.16~\kms. This line likely overlaps with CH$_{3}$OH (2$_{0}$--1$_{0}$ E) in \citet{2015PASJ...67...95N} due to their coarser spectral resolution. This leads to the fact that their derived velocity ($\sim$39.1~\kms) for CH$_{3}$OH (2$_{0}$--1$_{0}$ E) is more redshifted than our value (35.9$\pm$0.2~\kms; see Table~\href{https://zenodo.org/records/21234367}{A.10}), but the origin of this line remains elusive.

\begin{figure}[!htbp]
\centering
\includegraphics[width = 0.45 \textwidth]{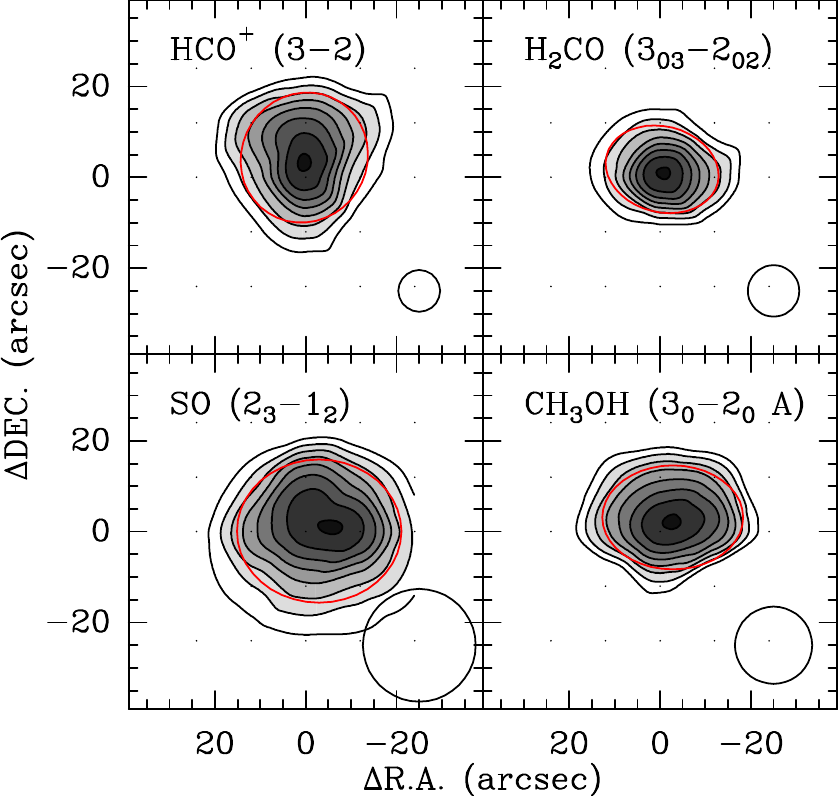}
\caption{{Raster maps of HCO$^{+}$ (3--2), H$_{2}$CO (3$_{0,3}$-2$_{0,2}$), SO (2$_{3}$--1$_{2}$), and CH$_{3}$OH (3$_{0}$-2$_{0}$ A$^{+}$) integrated over 33--40~\kms. The black dots represent the footprints of the grid. The distribution of each transition is fitted with a two-dimensional Gaussian model which is indicated by the red ellipse in each panel. The beam size is shown in the lower-right corner of each panel. The (0, 0) offset corresponds to ($\alpha_{\rm J2000}$=19$^{\rm h}$33$^{\rm m}$24$\rlap{.}^{\rm s}$40, $\delta_{\rm J2000}$=19\degr56\arcmin54$\rlap{.}^{\prime\prime}$8). In all panels, contour levels correspond to 30\%, 40\%, 50\%, 60\%, 70\%, 80\%, 90\%, 99\% of the peak integrated intensities which are 4.94, 1.62, 2.49, and 0.90 K~\kms\,for HCO$^{+}$ (3--2), H$_{2}$CO (3$_{0,3}$-2$_{0,2}$), SO (2$_{3}$--1$_{2}$), and CH$_{3}$OH (3$_{0}$-2$_{0}$ A$^{+}$), respectively.}\label{Fig:map}}
\end{figure}

Raster maps of HCO$^{+}$ (3--2), H$_{2}$CO (3$_{0,3}$--2$_{0,2}$), SO (2$_{3}$--1$_{2}$) and CH$_{3}$OH (3$_{0}$--2$_{0}$ A) are presented in Fig.~\ref{Fig:map}. Compact morphologies are seen in all four lines, but the peaks and sizes are slightly different from each other. To quantify the differences, we employed a two-dimensional Gaussian model to fit these distributions. The fitted results are summarized in Table~\ref{Tab:size} and displayed by red ellipses in Fig.~\ref{Fig:map}. These peak positions and deconvolved FWHM sizes are consistent with each other within 3$\sigma$.

\subsubsection{SMA results}\label{sec:sma-line}
\begin{figure*}[!htbp]
\centering
\includegraphics[width = 0.95 \textwidth]{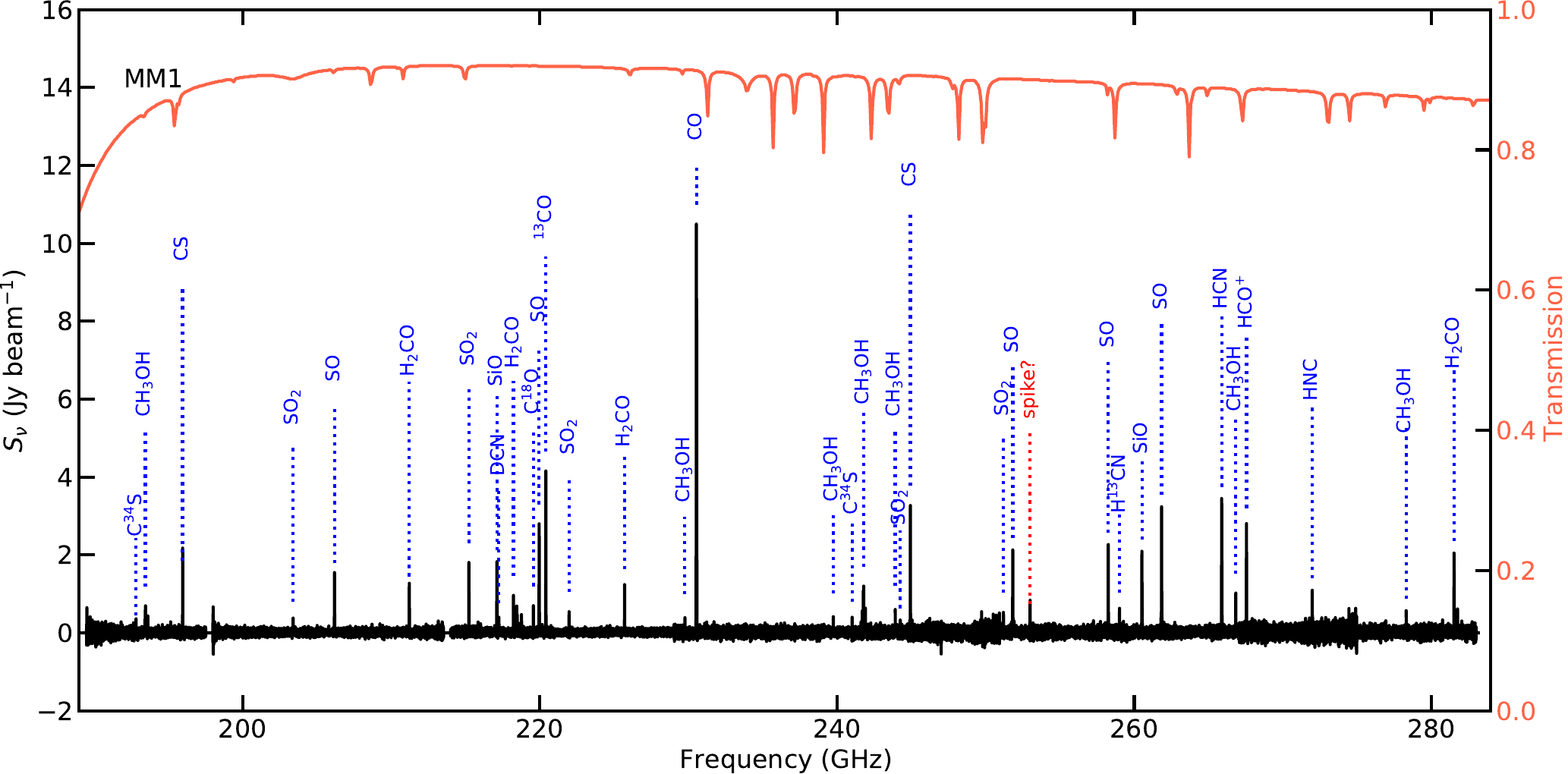}
\caption{{SMA view of detected molecular lines within the frequency range of 190--280~GHz toward MM1. The atmospheric transmission with a precipitable water vapor of 2~mm is shown in orange. Zoom-in plots are provided in Fig.~\href{https://zenodo.org/records/21234367}{C.1}.}\label{Fig:MM1-line}}
\end{figure*}

Due to the limited resolution of our single-dish observations ($\gtrsim$0.2~pc) in resolving molecular species distributions, we investigate the chemical compositions on a small scale of $\sim$0.06~pc using the SMA observations. An overview of the SMA spectra toward MM1 is presented in Fig.~\ref{Fig:MM1-line}. We detected 75 spectral features that can be assigned to 12 molecules (i.e., CO, CS, SO, SiO, SO$_{2}$, HCN, HNC, HCO$^{+}$, C$_{2}$H, H$_{2}$CO, HC$_{3}$N, and CH$_{3}$OH) and their isotopologues. The deuterated molecule DCN is also well detected by our SMA observations (see Fig.~\ref{Fig:SMA-dcn}), confirming the presence of deuterium close to the infrared source on scales of $\lesssim$0.1~pc. We have detected lines with upper energy levels of $>$100~K, indicative of warm gas. The spectral line density is 0.8 lines per GHz, well below the line confusion limit, which allows further investigations using more sensitive observations in the future. For lines exhibiting deviations from a single Gaussian profile, we derived only their integrated intensities. For blended lines, we calculated the integrated intensities of all blended components as the upper limits. For all other lines, we fitted a single Gaussian component to determine their observed characteristics. Observed properties of the spectral lines detected in our SMA observations are summarized in Table~\href{https://zenodo.org/records/21234367}{C.1}. 

\begin{figure*}[!htbp]
\centering
\includegraphics[width = 1.0 \textwidth]{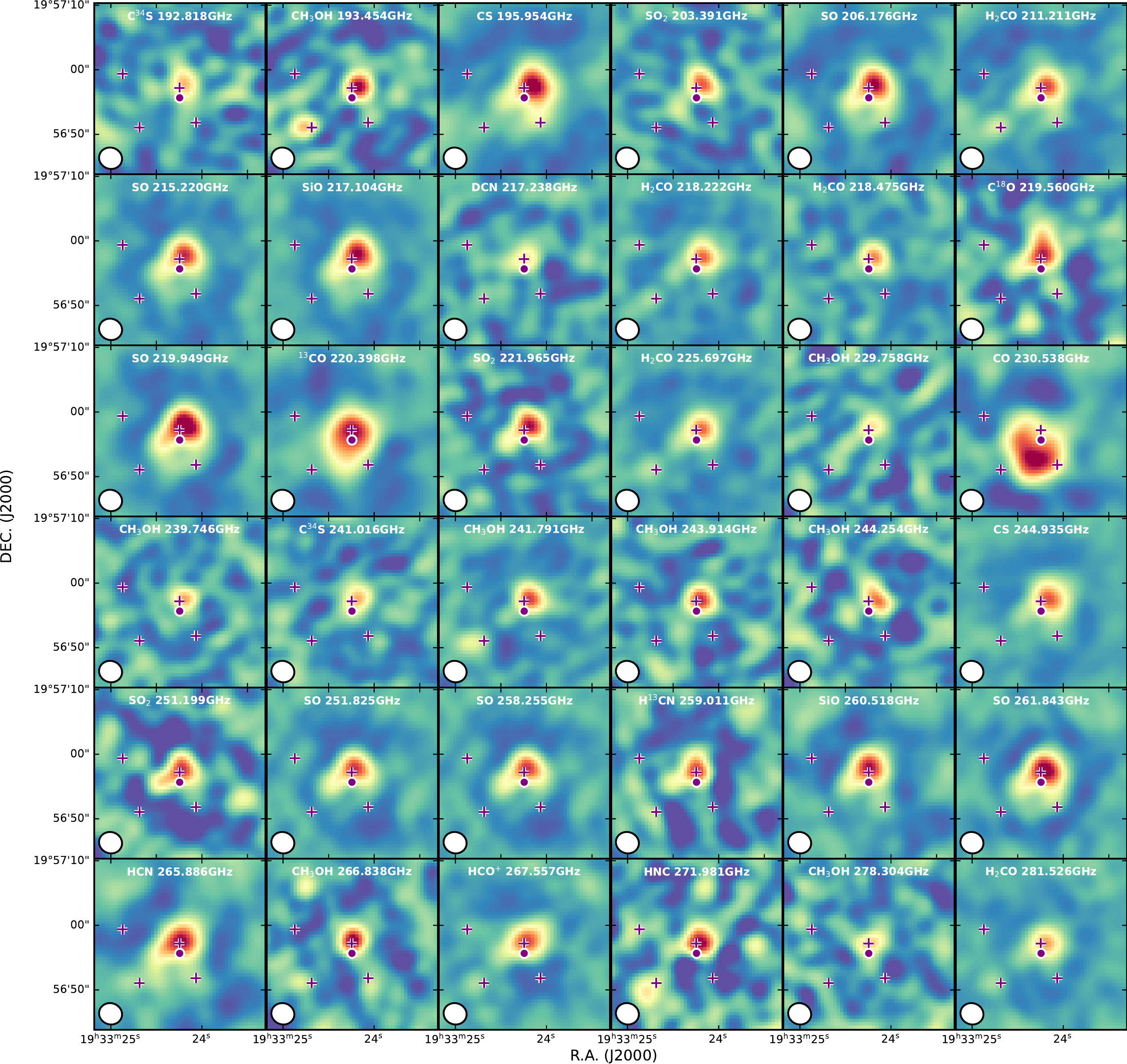}
\caption{{Distribution of the 36 selected molecular transitions observed by the SMA. The maps were integrated from 30~\kms\,to 45~\kms. The dust continuum peaks are indicated by the plus symbols, while the SiO maser position is indicated by the circle. The beam size is shown in the lower-left corner of each panel.}\label{Fig:SMA-mom0}}
\end{figure*}

\begin{figure}[!htbp]
\centering
\includegraphics[width = 0.49 \textwidth]{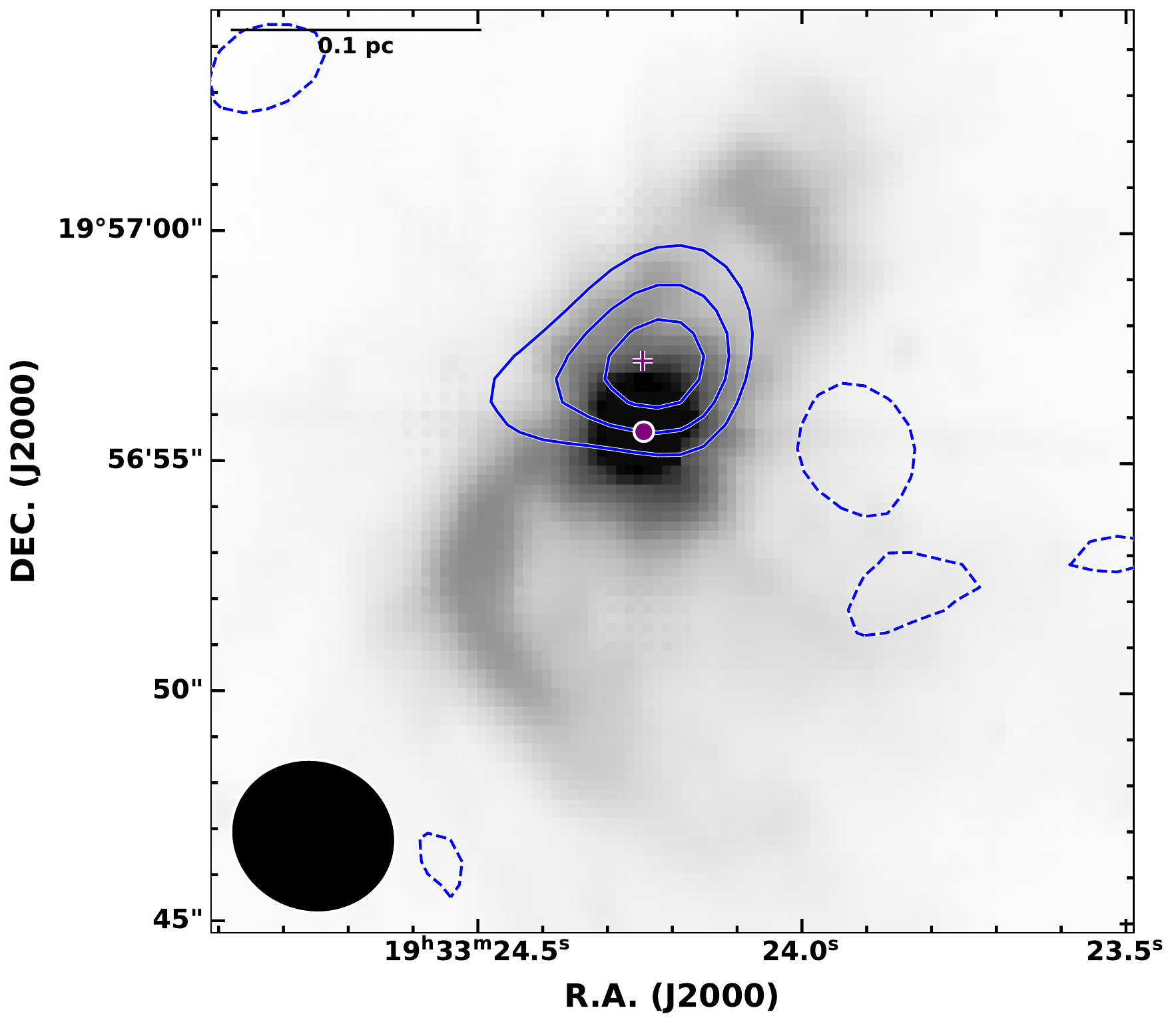}
\caption{{UKIRT 2.12~$\mu$m image overlaid with the SMA DCN (3--2) intensity contours integrated over 30~\kms\, to 45~\kms. The contours are $-$3$\sigma$, 3$\sigma$, 5$\sigma$, and 7$\sigma$ (1$\sigma$=0.3~Jy~beam$^{-1}$~\kms). The dust continuum peak is indicated by the plus symbol, while the SiO maser position is indicated by the circle. The beam size is shown in the lower-left corner of each panel.}\label{Fig:SMA-dcn}}
\end{figure}

Among the detected lines, we selected 36 bright molecular transitions to investigate their spatial distributions, as shown in Fig.~\ref{Fig:SMA-mom0}. We find that all the prominent emissions arise from a compact zone within a radius of $\lesssim$5\arcsec, suggesting that these emissions are likely primarily associated with MM1. However, the morphologies of different molecular transitions exhibit diversity and do not precisely coincide with the 1.27~mm dust continuum emission mapped by the SMA.

The distributions of CO (2--1) and $^{13}$CO (2--1) are more extended than other transitions due to their low energies and high abundances. CO (2--1) exhibits a distinct morphology with its peak located south of the dust peak, which can be attributed to the outflow enhancement and the presence of self-absorption (see Sect.~\ref{sec.outflow} for more details).  Peaks of several shock tracers, such as SiO (5--4), SO$_{2}$ ($11_{1,11}-10_{0,10}$), SO ($6_{5}-5_{4}$), and H$_{2}$CO ($3_{1,3}-2_{1,2}$), also do not coincide with the dust peak. Instead, they converge at a position $\sim$1\arcsec\, northwest of the dust peak.   

An elongation extending southeast from the dust peak is observed in the transitions of CS, SO, SO$_{2}$, H$_{2}$CO, SiO, HCN, and HCO$^{+}$. This elongation, oriented at a position angle of $\sim$135\degr, is similar to the NIR southeast-northwest elongation direction (see Sect.~\ref{sec.outflow} for instance). This pattern is reminiscent of jet-like features, but its nature is still to be investigated. 

\subsection{CO outflows revisited}\label{sec.outflow}
\begin{figure*}[!htbp]
\centering
\includegraphics[width = 0.95 \textwidth]{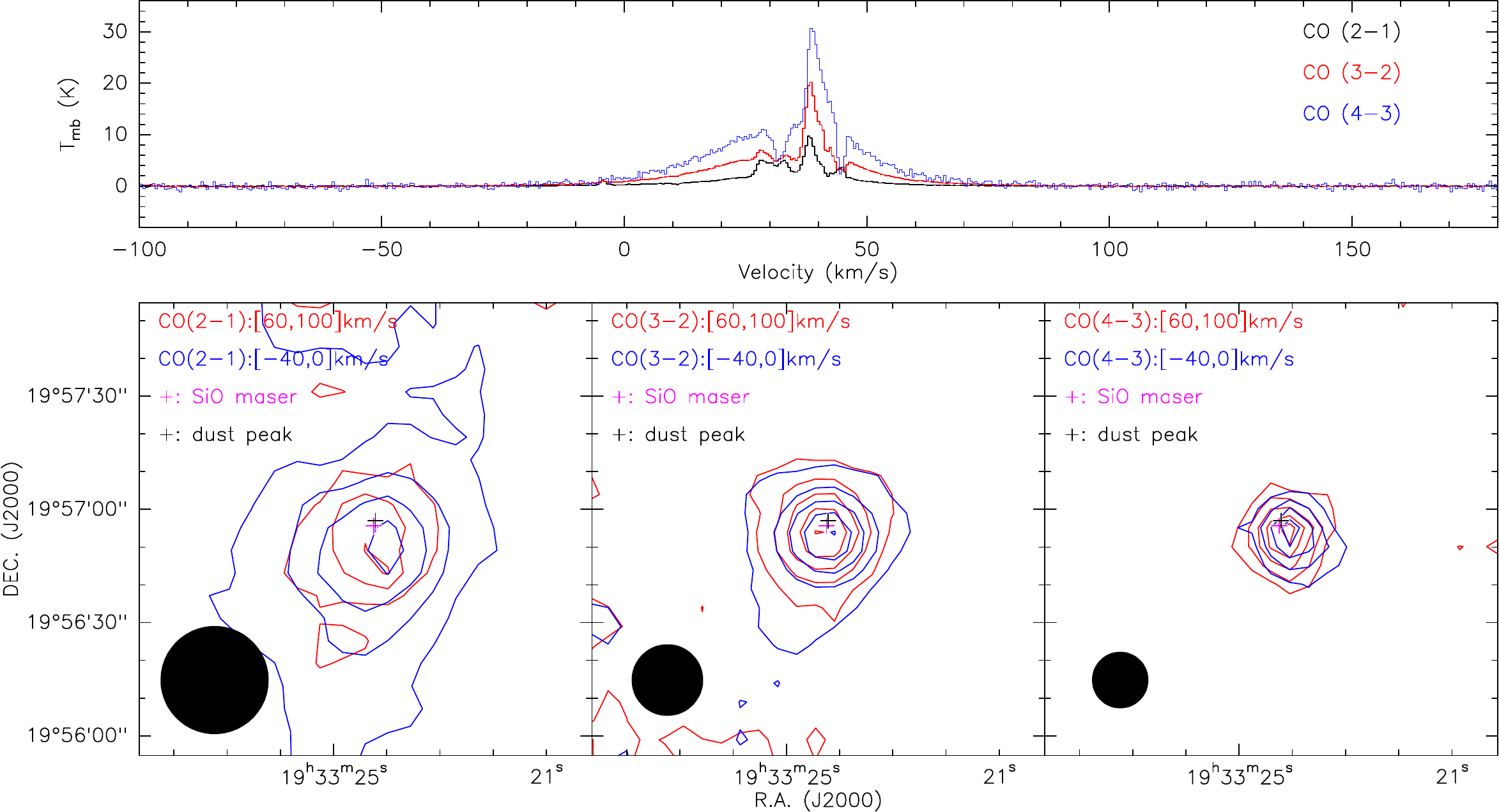}
\caption{{APEX view of the outflow associated with NDO. \textit{Top}: Observed CO (2--1), (3--2), and (4--3) spectra extracted at the position corresponding to the beam centered near the crosses shown in the bottom panels. \textit{Bottom}: CO outflow maps. The blueshifted emission integrated from $-$40 to 0~\kms\,is shown in blue contours. For CO (2--1), the contours start at 1~K~\kms, and increase by 1~K~\kms. For CO (3--2), the contours start at 1~K~\kms, and increase by 2~K~\kms. For CO (4--3), the contours start at 3~K~\kms, and increase by 3~K~\kms. The redshifted emission integrated from 60 to 100~\kms\,is shown in red contours. For CO (2--1), the contours start at 1~K~\kms, and increase by 1~K~\kms. For CO (3--2), the contours start at 1~K~\kms, and increase by 2~K~\kms. For CO (4--3), the contours start at 3~K~\kms, and increase by 3~K~\kms. The pink and black plus symbols mark the positions of the SiO maser and MM1, respectively. The beam size is shown in the lower-left corner of each panel.}\label{Fig:apex-outflow}}
\end{figure*}

Figure~\ref{Fig:apex-outflow} shows the APEX view of the molecular outflow from NDO in CO (2--1), (3--2), and (4--3). In the top panel, the high-velocity wings are evident in the CO transitions and the blueshifted wing is more prominent, which is consistent with the presence of outflows suggested by previous observations \citep[e.g.,][]{2004ApJ...610L..41N,2005ApJ...633..282N,2016ApJ...828...51C}. The blueshifted and redshifted outflow lobes appear coincident along the line of sight, typically attributed to the quasi-spherical outflows of AGB stars.

To characterize the outflow on a smaller scale, we constructed the SMA+APEX combined CO (2--1) maps. Figure~\href{https://zenodo.org/records/21234367}{C.2} 
compares the SMA-only and SMA+APEX CO (2--1) data, highlighting significant missing zero-spacing fluxes for the extended emission at LSR velocities near the systemic velocity of $\sim$36.5~\kms. However, this issue is negligible for the high-velocity outflow components at LSR velocities around $\sim$15~\kms\,away from the systemic velocity. To mitigate potential contamination from the extended molecular cloud component in our subsequent outflow analysis, we focus primarily on the high-velocity outflow components.

\begin{figure*}[!htbp]
\centering
\includegraphics[width = 1.0 \textwidth]{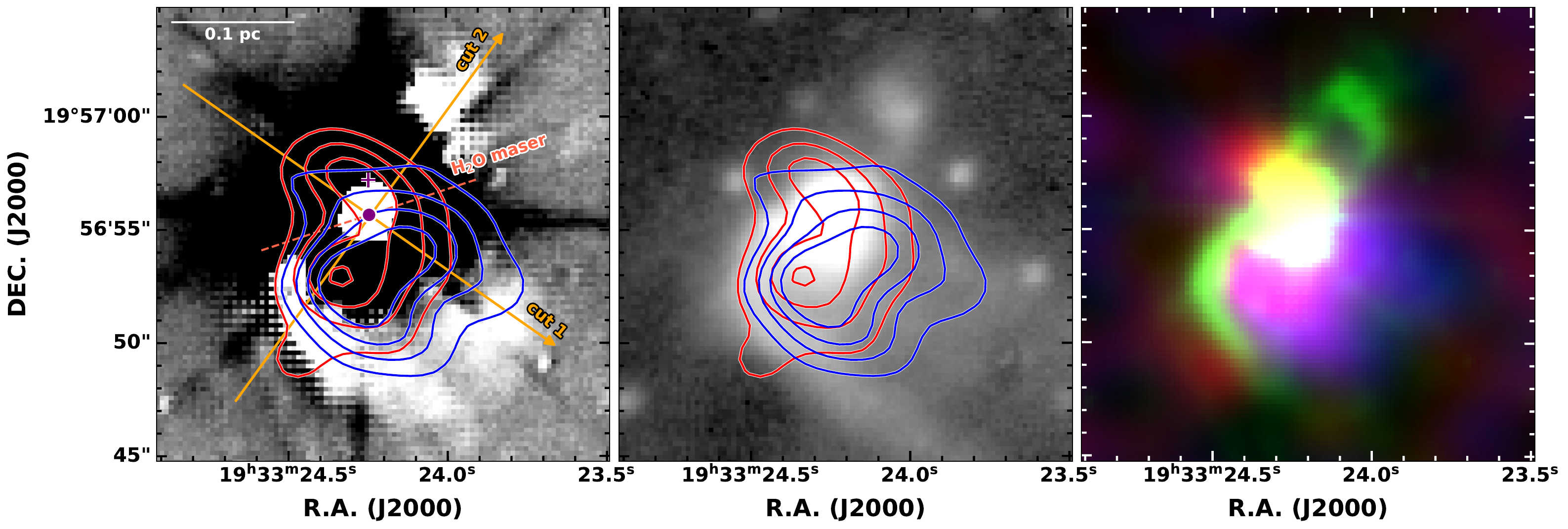}
\caption{{Overview of the outflow associated with NDO. \textit{Left:} UKIRT continuum-subtracted 2.12~$\mu$m H$_{2}$ v$=1-0$ S(1) image overlaid with the extremely redshifted and blueshifted CO (2--1) lobes. The redshifted lobe, shown in the red contours, is the CO (2--1) intensity map integrated from 60 to 100~\kms, while the blueshifted lobe, shown in the blue contours, is the CO (2--1) intensity map integrated from $-40$ to 0~\kms. The contours of both lobes start at 5~Jy~beam$^{-1}$~\kms, and increase by 5~Jy~beam$^{-1}$~\kms. The purple plus and circle symbols indicate the dust peak and the SiO maser position, respectively. The red dashed line marks the position angle defined by the H$_{2}$O masers \citep{2011ApJ...728...76N}, while the two orange arrows mark the cuts used to construct position-velocity diagrams (see Fig.~\ref{Fig:outflow-pv}). \textit{Middle:} Similar to the left panel but for the UKIRT $J$ band image. \textit{Right:} Three-color composite image of NDO with the SMA+APEX CO (2--1) redshifted lobe shown in red, the UKIRT 2.12~$\mu$m image in green, and the SMA+APEX CO (2--1) blueshifted lobe shown in blue. All the NIR images are based on the UKIRT Wide-field Infrared Survey for H$_{2}$ \citep[UWISH2,][]{2011MNRAS.413..480F}}\label{Fig:sma-outflow}}
\end{figure*}

Figure~\ref{Fig:sma-outflow} illustrates the SMA+APEX view of the outflow in CO (2--1). Owing to the higher angular resolution and sensitivity of our SMA observations, our outflow map surpasses previous interferometric observations \citep{2004ApJ...610L..41N,2005ApJ...633..282N}. In this figure, the dust peak is offset from the center of the outflow lobes. Instead, the NIR source associated with the SiO maser is closer to the centriod of the overlapping region of the blueshifted and redshifted lobes, suggesting that the NIR source is hiding the dynamical source of the outflows. This morphology is inconsistent with the spherical morphology of high-velocity components proposed by earlier studies \citep{2004PASJ...56..193N,2005ApJ...633..282N}, resembling more a wide-angle bipolar outflow with a direction different from the NIR's long-axis orientation and the position angle (i.e., $\sim$108\degree; see the red dashed line in Fig.~\ref{Fig:sma-outflow}) defined by the H$_{2}$O masers \citep{2011ApJ...728...76N}. This bipolar morphology is also different from the low-velocity bipolar structure reported by previous studies \citep[e.g., see Fig.~2 in][]{2004ApJ...610L..41N}. It should be noted that the low-velocity structure may be contaminated by extended cloud components. Such issues are not expected in the high-velocity components, which can therefore better trace the intrinsic outflow structure. Furthermore, the extended NIR emission traces heated dust on the surface of the molecular cloud, likely associated with the driving of the observed outflow. The continuum-subtracted UKIRT H$_2$ emission reveals a curved structure to the south and a more compact morphology to the north. The curved H$_2$ structure is better aligned with the southern outflow lobe than the northern one. This indicates that the southern lobes are directed toward the observer, whereas the northern counterpart is receding into denser molecular material. 

\begin{figure*}[!htbp]
\centering
\includegraphics[width = 0.95 \textwidth]{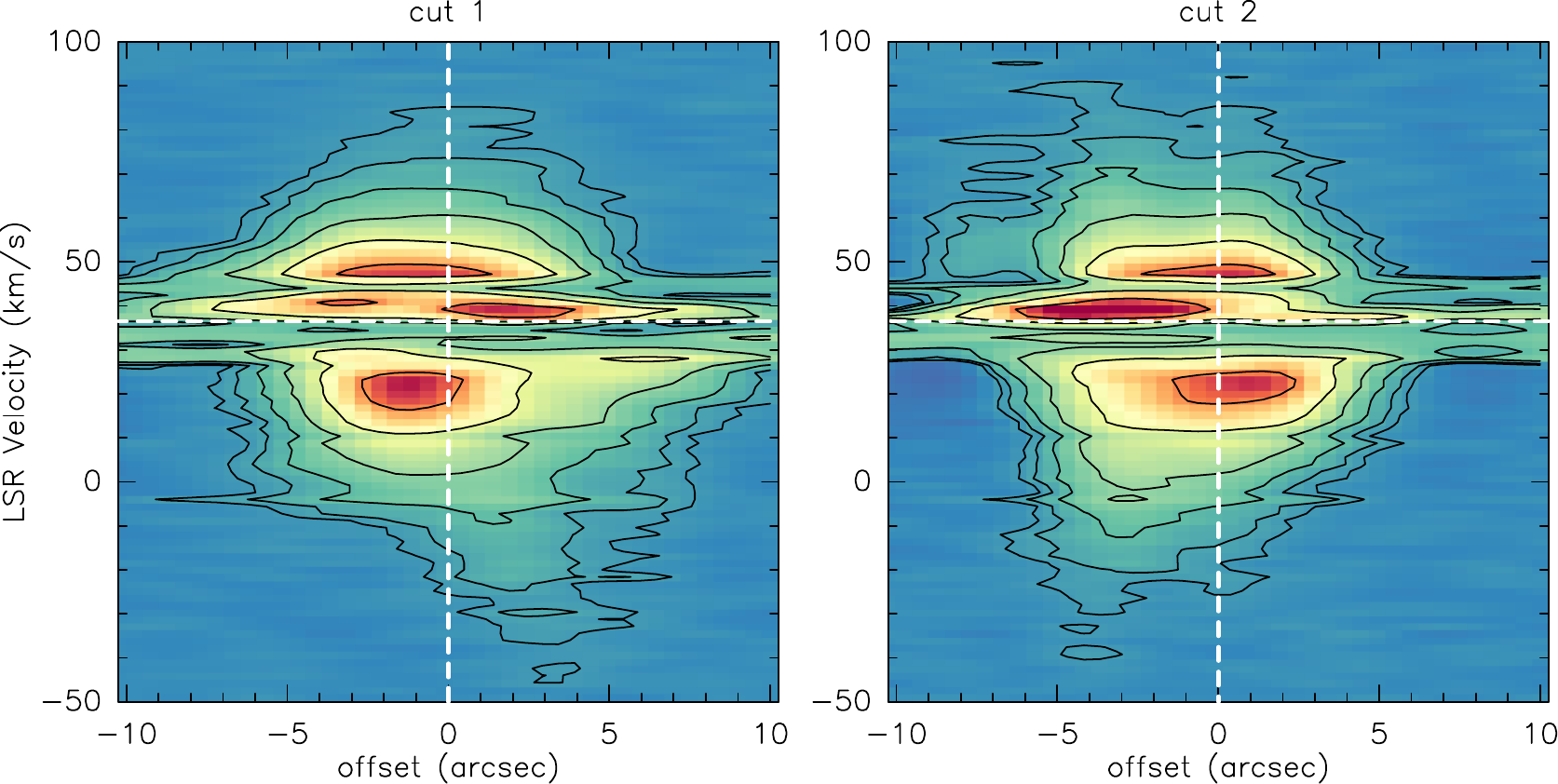}
\caption{{Position-velocity diagrams of the SMA+APEX CO (2--1) emission along the cuts indicated by the two orange arrows in the left panel of Fig.~\ref{Fig:sma-outflow}. The contours start at 0.18~Jy~beam$^{-1}$ and increase by 0.18~Jy~beam$^{-1}$ in both panels. The dashed horizontal line indicates the systemic velocity of NDO, and the dashed vertical line marks the SiO maser position. The interactive version of the 3D view of the outflow can be accessed via \href{https://gongyan2444.github.io/NDO-CO21-SMA-APEX-3D.html}{this link}.}\label{Fig:outflow-pv}}
\end{figure*}

Figure~\ref{Fig:outflow-pv} presents the position-velocity diagrams along the two cuts indicated in Fig.~\ref{Fig:sma-outflow}. An intensity drop is seen around the systemic velocity due to self-absorption. On the other hand, the outflow exhibits a velocity span of $\sim$130~\kms, but the high-velocity wing structures are only evident in the offset ranging from $-$5\arcsec to 5\arcsec\,(i.e., within a size of $\sim$0.2~pc). This suggests that the outflow is both fast and compact. The blueshifted wing structure is confirmed to be more powerful than its redshifted counterpart on the small scale. Additionally, the 1612 MHz OH masers are detected only at LSR velocities of 10--32~\kms\,\citep{2011ApJ...728...76N}, implying an association with the blueshifted wing structure and suggesting that they may be induced by outflow-driven shocks. The current angular resolution is insufficient to fully resolve the outflow morphology, highlighting the need for higher-resolution observations.

\subsection{Non-detection of radio continuum emission}\label{sec.radio}
Based on infrared observations, \citet{2016ApJ...828...51C} suggested that NDO should be classified as a massive YSO. Such massive YSOs could have already ionized ambient gas to form H{\scriptsize II} regions which are readily detected by radio continuum emission. However, \citet{2016ApJ...828...51C} reported a non-detection of radio continuum emission from the Co-Ordinated Radio 'N' Infrared Survey for High-mass star formation (CORNISH) survey, placing a 3$\sigma$ upper limit of 10$^{45}$~s$^{-1}$ on the total ionizing photon flux. In this study, we revisit the radio continuum emission using more sensitive radio continuum data from the GLObal view of STAR formation (GLOSTAR) survey \citep{2021A&A...651A..85B}. The radio continuum image reveals no point sources within a 2\arcmin\,field of view centered on NDO, confirming the absence of detectable H{\scriptsize II} regions in this area. The 1$\sigma$ sensitivity is 40~$\mu$Jy~beam$^{-1}$ for the GLOSTAR 5.8 GHz continuum emission. Assuming a typical electron temperature of 10$^{4}$~K and optically thin radio continuum emission, we derive a 3$\sigma$ upper limit of 1.8$\times$10$^{44}$~s$^{-1}$ for the Lyman ionizing photon flux from the GLOSTAR 5.8~GHz continuum data, which is five times lower than the value reported by \citet{2016ApJ...828...51C}. Our upper limit on the Lyman ionizing photon flux is lower than the expected value (4.5$\times$10$^{44}$~s$^{-1}$) for a single zero-age main-sequence star with a spectral type of B2 \citep{1973AJ.....78..929P}. We also do not detect mm-wave hydrogen recombination lines in our SMA spectra (see Fig.~\ref{Fig:MM1-line}).

\section{Discussion}\label{Sec:dis}
\subsection{Molecular column densities and fractional abundances}\label{Sec.col} 
We utilized the rotation-diagram method to estimate the rotational temperatures and column densities of the detected molecules. Assuming local thermodynamic equilibrium (LTE) conditions, we adopted the standard formula \citep[e.g.,][]{1999ApJ...517..209G,2009ARA&A..47..427H}:
\begin{equation}\label{f.rd}
    N_{\rm u}/g_{\rm u} = \frac{N_{\rm tot}}{Q(T_{\rm rot})}{\rm exp}(-E_{\rm u}/kT_{\rm rot}) = \frac{3k\int T_{\rm mb}{\rm d}\varv}{8\pi^3\nu\mu^2 S_{\rm ul}}
\end{equation}
where $N_{\rm tot}$ represents the molecular column density, $Q(T_{\rm rot})$ denotes the partition function at the rotational temperature $T_{\rm rot}$, $\int T_{\rm mb} {\rm d}\varv$ denotes the velocity-integrated intensity, $k$ represents the Boltzmann constant, $\nu$ stands for the transition's rest frequency, $\mu$ is the permanent dipole moment of the molecule, and $S_{\rm ul}$ represents the line strength. 

Due to the large beam sizes of our single-dish telescopes compared to the expected source size, the observed integrated intensities should be corrected by dividing by the beam dilution factor, $\frac{\theta_{\rm s}^{2}}{\theta_{\rm b}^{2}+\theta_{\rm s}^{2}}$, where $\theta_{\rm s}$ and $\theta_{\rm b}$ represent the FWHM source size and HPBW, respectively. We have adopted a value of 24\arcsec, for $\theta_{\rm s}$ based on the fitting results of our single-dish images (see Table~\ref{Tab:size}). The corresponding $\theta_{\rm b}$ for each transition, utilized in the calculations, is listed in Tables~\href{https://zenodo.org/records/21234367}{A.2--A.10}.
For the SMA data, beam-averaged column densities are derived because all data have been convolved to the same angular resolution.

Figures~\href{https://zenodo.org/records/21234367}{B.11 and B.12} 
present the rotational diagrams for different molecules detected by our single-dish and SMA observations, respectively. The fractional molecular abundances relative to H$_{2}$ are derived as the ratio between the molecular column density and the H$_{2}$ column density. Based on our SED fitting results in Sect.~\ref{Sec:dust}, we adopted the source-averaged (24\arcsec) H$_{2}$ column density of ($4.4\pm 0.6$)$\times 10^{22}$~cm$^{-2}$ to derive fractional molecular abundances for the single-dish observations. For MM1, the H$_{2}$ column density in Table~\ref{Tab:sma-dust} is used to derive molecular abundances. The derived rotational temperatures, column densities, and molecular abundances are listed in Table~\href{https://zenodo.org/records/21234367}{A.11}.

Except for C$^{18}$O, the rotational temperatures of other molecules detected by our single-dish observations are $\lesssim$15~K, which is comparable to the kinetic temperature of 16$\pm$3~K derived from ammonia observations in this source \citep{2012A&A...544A.146W}. In contrast, the derived rotational temperature of C$^{18}$O is higher, 29$\pm$2.8~K, which is close to the kinetic temperature of $\sim$36~K derived from a large velocity gradient analysis of multiple methanol transitions \citep{2015PASJ...67...95N}. We find that the rotational temperatures (up to $\sim$50~K) derived from our SMA data are generally higher compared to those from single-dish data. Given the different spatial scales probed by these observations, this suggests the presence of an internal heating source, which could be attributed to the radiation or outflows from NDO.

In Table~\href{https://zenodo.org/records/21234367}{A.11}, 
the molecular column densities range from $4.9\times 10^{11}$~cm$^{-2}$ to $6.7\times 10^{15}$~cm$^{-2}$, while the molecular abundances range from $1.1\times 10^{-11}$~cm$^{-2}$ to $1.5\times 10^{-7}$. Our column densities and abundances differ from those reported by  \citet{2023A&A...669A.121Q} for the same molecules. This discrepancy is mainly due to the different adopted source sizes and rotational temperatures, that is, \citet{2023A&A...669A.121Q} used a smaller source size of 15\arcsec\, and higher rotational temperatures (i.e., 15~K and 26.5~K). 

\begin{figure*}[!htbp]
\centering
\includegraphics[width = 0.95 \textwidth]{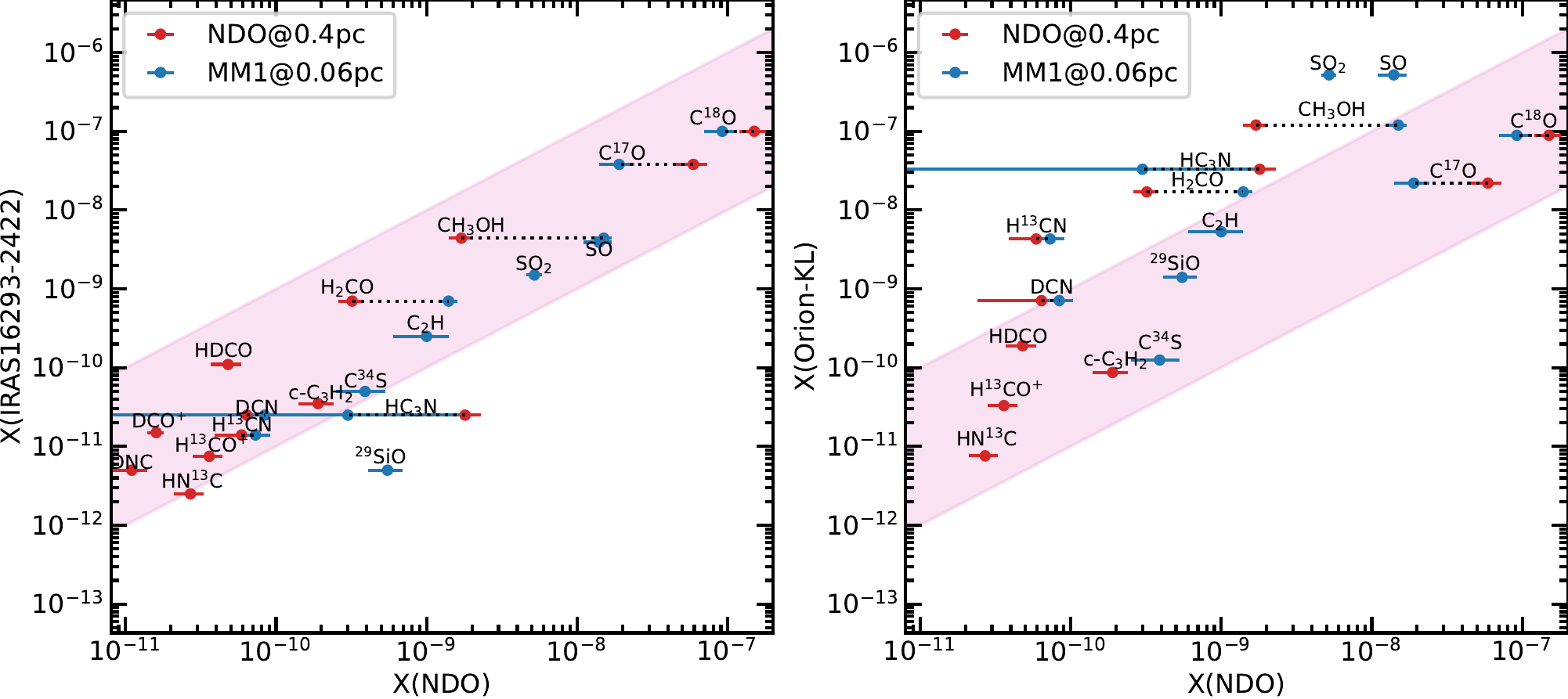}
\caption{{Comparison of molecular abundances of NDO with those in IRAS~16293$-$2422 and Orion~KL \citep[e.g.,][]{1987ApJ...315..621B,1994ApJ...428..680B,1995ApJ...447..760V,2011A&A...532A..23C,2011A&A...528A..26T}. The red and blue points indicate the abundances of NDO measured on 0.4 pc and 0.06 pc scales, respectively. Abundances of the same molecule measured on different scales are connected with black dotted lines. The pink-shaded region represents the abundance variation within an order of magnitude.}\label{Fig:abun}}
\end{figure*}

All molecules detected in NDO have previously been observed in well-known star-forming regions such as IRAS~16293$-$2422 (representing low-mass protostars) and Orion~KL (high/intermediate-mass protostars) \citep[e.g.,][]{1987ApJ...315..621B,1994ApJ...428..680B,1995ApJ...447..760V,2011A&A...532A..23C,2011A&A...528A..26T}. We therefore compare the molecular abundances in NDO with those in IRAS~16293$-$2422 and Orion~KL as a qualitative reference for its chemical 
composition relative to well-studied star-forming environments. The results are shown in Fig.~\ref{Fig:abun}. Given the potentially large uncertainties of molecular abundances, we only discuss the abundance differences exceeding an order of magnitude. 

The fractional abundances of most molecules agree with those in IRAS~16293$-$2422 and Orion~KL within an order of magnitude. However, the $^{29}$SiO abundance is higher in NDO than in IRAS~16293$-$2422. As the FWZPs (100--160~\kms) of the outflow in NDO are much broader than those ($<$30~\kms) in IRAS~16293$-$2422 \citep[e.g.,][]{2023A&A...673A.143K}, we infer that the high-velocity shocks in NDO are responsible for the elevated $^{29}$SiO abundance. 

Similarly, the abundances of SO, SO$_2$, CH$_3$OH, H$_2$CO, HC$_3$N, and H$^{13}$CN in Orion~KL are more than an order of magnitude higher than those in NDO. Orion~KL's more powerful shocks heat the surrounding gas to kinetic temperatures exceeding 100~K within a $\sim$0.1 pc region \citep[e.g.,][]{2018A&A...609A..16T}. In contrast, the lower rotational temperatures of $\lesssim 50$~K (see Table~\href{https://zenodo.org/records/21234367}{A.11}  
suggest that NDO likely harbors cooler gas on similar spatial scales. These high kinetic temperatures and outflow shocks enhance the SO, SO$_{2}$, CH$_{3}$OH, H$_{2}$CO, HC$_{3}$N, and H$^{13}$CN abundances, making these molecules more prominent in Orion KL. Therefore, we speculate that the observed abundance differences are primarily due to different velocities of the shocks exciting these environments. Alternatively, the differences may reflect different evolutionary timescales following recent shock events.

\subsection{Isotope ratios}
Isotope ratios serve as important tools for tracing stellar nucleosynthesis and assessing the impact of stellar ejecta \citep[e.g.,][]{1994ARA&A..32..191W}, and thus may provide clues to the nature of NDO. The detection of rare isotopologues allows us to revisit the isotope ratios of this source.

As HCN transitions can be optically thick, the integrated-intensity ratios between HCN and H$^{13}$CN transitions can place a lower limit on the $^{12}$C/$^{13}$C isotope ratio. The integrated-intensity ratios are found to be 13.9$\pm$6.6, 19.2$\pm$1.7, and 34.1$\pm$2.5 for the $J=1-0$, $J=2-1$, $J=3-2$, respectively. This implies a $^{12}$C/$^{13}$C ratio of $>$34. Similarly, the $^{12}$C/$^{13}$C isotope ratio are found to be $>$18 and $>$13 from HCO$^{+}$, HNC, and their $^{13}$C-bearing isotopologues. Our C$^{18}$O (2--1) and $^{13}$C$^{18}$O (2--1) measurements also indicate a $^{12}$C/$^{13}$C ratio of $>$35 (see Table~\ref{Tab:ratio}). These lower limits appear to be higher than some previously reported $^{12}$C/$^{13}$C ratios of 1.9--27 \citep{2004PASJ...56.1083D,2023A&A...669A.121Q}. Although our estimates do not provide precise isotope ratios, our robust lower limit of $^{12}$C/$^{13}$C $>35$ is difficult to reconcile with true ratios below $\sim$27 if the same gas component is being traced. The difference may therefore reflect optical-depth effects, sensitivity limitations, or different analysis assumptions. In particular, ratios derived from optically thick main isotopologue lines can be biased toward low values if the optical depths are not well constrained. The value reported by \citet{2004PASJ...56.1083D}, derived from C$^{18}$O (1--0) and $^{13}$C$^{18}$O (1--0), was also described as provisional because of the low signal-to-noise ratio of the $^{13}$C$^{18}$O (1--0) line. On the other hand, our observations yielded an $^{18}$O/$^{17}$O isotope ratio of $2.6\pm0.7$. The $^{32}$S/$^{34}$S and $^{28}$Si/$^{29}$Si ratios are derived to be $>$16.6 and $>$5, respectively. These values are consistent with early results reported by \citet{2023A&A...669A.121Q} within uncertainties. 

Thanks to our sensitive observations, we can now constrain the deuterium fraction in NDO for the first time. We assumed a $^{12}$C/$^{13}$C isotope ratio to calculate the deuterium fraction in NDO. Following the gradient in the Milky Way \citep[Eq.~(6) in][]{2023A&A...670A..98Y}, we estimate the $^{12}$C/$^{13}$C isotope ratio to be 53 at the Galactocentric distance of NDO. This value was applied to obtain HCO$^{+}$, HCN, and HNC column densities from their $^{13}$C-bearing counterparts. The deuterium fractions are derived from their respective column density ratios, and the results are presented in Table~\ref{Tab:ratio}. We note, however, that the source-specific $^{12}$C/$^{13}$C ratio may deviate from the Galactic mean trend. This assumption directly affects the derived deuterium fractions. For example, if we instead adopt $^{12}$C/$^{13}$C = 27, as reported by \citet{2004PASJ...56.1083D}, the resulting deuterium fractions would be about a factor of two higher than 
the values listed in Table~\ref{Tab:ratio}. Although the absolute deuterium fractions therefore depend on the assumed 
$^{12}$C/$^{13}$C ratio, adopting a lower ratio would not weaken the cold-gas interpretation, but would instead imply even higher levels of deuterium enrichment.

The deuterium fractions range from 0.5\% to 7.5\%\,for the different species, significantly higher than the cosmic [D/H] abundance of $1.5\times 10^{-5}$ \citep{2003SSRv..106...49L}. This indicates an efficient deuteration process, likely involving either H$_2$D$^{+}$ in cold prestellar gas or CH$_2$D$^{+}$ during a later warm phase \citep[e.g.,][]{2009A&A...508..737P,2014prpl.conf..859C}. The deuterium fraction derived from H$_{2}$CO is much higher than those of the other three species, which is also commonly observed in star-forming regions \citep[e.g.,][]{1995ApJ...447..760V,2014prpl.conf..859C}. The difference is attributed to the significant contribution of grain chemistry to the formation of H$_{2}$CO \citep[e.g.,][]{2014prpl.conf..859C}. These results demonstrate the presence of deuterium in this region. This practically excludes any possibility that this material was ejected from any star, as stars very quickly destroy any deuterium even in the protostellar phase \citep{1988ApJ...332..804S}. 

Table~\ref{Tab:ratio} compares the measured isotope ratios in NDO with those in two star-forming regions (i.e., IRAS 16293$-$2422 and Orion KL) and in three representative evolved objects: the C-rich AGB star IRC+10216, the O-rich AGB star IK~Tau, and the merger remnant CK~Vul, which likely originated from a red-giant-branch primary \citep{2017A&A...607A..78K}. As discussed earlier, the revised $^{12}$C/$^{13}$C isotope ratio is higher than previously reported by \citet{2023A&A...669A.121Q} and \citet{2004PASJ...56.1083D}. Consequently, the $^{12}$C/$^{13}$C isotope ratio ($>$35) for NDO exceeds those ($\lesssim$15) found in IK~Tau and CK~Vul. The derived $^{18}$O/$^{17}$O isotope ratio is higher than that (1.1$\pm$0.2) in IRC +10216, and roughly aligns with the prediction from the measured $^{18}$O/$^{17}$O gradient in the Galactic ISM \citep{2020ApJS..249....6Z,2023MNRAS.522..559O,2023ApJS..268...56Z}. For a wide range of initial stellar masses and solar metallicities, AGB winds are expected to exhibit $^{12}$C/$^{13}$C ratios spanning 3--270, while the $^{18}$O/$^{17}$O ratio remains constrained to 0.4--3.3 \citep{2016ApJ...825...26K}. The C and O isotope ratios measured in NDO 
therefore do not, by themselves, exclude material from AGB stars, even though these isotope ratios are consistent with values found in star-forming regions. 

A stronger constraint comes from the detection of deuterium fractionation. Deuterated molecules are commonly associated with star-forming regions \citep[e.g.,][]{2014prpl.conf..859C}, but not observed in circumstellar material produced by either low- or high-mass evolved stars. The deuterated species therefore strongly support the presence of cold interstellar material in the immediate vicinity of NDO, with the peak of the deuterated molecular emission lying at a projected offset of only $\lesssim$0.1~pc from the infrared object. Taken together, the chemical evidence favors the interpretation that a significant fraction of the molecular material around NDO is interstellar and related to a star-forming environment, but it does not by itself uniquely determine the nature of the central source.

\begin{table*}[!hbt]
\caption{Isotope ratios for NDO and other environments.}\label{Tab:ratio}
\small
\centering
\begin{tabular}{ccccccccc}
\hline \hline
                  &          & \multicolumn{2}{c}{Our measurements} & \multicolumn{2}{c}{star-forming regions} & \multicolumn{3}{c}{Evolved stars}   \\
\cmidrule(lr){3-4} \cmidrule(lr){5-6} \cmidrule(lr){7-9} 
ratio             & species  & NDO  & MM1  & IRAS 16293$-$2422\tablefootmark{a} & Orion KL\tablefootmark{b}  & IRC +10216\tablefootmark{c}  & IK Tau\tablefootmark{d} & CK Vul\tablefootmark{e} \\
\hline
D/H               & DCO$^{+}$/HCO$^{+}$        & ($0.8\pm 0.3$)\% & \nodata & 0.86\% & 0.2\%  & \nodata  & \nodata  & \nodata    \\
                  & DCN/HCN                    & ($2.0\pm 1.4$)\% & ($2.2\pm 0.7$)\% & 1.3\%  &   1\%--6\%    & \nodata  & \nodata  & \nodata    \\
                  & DNC/HNC                    & ($0.8\pm 0.2$)\% & \nodata  & 3\%   & 1.0\%  & \nodata  & \nodata  & \nodata    \\
                  & HDCO/H$_{2}$CO             & ($7.5\pm1.9$)\%\tablefootmark{f} & \nodata  & 14\%  & $0.66^{+0.68}_{-0.37}$\%            & \nodata  & \nodata  & \nodata    \\             
$^{12}$C/$^{13}$C & HCO$^{+}$/H$^{13}$CO$^{+}$ & $>$18            & $>$41 & $56^{+8}_{-11}$ &   55$\pm$13   & 49$\pm$9 & $\sim$15 & 3.8$\pm$1.0\\
                  & HCN/H$^{13}$CN             & $>$34            & $>$21 &        &               &          &          &  \\
                  & HNC/HN$^{13}$C             & $>$13            & $>$12 &        &               &          &          &  \\
                  & C$^{18}$O/$^{13}$C$^{18}$O & $>$35            & $>$8 &        &               &          &          &  \\
$^{18}$O/$^{17}$O & C$^{18}$O/C$^{17}$O        & 2.6$\pm$0.7      & 2.4$\pm$0.7 & $3.2^{+0.2}_{-0.3}$ &   4.1$\pm$0.1 & 1.1$\pm$0.2  &  \nodata &  $\gtrsim$5   \\
$^{32}$S/$^{34}$S & CS/C$^{34}$S               & $>$16.6          & $>$10.7       & \nodata & $21.2\pm 4.7$ & 18.9$\pm$1.3 & $\sim$15 & 14$\pm$3   \\
$^{28}$Si/$^{29}$Si & SiO/$^{29}$SiO           & $>$5             & $\gtrsim$21.9 & \nodata &  26$\pm$10    & 17.2$\pm$1.1 & $\sim$18 & 6.7$\pm$0.3\\
\hline
\end{tabular}
\tablefoot{Lower limits of the isotope ratios are derived from the 3$\sigma$ limits. Columns 3 and 4 present the results obtained from the single-dish observations and the SMA data, respectively. (a) IRAS 16293$-$2422 is a low-mass star-forming region, and the isotope ratios are taken from \citet{1995ApJ...447..760V} and \citet{2018A&A...610A..54P}; (b) Orion KL is thought to be a high-mass star-forming region, and its isotope ratios are based on studies of \citet{1992A&A...256..595S}, \citet{2011A&A...528A..26T}, \citet{2013ApJ...777...85N}, \citet{2023A&A...670A..98Y}, and \citet{2023ApJS..268...56Z}; (c) IRC +10216 is a C-rich star, and its isotope ratios measured in its CSE are based on the studies of \citet{2008ApJS..177..275H} and \citet{2015A&A...574A..56G}; (d) IK tau is a O-rich star, and its isotope ratios measured in its CSE are taken from \citet{2017A&A...597A..25V}. (e) CK Vul is thought to be a red nova remnant, and its isotope ratios are taken from \citet{2017A&A...607A..78K}. (f) The list D/H value taken as half of the column density ratio between HDCO and H$_2$CO, assuming $N$(HDCO)/$N$(H$_{2}$CO)=2D/H \citep[see Table B.1 in][]{2019A&A...623A..69M}.}
\normalsize
\end{table*}

\subsection{Astronomical nature of NDO}\label{sec.nature}
Since its discovery, different origins have been proposed to explain the observed properties of NDO: an analog of the Rotten Egg Nebula \citep{2000PASJ...52L..43N}, an oxygen-rich, dust-enshrouded AGB star
\citep{2007A&A...470..957M}, an evolved star like water fountain \citep{2011ApJ...728...76N,2025ApJ...981...41F}, a red supergiant \citep{2016ApJ...825...16N}, a red nova like CK~Vul \citep{2023A&A...669A.121Q}, and a massive YSO \citep{2016ApJ...828...51C}. In the following, we explore these scenarios and discuss a possible explanation for the origin of NDO.

One of our important discoveries is the detection of deuterated molecules. The presence of deuterated molecules in space is commonly attributed to the deuterium fractionation in cold or warm molecular clouds \citep[e.g.,][]{2009A&A...508..737P,2014prpl.conf..859C}. Deuterium was primarily produced during Big Bang nucleosynthesis \citep[e.g.,][]{2016RvMP...88a5004C,2018ApJ...855..102C}. Although it can be synthesized in stellar interiors, it is immediately consumed through reactions that form $^{3}$He \citep[e.g.,][]{2007hic..book.....C}. Consequently, deuterated molecules have never been detected in the circumstellar envelopes of evolved stars \citep[e.g.,][]{2015A&A...574A..56G}. Their detection toward NDO therefore provides strong evidence for the presence of cold interstellar molecular material. This argues against a scenario in which all of the molecular gas around NDO is produced solely by the circumstellar envelope or wind of a single evolved star. However, our SMA observations have shown that the emission of deuterated molecules does not peak toward NDO (see Fig.~\ref{Fig:SMA-dcn}). Such a morphology does not exclude more complex scenarios in which an evolved object is embedded in, or interacting with, surrounding interstellar material.

The distribution of CH$_3$OH provides another important constraint. CH$_3$OH is uncommon in the circumstellar envelopes of evolved stars 
\citep{2017A&A...603L...2O,2018A&A...618A.164S,2020A&A...644A..59K}. Previous single-dish detection of CH$_{3}$OH in NDO was ascribed to the interaction of NDO and ambient clouds \citep{2015PASJ...67...95N}, similar to scenarios used to explain the detected CH$_{3}$OH emission in AGB-related objects \citep{2017A&A...603L...2O,2018A&A...618A.164S,2020A&A...644A..59K}. Our high-resolution SMA maps suggest that the prominent CH$_{3}$OH emission is closely associated with the central object. This spatial association is naturally explained if CH$_3$OH is released from icy grain mantles by heating or shocks associated with a young embedded source. Nevertheless, we cannot completely exclude the possibility from interactions between a peculiar evolved star and the surrounding dense material.

NDO also shows several properties suggestive of high-mass star formation. Although NDO shares similarities with low-mass star-forming regions like IRAS~16293$-$2422, SiO masers have never been detected in low-mass star-forming regions. Their presence in NDO therefore points toward a high-mass star-forming region, consistent with the scenario of \citet{2016ApJ...828...51C}. On the other hand, the non-detection of radio continuum emission further indicates the absence of a detectable H{\scriptsize II} region created by an ongoing massive YSO (see Sect.~\ref{sec.radio}). Our upper limit on the Lyman ionizing photon flux indicates a single zero-age main sequence star with a spectral type later than B2, implying that NDO is at an evolutionary stage prior to the formation of a classical H{\scriptsize II} region. Given its high luminosity of $\sim 2\times 10^{4}$~\Lsun\,and its distinctive far-IR spectral features, it is proposed that NDO might lie at a rare phase of massive star formation \citep{2016ApJ...828...51C}. This is comparable to the situation in the Orion~KL region, where the runaway stars are estimated to have masses of 8--10~$M_\odot$, corresponding to early B-type stars that do not generate extensive ionized nebulae, and where Source~I exhibits a cool, red-supergiant–like photosphere \citep{2010A&A...522A..44T} that likewise does not produce a detectable H{\scriptsize II} region. 
Such luminous massive protostars may fail to produce observable H{\scriptsize II} regions when undergoing high mass accretion. In this regime, protostellar bloating results in enlarged stellar radii and consequently low effective temperatures, yielding an insufficient flux of ionizing photons to ionize the surrounding medium \citep{2009ApJ...691..823H}.

Several observed properties in NDO resemble those of the massive protostar Source I in Orion KL. NDO is known to possess H$_{2}$O, SiO, OH (1612~MHz and 1665~MHz), and possible class I CH$_{3}$OH masers \citep{2000PASJ...52L..43N,2011ApJ...728...76N,2015PASJ...67...95N}, but lacks class II CH$_{3}$OH masers which are good signposts of massive star formation \citep[e.g.,][]{1991ApJ...380L..75M,2013MNRAS.431.1752U}. Astrometric observations of H$_{2}$O masers from this source yielded a large space motion that deviates by $\sim$40~km~s$^{-1}$ from the Galactic rotation \citep{2011PASJ...63...81I}. By analogy to Orion KL's Source I, the large space motion of NDO may be caused by dynamical interaction with other YSOs, although this remains speculative. The similarity is further suggested by the presence of a double-peaked SiO maser emission in both sources. Despite substantial searches, SiO masers were only detected in seven massive star-forming regions including NDO \citep{2000PASJ...52L..43N,2009ApJ...691..332Z,2015A&A...584L...7G,2016ApJ...826..157C}. NDO and Source I within Orion KL are the only two known maser sources exhibiting double-peaked SiO profiles. In Source I, this kinematical feature is attributed to the disk–wind origin of the peculiar masers \citep{2007ApJ...664..950R,2010ApJ...708...80M,2017A&A...606A.126I}. A similar geometry may therefore apply to NDO, but this hypothesis requires future high-angular-resolution maser observations. In addition, OH 1612 MHz masers are also unusual in star-forming regions \citep{1999MNRAS.308..683C}, but are detected in both NDO and Orion KL \citep{2011ApJ...728...76N,1983ApJ...267..625H,2006MNRAS.367..541C}. Our analysis indicates that OH 1612 MHz masers could be excited by outflows (see Sect.~\ref{sec.outflow}), similar to that in Orion KL \citep{2006MNRAS.367..541C}.

Additional similarities with Orion~KL include a luminous infrared nebula, highly excited H$_2$ emission, and a powerful wide-angle outflow. Both sources are embedded in an extended luminous infrared nebula. The narrow-band H$_{2}$ image suggests enhanced emission of highly excited H$_{2}$ around NDO (see Fig.~\ref{Fig:sma-outflow}). Our data have suggested that its associated outflow, exhibiting a wide opening angle, is exceptionally powerful with broad line wings spanning a range of $\sim$150~km~s$^{-1}$, reminiscent of the explosive outflow in Orion KL \citep[e.g.,][]{2017ApJ...837...60B}. Our dust continuum observations suggest a total gas mass of $\sim$200~$M_{\odot}$ for the dense clump where NDO is located (see also \citealt{2016ApJ...825...16N}). The total mass is comparable to the $\sim$100~$M_{\odot}$ of OMC1 in which Orion KL resides \citep{1989ARA&A..27...41G}. 

We therefore suggest that NDO may represent an Orion~KL/Source~I-like object viewed at a distance about ten times farther away. This scenario provides a coherent explanation for several observed properties, including the dense interstellar environment, deuterated molecules, CH$_3$OH emission, rare maser species, luminous infrared nebula, and powerful outflow. However, this analogy should be regarded as a plausible hypothesis rather than a unique identification. In particular, we cannot exclude a more complex scenario in which a peculiar infrared object is embedded in, or interacting with, surrounding interstellar material. Future high-angular-resolution observations will be essential for testing whether the central infrared object is indeed analogous to Source~I or whether NDO represents a more complex system involving both a peculiar central object and surrounding interstellar material.

\section{Summary}\label{Sec:sum}
We have performed an observational campaign to study the dust and molecular environment of NDO using the IRAM-30m and APEX telescopes as well as the SMA. Our main results are summarized as follows:
\begin{itemize}
    \item Based on the SED fitting of the single-dish observations, we find that the clump hosting NDO is characterized by a dust temperature of $25\pm2$~K, a source-averaged (24\arcsec) H$_2$ column density of ($4.4\pm 0.6$)$\times 10^{22}$ cm$^{-2}$, a dust emissivity index of 1.7$\pm$0.2, and a total gas mass of 220$\pm$21~M$_{\odot}$. Our SMA dust continuum observations further resolve the clump into at least four dust continuum cores, among which the centrally peaked core MM1 appears to be associated with the SiO maser and is the main mass reservoir. \\
    
    \item Our single-dish observations of NDO led to the detection of 59 spectral lines which are assigned to 12 species and their isotopologues. This includes the first detection of deuterated molecules, DCO$^{+}$, DCN, DNC, and HDCO, which demonstrates the presence of deuterium fractionation toward NDO. \\

    \item Our SMA observations resulted in the detection of 75 spectral lines toward the most prominent dust core MM1. These lines are also assigned to 12 species and their isotopologues. Their morphologies exhibit diversity on a scale of 0.1~pc, which is likely due to the shock feedback. Furthermore, the high-velocity outflow slightly resolved by our SMA observations does not have a spherical morphology and resembles more a wide-angle bipolar outflow. \\

    \item Our observations suggest that the surrounding material of NDO is more naturally associated with a dense star-forming environment than with a purely evolved-star circumstellar envelope. Based on the observed similarity between NDO and Orion KL, we postulate that NDO could be an Orion KL analog about ten times farther away.
    
\end{itemize}

Our observations probe the physics and chemistry of NDO on scales of $\gtrsim$0.06~pc and reveal that it is embedded in a dense interstellar environment. The detected deuterated molecules, rich molecular inventory, cold dust, outflow activity, and rare maser species together favor an association with star-forming material. However, the nature of the central infrared source remains uncertain. In particular, the molecular outflow and the source associated with the SiO maser are still poorly resolved. Higher angular resolution observations are therefore essential to test whether NDO is indeed an Orion~KL-like system, or whether it represents a more complex object involving a peculiar central source and surrounding interstellar 
material.

\section{Data availability}
Appendix A, B, and C are available on Zenodo, at \url{https://zenodo.org/records/21234367}.

\begin{acknowledgements}
We thank the IRAM, APEX, and SMA staff for their assistance with our observations. Y.G. is supported by the Strategic Priority Research Program of the Chinese Academy of Sciences, Grant No. XDB0800301. T.K. acknowledges funding from grant SONATA BIS no. 2018/30/E/ST9/00398 from the Polish National Science Center. FN thanks the Brazilian National Council for Scientific and Technological Development (CNPq) for the financial support through Grant number 303093/2025-0. This work is based on observations carried out with the IRAM 30m telescope. IRAM is supported by INSU/CNRS (France), MPG (Germany) and IGN (Spain). The research leading to these results has received funding from the European Union's Horizon 2020 research and innovation program under grant agreement No 730562 [RadioNet]. This publication is based on data acquired with the Atacama Pathfinder Experiment (APEX). APEX is a collaboration between the Max-Planck-Institut fur Radioastronomie, the European Southern Observatory, and the Onsala Space Observatory. The Submillimeter Array is a joint project between the Smithsonian Astrophysical Observatory and the Academia Sinica Institute of Astronomy and Astrophysics and is funded by the Smithsonian Institution and the Academia Sinica. This research has made use of NASA's Astrophysics Data System. This work also made use of the Cube Analysis and Rendering Tool for Astronomy (CARTA) software \citep{2021zndo...3377984C} and Python libraries including Astropy\footnote{\url{https://www.astropy.org/}} \citep{2013A&A...558A..33A}, NumPy\footnote{\url{https://www.numpy.org/}} \citep{5725236}, SciPy\footnote{\url{https://www.scipy.org/}} \citep{jones2001scipy}, Matplotlib\footnote{\url{https://matplotlib.org/}} \citep{Hunter:2007}, and APLpy \citep{2012ascl.soft08017R}. We would like to thank the anonymous referee for the helpful comments. 

\end{acknowledgements}

\bibliographystyle{aa}
\bibliography{references}

\clearpage
\begin{appendix}

\section{Tables}\label{app.a}

\begin{table}[!hbt]
\caption{Flux densities of different bands used in the SED fitting.}\label{Tab:sed}
\small
\centering
\begin{tabular}{cccccccc}
\hline \hline
$\lambda$              & $\int S_{\nu} {\rm d}\Omega$     &  instrument  & reference \\ 
 ($\mu$m)              & (Jy)         &               & \\ 
\hline                 
65                     & 313.10$\pm$29.40   &  AKARI/FIS     & \citet{2007PASJ...59S.389K} \\
70                     & 359.57$\pm$0.61    &  Hershel/PACS  & \citet{2016AA...591A.149M} \\  
90                     & 176.30$\pm$38.90   &  AKARI/FIS     & \citet{2007PASJ...59S.389K} \\ 
140                    & 302.90$\pm$42.00   &  AKARI/FIS     & \citet{2007PASJ...59S.389K} \\ 
160                    & 275.10$\pm$31.80   &  AKARI/FIS     & \citet{2007PASJ...59S.389K} \\
160                    & 231.15$\pm$0.85    &  Hershel/PACS  & \citet{2016AA...591A.149M} \\ 
250                    & 127.19$\pm$0.91    &  Hershel/SPIRE & \citet{2016AA...591A.149M} \\ 
350                    & 44.92$\pm$0.90     &  Hershel/SPIRE & \citet{2016AA...591A.149M} \\ 
500                    & 23.14$\pm$0.99     &  Hershel/SPIRE & \citet{2016AA...591A.149M} \\ 
870                    & 2.96$\pm$0.21      &  APEX/LABOCA   & this study                  \\
1100                   & 1.78$\pm$0.37      &  CSO/BOLOCAM   & \citet{2013ApJS..208...14G} \\
1200                   & 1.39$\pm$0.10      &  IRAM/MAMBO2   & this study                  \\ 
\hline
\end{tabular}
\normalsize
\end{table}

\begin{table*}[!hbt]
\caption{Observed properties of CO transitions.}\label{Tab:co}
\small
\centering
\begin{tabular}{ccccccccccc}
\hline \hline
Line                   & Frequency        & Telescope  & HPBW       & $E_{\rm u}/k$& $\int T_{\rm mb}{\rm d}\varv$ & $\varv$ & $\Delta \varv$ & $T_{\rm mb}$&$\sigma$ & Note      \\ 
                       & (MHz)            &            & (\arcsec)  &  (K)       & (K~\kms)                     & (\kms)  & (\kms)         & (K)       &(K)      &            \\
\hline                 
CO (1--0)              & 115271.202(1)    & IRAM-30m   & 21.3       & 5.5          & 62.350$\pm$1.820 & 33.5$\pm$0.2 & 25.0$\pm$0.6 & 2.350$\pm$0.194  & 0.201 & B   \\ 
                       &                  &            &            &              & 3.620$\pm$0.195  &$-$4.4$\pm$0.1& 1.3$\pm$0.1  & 2.650$\pm$0.244  & 0.201 &     \\   
                       &                  &            &            &              & 26.810$\pm$0.469 & 29.1$\pm$0.1 & 3.4$\pm$0.1  & 7.400$\pm$0.244  & 0.201 & N1  \\  
                       &                  &            &            &              & 17.320$\pm$0.507 & 33.1$\pm$0.1 & 2.6$\pm$0.1  & 6.240$\pm$0.244  & 0.201 &     \\  
                       &                  &            &            &              & 38.880$\pm$0.375 & 38.0$\pm$0.1 & 3.5$\pm$0.1  & 10.500$\pm$0.244 & 0.201 & N2  \\ 
                       &                  &            &            &              & 15.010$\pm$0.277 & 44.5$\pm$0.1 & 2.4$\pm$0.1  & 5.950$\pm$0.244  & 0.201 & N3  \\  
CO (2--1)              & 230538.000(1)    & IRAM-30m   & 10.7       & 16.6         & 339.500$\pm$1.280& 33.4$\pm$0.1 & 38.1$\pm$0.2 & 8.360$\pm$0.123  & 0.087 & B   \\
                       &                  &            &            &              & 51.030$\pm$1.890 & 38.8$\pm$0.1 & 2.6$\pm$0.1  & 18.500$\pm$1.260 & 0.087 & N2  \\
CO (3--2)              & 345795.990(1)    & APEX       & 18.0       & 33.2         & 291.900$\pm$1.790& 32.7$\pm$0.1 & 40.1$\pm$0.3 & 6.830$\pm$0.176  & 0.122 & B   \\
                       &                  &            &            &              & 55.860$\pm$0.998 & 38.8$\pm$0.1 & 3.1$\pm$0.1  & 16.900$\pm$0.562 & 0.122 & N2  \\
CO (5--4)              & 576267.931(1)    & Herschel   & 37.4       & 83.0         & 82.290$\pm$0.170 & 32.2$\pm$0.0 & 37.9$\pm$0.1 & 2.040$\pm$0.023  & 0.011 & B   \\
                       &                  &            &            &              & 34.260$\pm$0.121 & 37.4$\pm$0.0 & 7.7$\pm$0.1  & 4.200$\pm$0.057  & 0.011 & N2  \\
CO (6--5)              & 691473.076(1)    & Herschel   & 31.2       & 116.2        & 99.880$\pm$0.190 & 32.3$\pm$0.0 & 36.8$\pm$0.1 & 2.550$\pm$0.028  & 0.012 & B   \\
                       &                  &            &            &              & 49.320$\pm$0.025 & 36.5$\pm$0.0 & 10.2$\pm$0.1 & 4.530$\pm$0.052  & 0.012 & N2  \\
$^{13}$CO (1--0)        & 110201.354 (5)  & IRAM-30m   & 22.3       & 5.3          & 8.394$\pm$0.096  & 29.1$\pm$0.1 & 2.1$\pm$0.1  & 3.800$\pm$0.093  & 0.088 & N1  \\   
                       &                  &            &            &              & 33.270$\pm$0.128 & 35.8$\pm$0.1 & 4.0$\pm$0.1  & 7.880$\pm$0.093  & 0.088 & N2  \\  
                       &                  &            &            &              & 5.025$\pm$0.065  & 44.6$\pm$0.1 & 1.0$\pm$0.1  & 4.860$\pm$0.093  & 0.088 & N3  \\
$^{13}$CO (2--1)        & 220398.684(1)   & IRAM-30m   & 11.2       & 15.9         & 6.193$\pm$0.598  & 29.2$\pm$0.1 & 2.2$\pm$0.3  & 2.600$\pm$0.358  & 0.377 & N1  \\  
                       &                  &            &            &              & 42.380$\pm$0.731 & 36.1$\pm$0.1 & 4.2$\pm$0.1  & 9.370$\pm$0.358  & 0.377 & N2  \\ 
                       &                  &            &            &              & 3.521$\pm$0.397  & 44.6$\pm$0.1 & 1.1$\pm$0.2  & 2.980$\pm$0.358  & 0.377 & N3  \\ 
$^{13}$CO (3--2)        & 330587.965(1)   & APEX       & 18.9       & 31.7         & 19.810$\pm$0.645 & 36.1$\pm$0.2 & 14.9$\pm$0.5 & 1.250$\pm$0.094  & 0.090 & B   \\  
                       &                  &            &            &              &  2.677$\pm$0.122 & 29.2$\pm$0.1 & 1.7$\pm$0.1  & 1.490$\pm$0.097  & 0.090 & N1  \\    
                       &                  &            &            &              &  39.600$\pm$0.179& 36.1$\pm$0.1 & 3.2$\pm$0.1  & 11.500$\pm$0.097 & 0.090 & N2  \\  
                       &                  &            &            &              &  0.599$\pm$0.078 & 44.5$\pm$0.1 & 0.7$\pm$0.1  & 0.803$\pm$0.097  & 0.090 & N3  \\
$^{13}$CO (6--5)        & 661067.277(1)   & Herschel   & 32.6       & 111.1        & 7.075$\pm$0.056  & 36.5$\pm$0.1 & 12.7$\pm$0.2 & 0.525$\pm$0.016  & 0.014 & B   \\
                       &                  &            &            &              & 6.709$\pm$0.024  & 36.2$\pm$0.0 & 2.8$\pm$0.0  & 2.210$\pm$0.018  & 0.014 & N2  \\
$^{13}$CO (10-9)        & 1101349.597(2)  & Herschel   & 19.5       & 290.8        &  2.762$\pm$0.083 & 35.7$\pm$0.3 & 18.9$\pm$1.1 & 0.137$\pm$0.030  & 0.031 & B   \\           
                       &                  &            &            &              &  1.656$\pm$0.033 & 36.5$\pm$0.0 & 3.5$\pm$0.1  & 0.441$\pm$0.031  & 0.031 & N2  \\  
C$^{17}$O (2--1)        & 224714.385(3)   & IRAM-30m   & 10.9       & 16.2         & 3.573$\pm$0.188  & 36.4$\pm$0.1 & 2.7$\pm$0.2  & 1.270$\pm$0.108  & 0.121 & N2  \\
C$^{18}$O (2--1)        & 219560.354(1)   & IRAM-30m   & 11.2       & 15.8         & 0.815$\pm$0.049  & 29.1$\pm$0.1 & 0.9$\pm$0.1  & 0.830$\pm$0.053  & 0.046 & N1  \\ 
                       &                  &            &            &              &  8.966$\pm$0.077 & 36.6$\pm$0.1 & 2.1$\pm$0.1  & 4.110$\pm$0.053  & 0.046 & N2  \\ 
                       &                  &            &            &              &  0.472$\pm$0.048 & 44.8$\pm$0.1 & 0.8$\pm$0.1  & 0.530$\pm$0.053  & 0.046 & N3  \\
C$^{18}$O (6--5)        & 658553.278(1)   & Herschel   & 32.7       & 110.6        & 1.331$\pm$0.018  & 36.3$\pm$0.0 & 3.4$\pm$0.1  & 0.367$\pm$0.012  & 0.012 & N2  \\    
$^{13}$C$^{18}$O (1--0)  & 104711.404(6)  & IRAM-30m   & 23.5       & 5.0          & \nodata          & \nodata      & \nodata      & $<$0.048         & 0.016 &     \\
$^{13}$C$^{18}$O (2--1)  & 209419.172(5)  & IRAM-30m   & 11.7       & 15.1         & \nodata          & \nodata      & \nodata      & $<$0.117            & 0.039 &     \\

\hline
\end{tabular}
\tablefoot{The last column lists the velocity component associated with each row. ``B" indicates the broad component, while ``N1", ``N2", and ``N3" correspond to the 33.1, 36.5, and 44.5~\kms\,components, respectively.}
\normalsize
\end{table*}

\begin{table*}[!hbt]
\caption{Observed properties of HCN transitions.}\label{Tab:hcn}
\scriptsize
\centering
\begin{tabular}{cccccccccccc}
\hline \hline
Line                   & Frequency & Telescope & HPBW        & $E_{\rm u}/k$ &$\tau$\tablefootmark{*} & $\int T_{\rm mb}{\rm d}\varv$ & $\varv$ & $\Delta \varv$ & $T_{\rm mb}$&$\sigma$ & note  \\ 
                       & (MHz)     &           & (\arcsec)   &  (K)        &       & (K~\kms)                    & (\kms)  & (\kms)         & (K)       &(K)      &        \\ 
\hline                 
HCN (1--0)            &  88631.6023(2) & IRAM-30m &   27.8      & 4.3         & \nodata    & 7.183$\pm$0.145  & 32.0$\pm$0.2 & 31.3$\pm$0.7 & 0.215$\pm$0.015 & 0.016 & B     \\
                      &                &          &             &             &4.5$\pm$0.3 & 4.385$\pm$0.394  & 36.2$\pm$0.1 & 2.1$\pm$0.1  & 0.711$\pm$0.029 & 0.016 & N2    \\
HCN (2--1)            & 177261.1115(5) & IRAM-30m &   13.9      & 12.8        & \nodata    & 19.770$\pm$0.452 & 31.5$\pm$0.3 & 28.6$\pm$0.7 & 0.650$\pm$0.115 & 0.119 & B     \\
                      &                & IRAM-30m &             & 0.37$\pm$0.05     & 5.596$\pm$0.207  & 37.3$\pm$0.1 & 1.79$\pm$0.1 & 1.156$\pm$0.181 & 0.119 & N2    \\
HCN (3--2)            & 265886.4343(7) & IRAM-30m &   9.3       & 25.5        & \nodata    & 32.040$\pm$0.451 & 31.1$\pm$0.1 & 26.2$\pm$0.4 & 1.150$\pm$0.062 & 0.058 & B     \\
                      &                &          &             &             & \nodata    & 13.260$\pm$0.162 & 36.3$\pm$0.0 & 6.9$\pm$0.1  & 1.800$\pm$0.066 & 0.058 & N2    \\
HCN (4--3)            & 354505.4779(9) & APEX     & 17.6       & 42.5        & \nodata    & 14.890$\pm$0.455 & 32.8$\pm$0.3 & 21.9$\pm$0.8 & 0.640$\pm$0.074 & 0.076 & B     \\
                      &                &          &            &             & \nodata    & 5.713$\pm$0.161  & 36.5$\pm$0.1 & 5.2$\pm$0.2  & 1.020$\pm$0.070 & 0.074 & N2    \\
HCN (7-6)             & 620304.00(1)   & Herschel & 34.7       & 119.1       & \nodata    & 1.278$\pm$0.064  & 31.6$\pm$0.8 & 30.6$\pm$2.2 & 0.039$\pm$0.014 & 0.014 & B     \\
                      &                &          &            &             & \nodata    & 0.660$\pm$0.025  & 35.6$\pm$0.1 & 6.9$\pm$0.3  & 0.090$\pm$0.013 & 0.013 & N2    \\
H$^{13}$CN (1--0)      &  86339.9214(1) & IRAM-30m &  28.5      & 4.1         & $<$0.1     & 0.831$\pm$0.394  & 36.2$\pm$0.1 & 2.7$\pm$0.1  & 0.153$\pm$0.015 & 0.016 & N2    \\
H$^{13}$CN (2--1)      & 172677.8512(1) & IRAM-30m &  14.2      & 12.4        & $<$0.1     & 1.325$\pm$0.113  & 36.3$\pm$0.2 & 2.6$\pm$0.4  & 0.151$\pm$0.032 & 0.033 & N2    \\ 
H$^{13}$CN (3--2)      & 259011.7976(1)  & IRAM-30m &  9.5       & 24.9        & \nodata    &  1.329$\pm$0.096 & 35.9$\pm$0.3 & 7.8$\pm$0.8  & 0.160$\pm$0.034 & 0.033 & N2    \\
H$^{13}$CN (4--3)      & 345339.7693(2)  & APEX     & 18.1       & 41.4        & \nodata    &  \nodata         & \nodata      &  \nodata     & $<$0.156        & 0.052 &       \\ 
DCN (2--1)            & 144828.001((3.379)& IRAM-30m &   17.0      & 10.4        & $<$0.1     & 1.176$\pm$0.424  & 36.3$\pm$0.1 & 1.4$\pm$0.1  & 0.289$\pm$0.029 & 0.011 & N2    \\
DCN (3--2)            & 217238.538(3.379)  & IRAM-30m &  11.3       & 20.9        & \nodata    & 0.444$\pm$0.026  & 36.9$\pm$0.0 & 1.8$\pm$0.1  & 0.233$\pm$0.018 & 0.018 & N2    \\
DCN (5--4)            & 362045.754(3.379)  & APEX     & 17.2       & 52.1        & \nodata    & \nodata          & \nodata      &  \nodata     & $<$0.264        & 0.088 &       \\ 
\hline
\end{tabular}
\normalsize
\tablefoot{\tablefoottext{*}{Total optical depth. The values are derived in the residual spectra where the broad component has been subtracted.} The last column lists the velocity component associated with each row. ``B" indicates the broad component, while ``N1", ``N2", and ``N3" correspond to the 33.1, 36.5, and 44.5~\kms\,components, respectively.}

\end{table*}

\begin{table*}[!hbt]
\caption{Observed properties of HNC transitions.}\label{Tab:hnc}
\scriptsize
\centering
\begin{tabular}{ccccccccccc}
\hline \hline
Line                   & Frequency    & Telescope  &  HPBW       & $E_{\rm u}/k$ & $\int T_{\rm mb}{\rm d}\varv$ & $\varv$ & $\Delta \varv$ & $T_{\rm mb}$&$\sigma$ & Note        \\ 
                       & (MHz)        &            &  (\arcsec)  &  (K)        & (K~\kms)                     & (\kms)  & (\kms)         & (K)       &(K)      &             \\ 
\hline                 
HNC (1--0)             & 90663.568(4) & IRAM-30m   &  27.1       & 4.4         & 0.205$\pm$0.018 & 29.5$\pm$0.1 & 1.5$\pm$0.2 & 0.125$\pm$0.038 & 0.038 & N1    \\
                       &              &            &             &             & 3.146$\pm$0.024 & 36.4$\pm$0.1 & 2.4$\pm$0.1 & 1.220$\pm$0.040 & 0.038 & N2    \\
HNC (3--2)             & 271981.142(20) & IRAM-30m &  9.0        & 26.1        & 2.556$\pm$0.093 & 36.3$\pm$0.1 & 2.4$\pm$0.1 & 1.000$\pm$0.067 & 0.064 & N2    \\
HN$^{13}$C (1--0)       & 87090.85(5)   & IRAM-30m &  28.2       & 4.2         & 0.245$\pm$0.008 & 36.5$\pm$0.1 & 2.2$\pm$0.1 & 0.106$\pm$0.007 & 0.006 & N2    \\
HN$^{13}$C (3--2)       & 261263.31(4)  & IRAM-30m &  9.4        & 25.1        & \nodata         & \nodata      & \nodata     & $<$0.099        & 0.033 &       \\
DNC (3--2)             & 228910.489(30) & IRAM-30m &  10.7       & 22.0        & 0.296$\pm$0.035 & 36.5$\pm$0.1 & 1.5$\pm$0.2 & 0.189$\pm$0.031 & 0.030 & N2    \\
\hline
\end{tabular}
\tablefoot{The last column lists the velocity component associated with each row. ``B" indicates the broad component, while ``N1", ``N2", and ``N3" correspond to the 33.1, 36.5, and 44.5~\kms\,components, respectively.}
\normalsize
\end{table*}

\begin{table*}[!hbt]
\caption{Observed properties of CS transitions.}\label{Tab:cs}
\scriptsize
\centering
\begin{tabular}{ccccccccccc}
\hline \hline
Line                   & Frequency   & Telescope &  HPBW    & $E_{\rm u}/k$ & $\int T_{\rm mb}{\rm d}\varv$ & $\varv$ & $\Delta \varv$ & $T_{\rm mb}$&$\sigma$ & note   \\ 
                       & (MHz)       &           &(\arcsec) &  (K)        & (K~\kms)                     & (\kms)  & (\kms)         & (K)       &(K)      &        \\ 
\hline                 
CS (2--1)              & 97980.953(2) & IRAM-30m & 25.1     & 7.1    &  0.418$\pm$0.023 & 29.2$\pm$0.1 & 2.1$\pm$0.2 & 0.190$\pm$0.024 & 0.023 & N1    \\
                       &              &          &          &        &  6.107$\pm$0.024 & 36.2$\pm$0.1 & 2.7$\pm$0.1 & 2.120$\pm$0.027 & 0.023 & N2    \\
                       &              &          &          &        &  0.189$\pm$0.015 & 44.6$\pm$0.1 & 1.3$\pm$0.1 & 0.141$\pm$0.023 & 0.023 & N3    \\
CS (3--2)              & 146969.029(3)& IRAM-30m & 16.7     & 14.1   &  3.747$\pm$0.174 & 34.5$\pm$0.4 & 15.8$\pm$1.0& 0.223$\pm$0.055 & 0.055 &  B    \\
                       &              &          &          &        &  6.450$\pm$0.056 & 36.3$\pm$0.1 & 2.7$\pm$0.1 & 2.230$\pm$0.056 & 0.055 &  N2   \\
CS (5--4)              & 244935.557(3)& IRAM-30m & 10.1     & 35.3   &  4.295$\pm$0.246 & 33.7$\pm$0.6 & 18.9$\pm$1.6& 0.213$\pm$0.050 & 0.046 & B     \\
                       &              &          &          &        &  5.293$\pm$0.085 & 36.7$\pm$0.1 & 3.0$\pm$0.1 & 1.640$\pm$0.049 & 0.046 & N2    \\
CS (7--6)              & 342882.850(5)& APEX     & 18.2     &  65.8  &  6.608$\pm$0.264 & 36.2$\pm$0.1 & 4.7$\pm$0.3 & 1.320$\pm$0.106 & 0.102 & N2    \\ 
C$^{34}$S (5--4)       & 241016.089(4)& IRAM-30m & 10.2     & 27.8   &  0.578$\pm$0.108 & 36.3$\pm$0.3 & 4.3$\pm$1.1 & 0.126$\pm$0.050 & 0.056 & N2    \\
\hline
\end{tabular}
\tablefoot{The last column lists the velocity component associated with each row. ``B" indicates the broad component, while ``N1", ``N2", and ``N3" correspond to the 33.1, 36.5, and 44.5~\kms\,components, respectively.}
\normalsize
\end{table*}

\begin{table*}[!hbt]
\caption{Observed properties of SiO transitions.}\label{Tab:sio}
\scriptsize
\centering
\begin{tabular}{ccccccccccc}
\hline \hline
Line              & Frequency & Telescope & HPBW  & $E_{\rm u}/k$ & $\int T_{\rm mb}{\rm d}\varv$ & $\varv$ & $\Delta \varv$ & $T_{\rm mb}$&$\sigma$ & note   \\ 
                  & (MHz)     &           &       &  (K)        & (K~\kms)                     & (\kms)  & (\kms)         & (K)       &(K)      &         \\ 
\hline
SiO (2--1)        & 86846.96(5)& IRAM-30m & 28.3  &  6.3        & 3.423$\pm$0.150 & 31.3$\pm$0.7 & 31.8$\pm$1.9 & 0.101$\pm$0.025 & 0.024 &    B  \\ 
                  &            &          &       &             & 1.392$\pm$0.065 & 36.8$\pm$0.2 & 9.9$\pm$0.6  & 0.132$\pm$0.025 & 0.024 &    N2 \\
SiO (4--3)        & 173688.31(8)& IRAM-30m & 14.2  &  20.8       & 10.060$\pm$0.194& 31.4$\pm$0.3 & 29.4$\pm$0.6 & 0.322$\pm$0.046 & 0.048 &    B  \\
                  &            &          &       &             & 3.212$\pm$0.070 & 37.0$\pm$0.1 & 8.3$\pm$0.2  & 0.363$\pm$0.043 & 0.048 &    N2 \\
SiO (5--4)        & 217104.98(8)& IRAM-30m & 11.3  &  31.3       & 10.320$\pm$0.241& 30.9$\pm$0.4 & 33.0$\pm$0.9 & 0.294$\pm$0.032 & 0.030 &    B  \\
                  &            &           &       &             & 2.817$\pm$0.098 & 37.1$\pm$0.2 & 9.4$\pm$0.4  & 0.282$\pm$0.032 & 0.030 &    N2 \\
SiO (14-13)       & 607607.58(7) & Herschel & 35.4  &  218.8      & 0.156$\pm$0.021 & 36.1$\pm$0.2 & 3.2$\pm$0.6  & 0.046$\pm$0.014 & 0.014 &   N2  \\
$^{29}$SiO (5--4) & 214385.76(1) & IRAM-30m & 11.5  &  30.9       & \nodata         & \nodata      & \nodata      & $<$0.084        & 0.027 &       \\
  
\hline
\end{tabular}
\tablefoot{The last column lists the velocity component associated with each row. ``B" indicates the broad component, while ``N1", ``N2", and ``N3" correspond to the 33.1, 36.5, and 44.5~\kms\,components, respectively.}
\normalsize
\end{table*}

\begin{table*}[!hbt]
\caption{Observed properties of SO, SO$_{2}$, N$_{2}$H$^{+}$, HDO, c-C$_{3}$H$_{2}$, HC$_{3}$N, and C$_{4}$H transitions.}\label{Tab:others}
\scriptsize
\centering
\begin{tabular}{ccccccccccc}
\hline \hline
Line                   & Frequency  & Telescope & HPBW     & $E_{\rm u}/k$ & $\int T_{\rm mb}{\rm d}\varv$ & $\varv$ & $\Delta \varv$ & $T_{\rm mb}$&$\sigma$& note  \\ 
                       & (MHz)      &  &(\arcsec) &  (K)        &  (K~\kms)                     & (\kms)  & (\kms)         & (K)       &(K)      &            \\ 
\hline                 
SO (3$_{2}$--2$_{1}$)           & 99299.87(10) & IRAM-30m   &  24.8     & 9.2         & 2.225$\pm$0.143 & 32.9$\pm$0.2 & 28.3$\pm$0.7 & 0.074$\pm$0.008 & 0.008 &  B   \\ 
                                &              &            &           &              & 0.121$\pm$0.008 & 29.2$\pm$0.1 & 1.2$\pm$0.1 & 0.093$\pm$0.008 & 0.008 &  N1   \\
                               &               &            &           &              & 2.265$\pm$0.016 & 36.2$\pm$0.1 & 4.1$\pm$0.1 & 0.522$\pm$0.010 & 0.008 &  N2   \\
SO (4$_{3}$--3$_{2}$)           &  138178.60(3) & IRAM-30m   &  18.3     & 15.9        & 4.129$\pm$0.176 & 33.1$\pm$0.5 & 27.1$\pm$1.3 & 0.143$\pm$0.037 & 0.038 & B    \\
                               &             &              &           &              & 3.288$\pm$0.061 & 36.5$\pm$0.1 & 5.6$\pm$0.1 & 0.554$\pm$0.038 & 0.038 & N2    \\
SO$_{2}$ (3$_{2,2}$--2$_{1,1}$)   &   208700.336(1) & IRAM-30m &  11.8     &  15.3       & \nodata         & \nodata      & \nodata     & $<$0.060        & 0.020 &       \\
SO$_{2}$ (3$_{3,1}$--3$_{2,2}$)   &   255958.044(1) & IRAM-30m &  9.6      &  27.6       & \nodata         & \nodata      & \nodata     & $<$0.060        & 0.020 &       \\
N$_{2}$H$^{+}$ (3--2)             & 279511.749(3)  & IRAM-30m  & 8.8  &   26.8       & \nodata & \nodata & \nodata & $<$0.339 & 0.113 &     \\
HDO (1$_{1,0}$--1$_{1,1}$)        & 80578.295(20)  & IRAM-30m & 30.5 &  46.8        & \nodata & \nodata & \nodata & $<$0.069 & 0.023 &     \\
c-C$_{3}$H$_{2}$ (3$_{1,2}$--2$_{2,1}$)     & 145089.606(2) & IRAM-30m  & 17.0 &  16.0        & 0.490$\pm$0.036 & 36.8$\pm$0.1 & 1.6$\pm$0.1 & 0.291$\pm$0.034 & 0.047 & N2, ortho  \\
HC$_{3}$N (16--15)                            & 145560.960(10) & IRAM-30m  & 16.9 &  59.4        & 0.587$\pm$0.041 & 36.5$\pm$0.1 & 2.1$\pm$0.2 & 0.257$\pm$0.042 & 0.045 & N2  \\
stacked C$_{4}$H ($N$=11--10)                 & 104666.565(3)  & IRAM-30m & 23.5 &  30.1        & 0.071$\pm$0.008 & 36.7$\pm$0.1 & 1.2$\pm$0.2 & 0.056$\pm$0.014 & 0.010 & N2  \\
\hline
\end{tabular}
\tablefoot{The last column lists the velocity component associated with each row. ``B" indicates the broad component, while ``N1", ``N2", and ``N3" correspond to the 33.1, 36.5, and 44.5~\kms\,components, respectively.}
\normalsize
\end{table*}


\begin{table*}[!hbt]
\caption{Observed properites of HCO$^{+}$ transitions.}\label{Tab:hcop}
\scriptsize
\centering
\begin{tabular}{ccccccccccc}
\hline \hline
Line                   & Frequency    & Telescope  &  HPBW        & $E_{\rm u}/k$& $\int T_{\rm mb}{\rm d}\varv$ & $\varv$ & $\Delta \varv$ & $T_{\rm mb}$&$\sigma$  & Note  \\ 
                       & (MHz)        &  &  (\arcsec)   &  (K)       &  (K~\kms)                     & (\kms)  & (\kms)         & (K)       &(K)  &        \\ 
\hline                 
HCO$^{+}$ (1--0)        & 89188.525(4)& IRAM-30m   &  27.6   &  4.3        & 8.450$\pm$0.112 & 33.5$\pm$0.2 & 43.6$\pm$0.8 & 0.182$\pm$0.015 & 0.014 & B    \\
                        &             &            &         &             & 2.286$\pm$0.034 & 37.1$\pm$0.1 & 2.1$\pm$0.1 & 1.020$\pm$0.026 & 0.014 & N2    \\
HCO$^{+}$ (2--1)        & 178375.056(8)& IRAM-30m  &  13.8   & 12.8        & 17.530$\pm$1.260 & 35.2$\pm$1.1 & 40.5$\pm$3.7 & 0.406$\pm$0.223 & 0.213 & B    \\
                        &              &           &         &             & 4.677$\pm$0.213 & 37.2$\pm$0.1 & 2.3$\pm$0.2 & 1.900$\pm$0.222 & 0.213 & N2     \\
HCO$^{+}$ (3--2)        & 267557.626(1)& IRAM-30m  &  9.2    &  25.7       & 12.450$\pm$0.235 & 34.4$\pm$0.2 & 26.6$\pm$0.7 & 0.440$\pm$0.045 & 0.045 & B    \\
                        &              &           &         &             & 7.298$\pm$0.084 & 36.5$\pm$0.1 & 3.0$\pm$0.1 & 2.250$\pm$0.052 & 0.044 & N2    \\ 
HCO$^{+}$ (4--3)        & 356734.223(2)& APEX      &  17.5   &  42.8       & 4.722$\pm$0.172 & 36.4$\pm$0.1 & 3.4$\pm$0.2 & 1.310$\pm$0.090 & 0.084 & N2    \\
H$^{13}$CO$^{+}$ (1--0) &  86754.288(5)& IRAM-30m   &  28.4   &  4.2        & 0.045$\pm$0.005 & 29.4$\pm$0.1 & 0.6$\pm$0.1 & 0.072$\pm$0.012 & 0.012 & N1    \\
                        &              &            &         &             & 1.130$\pm$0.009 & 36.5$\pm$0.1 & 1.9$\pm$0.1 & 0.571$\pm$0.012 & 0.012 & N2    \\
H$^{13}$CO$^{+}$ (2--1) & 173506.700(7)& IRAM-30m   &  14.2   & 12.5        & 0.991$\pm$0.063 & 36.6$\pm$0.1 & 1.7$\pm$0.1 & 0.539$\pm$0.081 & 0.080 & N2    \\
H$^{13}$CO$^{+}$ (3--2) & 260255.339(35)& IRAM-30m  &  9.5    & 25.0        & 1.131$\pm$0.046 & 37.0$\pm$0.1 & 1.2$\pm$0.1 & 0.855$\pm$0.045 & 0.044 & N2    \\
DCO$^{+}$ (2--1)        & 144077.289(5) & IRAM-30m  &  17.1   & 10.4        & 0.658$\pm$0.019 & 36.5$\pm$0.1 & 1.7$\pm$0.1 & 0.367$\pm$0.033 & 0.033 & N2    \\
DCO$^{+}$ (3--2)        & 216112.582(5) & IRAM-30m  &  11.4   & 20.7        & 0.515$\pm$0.022 & 36.6$\pm$0.1 & 1.4$\pm$0.1 & 0.335$\pm$0.018 & 0.018 & N2    \\
\hline
\end{tabular}
\tablefoot{The last column lists the velocity component associated with each row. ``B" indicates the broad component, while ``N1", ``N2", and ``N3" correspond to the 33.1, 36.5, and 44.5~\kms\,components, respectively.}
\normalsize
\end{table*}

\begin{table*}[!hbt]
\caption{Observed properties of H$_{2}$CO transitions.}\label{Tab:h2co}
\scriptsize
\centering
\begin{tabular}{ccccccccccc}
\hline \hline
Line                   & Frequency  & Telescope  & HPBW      & $E_{\rm u}/k$ & $\int T_{\rm mb}{\rm d}\varv$ & $\varv$ & $\Delta \varv$ & $T_{\rm mb}$&$\sigma$ & note     \\ 
                       & (MHz)      &            & (\arcsec)  &  (K)       & (K~\kms)                     & (\kms)  & (\kms)         & (K)       &(K)      &          \\ 
\hline                 
H$_{2}$CO (2$_{0,2}$--1$_{0,1}$)  & 145602.949(10)& IRAM-30m & 16.9        &  10.5       & 1.174$\pm$0.128 & 34.2$\pm$0.7 & 13.5$\pm$1.2 & 0.082$\pm$0.035 & 0.035 & B, para \\ 
                                 &                &          &             &             & 0.066$\pm$0.011 & 29.3$\pm$0.0 & 0.3$\pm$0.1 & 0.184$\pm$0.035 & 0.035 & N1, para  \\
                                 &                &          &             &             & 2.368$\pm$0.035 & 36.4$\pm$0.0 & 2.4$\pm$0.0 & 0.939$\pm$0.036 & 0.035 & N2, para  \\
H$_{2}$CO (2$_{1,1}$--1$_{1,0}$)    & 150498.334(10)& IRAM-30m & 16.3        &  22.6       & 1.954$\pm$0.538 & 36.5$\pm$0.8 & 9.5$\pm$2.3 & 0.193$\pm$0.049 & 0.048 & B, ortho   \\ 
                                 &            & &            &             & 0.179$\pm$0.029 & 29.2$\pm$0.1 & 1.1$\pm$0.2 & 0.155$\pm$0.049 & 0.048 & N1, ortho  \\
                                 &            &  &           &             & 2.220$\pm$0.046 & 36.5$\pm$0.0 & 2.2$\pm$0.1 & 0.942$\pm$0.050 & 0.048 & N2, ortho  \\
H$_{2}$CO (3$_{0,3}$--2$_{0,2}$)    & 218222.192(10) & IRAM-30m & 11.3        &  21.0       & 1.600$\pm$0.065 & 36.1$\pm$0.2 & 9.1$\pm$0.8 & 0.166$\pm$0.018 & 0.017 & B, para   \\
                                 &            &             &             & 1.154$\pm$0.027 & 36.5$\pm$0.0 & 2.0$\pm$0.1 & 0.533$\pm$0.019 & 0.017 & N2, para  \\
H$_{2}$CO (3$_{1,2}$--2$_{1,1}$)    & 225697.775(10) & IRAM-30m & 10.9        &  33.4       & 1.559$\pm$0.073 & 35.9$\pm$0.2 & 9.2$\pm$0.7 & 0.159$\pm$0.021 & 0.020 & B, ortho   \\
                                 &            &             &             & 1.320$\pm$0.029 & 36.6$\pm$0.0 & 1.9$\pm$0.1 & 0.656$\pm$0.021 & 0.021 & N2, ortho  \\
HDCO (3$_{1,2}$--2$_{1,1}$)        & 201341.362(50)  & IRAM-30m & 12.2       &  27.3       & 0.150$\pm$0.016 & 36.6$\pm$0.1 & 1.3$\pm$0.1 & 0.109$\pm$0.014 & 0.015 & N2  \\       
HDCO (1$_{1,1}$--0$_{0,0}$)        & 227668.453(50)  & IRAM-30m & 10.8       &  10.9       & \nodata         & \nodata      & \nodata      & $<$0.163       & 0.021 &     \\ 
\hline
\end{tabular}
\tablefoot{The last column lists the velocity component associated with each row. ``B" indicates the broad component, while ``N1", ``N2", and ``N3" correspond to the 33.1, 36.5, and 44.5~\kms\,components, respectively.}
\normalsize
\end{table*}

\begin{table*}[!hbt]
\caption{Observed properties of CH$_{3}$OH transitions.}\label{Tab:ch3oh}
\scriptsize
\centering
\begin{tabular}{ccccccccccc}
\hline \hline
Line                   & Frequency  & Telescope & HPBW         & $E_{\rm u}/k$ & $\int T_{\rm mb}{\rm d}\varv$ & $\varv$ & $\Delta \varv$ & $T_{\rm mb}$&$\sigma$ & Note  \\ 
                       & (MHz)      &           & (\arcsec)    &  (K)        & (K~\kms)                     & (\kms)  & (\kms)         & (K)       &(K)      &            \\ 
\hline
\multicolumn{10}{c}{E-type}\\                 
\hline
CH$_{3}$OH (2$_{-1}$--1$_{-1}$ E)    &  96739.358(2)  & IRAM-30m  & 25.4     &  12.5       & 0.351$\pm$0.013 & 35.7$\pm$0.1 & 2.6$\pm$0.1 & 0.127$\pm$0.010 & 0.010 & N2    \\ 
CH$_{3}$OH (2$_{0}$--1$_{0}$ E)      &  96744.545(2)  & IRAM-30m   & 25.4     &   20.1      & 0.182$\pm$0.018 & 35.9$\pm$0.2 & 3.7$\pm$0.4 & 0.047$\pm$0.010 & 0.010 &  N2   \\
CH$_{3}$OH (2$_{1}$--1$_{1}$ E)      &  96755.501(2)  & IRAM-30m   & 25.4     &   28.0      & 0.098$\pm$0.014 & 36.2$\pm$0.3 & 4.1$\pm$0.7 & 0.022$\pm$0.010 & 0.010 &  N2   \\
CH$_{3}$OH (3$_{0}$--2$_{0}$ E)      &  145093.754(3) & IRAM-30m   & 17.0     &   27.1      & 0.425$\pm$0.031 & 36.2$\pm$0.1 & 3.8$\pm$0.4 & 0.107$\pm$0.024 & 0.024 &  N2   \\
CH$_{3}$OH (3$_{-1}$--2$_{-1}$ E)    &  145097.435(3) & IRAM-30m   & 17.0     &    19.5     & 0.820$\pm$0.027 & 36.1$\pm$0.1 & 3.0$\pm$0.1 & 0.255$\pm$0.025 & 0.042 &  N2   \\
CH$_{3}$OH (3$_{2}$--2$_{2}$ E)      &  145126.191(3) & IRAM-30m   & 17.0     &   36.2      & \nodata         & \nodata      & \nodata     & $<$0.075        & 0.025 &  N2   \\
CH$_{3}$OH (3$_{-2}$--2$_{-2}$ E)    &  145126.386(3) & IRAM-30m   &  17.0    &    39.8     & \nodata         & \nodata      & \nodata     & $<$0.075        & 0.025 &  N2   \\
CH$_{3}$OH (3$_{1}$--2$_{1}$ E)      &  145131.864(3) & IRAM-30m   & 17.0     &    35.0     & 0.226$\pm$0.032 & 36.2$\pm$0.3 & 4.3$\pm$0.8 & 0.049$\pm$0.022 & 0.023 &  N2   \\
CH$_{3}$OH (5$_{-1}$--4$_{-1}$ E)    &  241767.234(4) & IRAM-30m   &  10.2    &    40.4     & 0.416$\pm$0.036 & 36.3$\pm$0.1 & 3.2$\pm$0.4 & 0.121$\pm$0.020 & 0.020 &  N2   \\
\hline                                                     
\multicolumn{10}{c}{A-type}\\                              
\hline                                                     
CH$_{3}$OH (2$_{0}$--1$_{0}$ A$^{+}$) &  96741.371(2) & IRAM-30m    &  25.4    &   7.0       & 0.533$\pm$0.013 & 35.9$\pm$0.1 & 2.6$\pm$0.1 & 0.190$\pm$0.010 & 0.010 & N2    \\
CH$_{3}$OH (3$_{0}$--2$_{0}$ A$^{+}$) &  145103.185(3)& IRAM-30m    &  17.0    &  13.9       & 0.985$\pm$0.050 & 36.1$\pm$0.1 & 3.0$\pm$0.2 & 0.308$\pm$0.045 & 0.046 & N2    \\ 
CH$_{3}$OH (5$_{0}$--4$_{0}$ A$^{+}$) &  241791.352(4)& IRAM-30m    &  10.2    &  34.8       & 0.304$\pm$0.034 & 36.1$\pm$0.1 & 2.4$\pm$0.4 & 0.117$\pm$0.020 & 0.020 & N2    \\
\hline
\end{tabular}
\tablefoot{The last column lists the velocity component associated with each row. ``B" indicates the broad component, while ``N1", ``N2", and ``N3" correspond to the 33.1, 36.5, and 44.5~\kms\,components, respectively.}
\normalsize
\end{table*}


\begin{table*}[!hbt]
\caption{Excitation temperatures and column densities of selected molecules toward NDO.}\label{Tab:colum}
\normalsize
\centering
\begin{tabular}{cccccc}
\hline \hline
Molecule            & $N_{\rm det}$      &  Size       & $T_{\rm e}$ & $N$ & $X=N/N_{\rm H_2}$ \\ 
                    &           &  (\arcsec)  &  (K)        & (cm$^{-2}$)  & \\ 
 (1)                &   (2)     &  (3)  &  (4)        & (5)  & (6)\\                     
\hline   
\multicolumn{6}{c}{Single dish: $N_{\rm H_2}=(4.4\pm0.6)\times 10^{22}$~cm$^{-2}$}\\
\hline
CH$_{3}$OH            & 10     &  24       & $11.5\pm0.9$  & $(7.5\pm0.6)\times 10^{13}$  & $(1.7\pm 0.3)\times 10^{-9}$  \\
H$_{2}$CO             & 4      &  24       & $9.0\pm 1.1$  & $(1.4\pm0.2)\times 10^{13}$  & $(3.2\pm 0.6)\times 10^{-10}$ \\
HDCO\tablefootmark{*} & 1      &  24       & 10            & $(2.1\pm 0.4)\times 10^{12}$ & $(4.8\pm 1.1)\times 10^{-11}$ \\
H$^{13}$CN            & 3      &  24       & $8.7\pm 2.1$  & $(2.6\pm 0.8)\times 10^{12}$ & $(5.9\pm 2.0)\times 10^{-11}$ \\
DCN                   & 2      &  24       & $5.2\pm 1.4$  & $(2.8\pm 1.7)\times 10^{12}$ & $(6.4\pm 4.0)\times 10^{-11}$ \\
HN$^{13}$C\tablefootmark{*}& 1 &  24       & 10            & $(1.2\pm 0.2)\times 10^{12}$ & $(2.7\pm 0.6)\times 10^{-11}$ \\
DNC\tablefootmark{*}  & 1      &  24       & 10            & $(4.9\pm 1.1)\times 10^{11}$ & $(1.1\pm 0.3)\times 10^{-11}$ \\
H$^{13}$CO$^{+}$      & 3      &  24       & $7.4\pm 0.8$  & $(1.6\pm 0.3)\times 10^{12}$ & $(3.6\pm 0.8)\times 10^{-11}$ \\
DCO$^{+}$             & 2      &  24       & $8.2\pm 2.0$  & $(6.9\pm 1.8)\times 10^{11}$ & $(1.6\pm 0.2)\times 10^{-11}$ \\
C$^{18}$O             & 2      &  24       & $29.0\pm 2.8$ & $(6.7\pm 1.2)\times 10^{15}$ & $(1.5\pm 0.3)\times 10^{-7}$  \\
C$^{17}$O\tablefootmark{*}             & 1      &  24       & 29            & $(2.6\pm 0.5)\times 10^{15}$ & $(5.9\pm 1.4)\times 10^{-8}$  \\
c-C$_{3}$H$_{2}$\tablefootmark{*}      & 1      &  24       & 10            & $(8.4\pm 1.8)\times 10^{12}$ & $(1.9\pm 0.5)\times 10^{-10}$ \\
C$_{4}$H\tablefootmark{*}              & 1      &  24       & 10            & $(1.6\pm 0.4)\times 10^{13}$ & $(3.6\pm 1.0)\times 10^{-10}$ \\
HC$_{3}$N\tablefootmark{*}             & 1      &  24       & 10            & $(7.9\pm 1.7)\times 10^{13}$ & $(1.8\pm 0.5)\times 10^{-9}$ \\
\hline
\multicolumn{6}{c}{SMA\tablefootmark{\S}: $N_{\rm H_2}=(6.4\pm0.5)\times 10^{22}$~cm$^{-2}$} \\
\hline
CH$_{3}$OH           & 23      &  -       & 41.7$\pm$4.5  & $(9.9\pm0.8)\times 10^{14}$  & $(1.5\pm 0.2)\times 10^{-8}$ \\
H$_{2}$CO            & 6       &  -       & 20.6$\pm$5.4  & $(8.9\pm1.0)\times 10^{13}$  & $(1.4\pm 0.2)\times 10^{-9}$ \\
C$^{34}$S            & 2       &  -       & 13.1$\pm$5.5  & $(2.5\pm0.9)\times 10^{13}$  & $(3.9\pm 1.4)\times 10^{-10}$ \\
SO$_{2}$             & 10      &  -       & 53.3$\pm$8.4  & $(3.3\pm0.3)\times 10^{14}$  & $(5.2\pm 0.6)\times 10^{-9}$ \\
SO                   & 6       &  -       & 22.0$\pm$6.1  & $(9.1\pm2.1)\times 10^{14}$  & $(1.4\pm 0.3)\times 10^{-8}$ \\
HC$_{3}$N            & 3       &  -       & 37.9$\pm$24.9 & $(1.8\pm9.7)\times 10^{13}$  & $(0.3\pm 1.5)\times 10^{-9}$ \\
C$^{18}$O\tablefootmark{*} & 1 &  -       & 40            & $(5.9\pm 1.3)\times 10^{15}$ & $(9.2\pm 2.2)\times 10^{-8}$ \\
C$^{17}$O\tablefootmark{*} & 1 &  -       & 40            & $(1.2\pm 0.3)\times 10^{15}$ & $(1.9\pm 0.5)\times 10^{-8}$ \\
H$^{13}$CN\tablefootmark{*}& 1 &  -       & 40            & $(4.7\pm 1.1)\times 10^{12}$ & $(7.3\pm 1.8)\times 10^{-11}$ \\
DCN\tablefootmark{*}       & 1 &  -       & 40            & $(5.4\pm 1.2)\times 10^{12}$ & $(8.4\pm 2.0)\times 10^{-11}$ \\
$^{29}$SiO\tablefootmark{*}& 1 &  -       & 40            & $(3.5\pm 0.9)\times 10^{12}$ & $(5.5\pm 1.4)\times 10^{-10}$ \\
C$_{2}$H\tablefootmark{*}  & 1 &  -       & 40            & $(6.7\pm 2.4)\times 10^{13}$ & $(1.0\pm 0.4)\times 10^{-9}$ \\
\hline
\end{tabular}
\normalsize
\tablefoot{(1) Molecule. (2) Number of detected lines. (3) Source size. (4) Excitation temperature. (5) Molecular column density. (6) Molecular abundance relative to H$_{2}$. \tablefoottext{*}{For molecules with only one detected transition, the excitation temperature was fixed (10~K for single-dish data and 40~K for SMA data) to estimate the molecular column density.} \tablefoottext{\S}{The column densities derived from the SMA data are beam-averaged values at a restoring beam of 3\rlap{.}\arcsec60$\times$3\rlap{.}\arcsec29.} }
\end{table*}

\section{Figures}\label{app.b}
 
\begin{figure}[!htbp]
\centering
\includegraphics[width = 0.45 \textwidth]{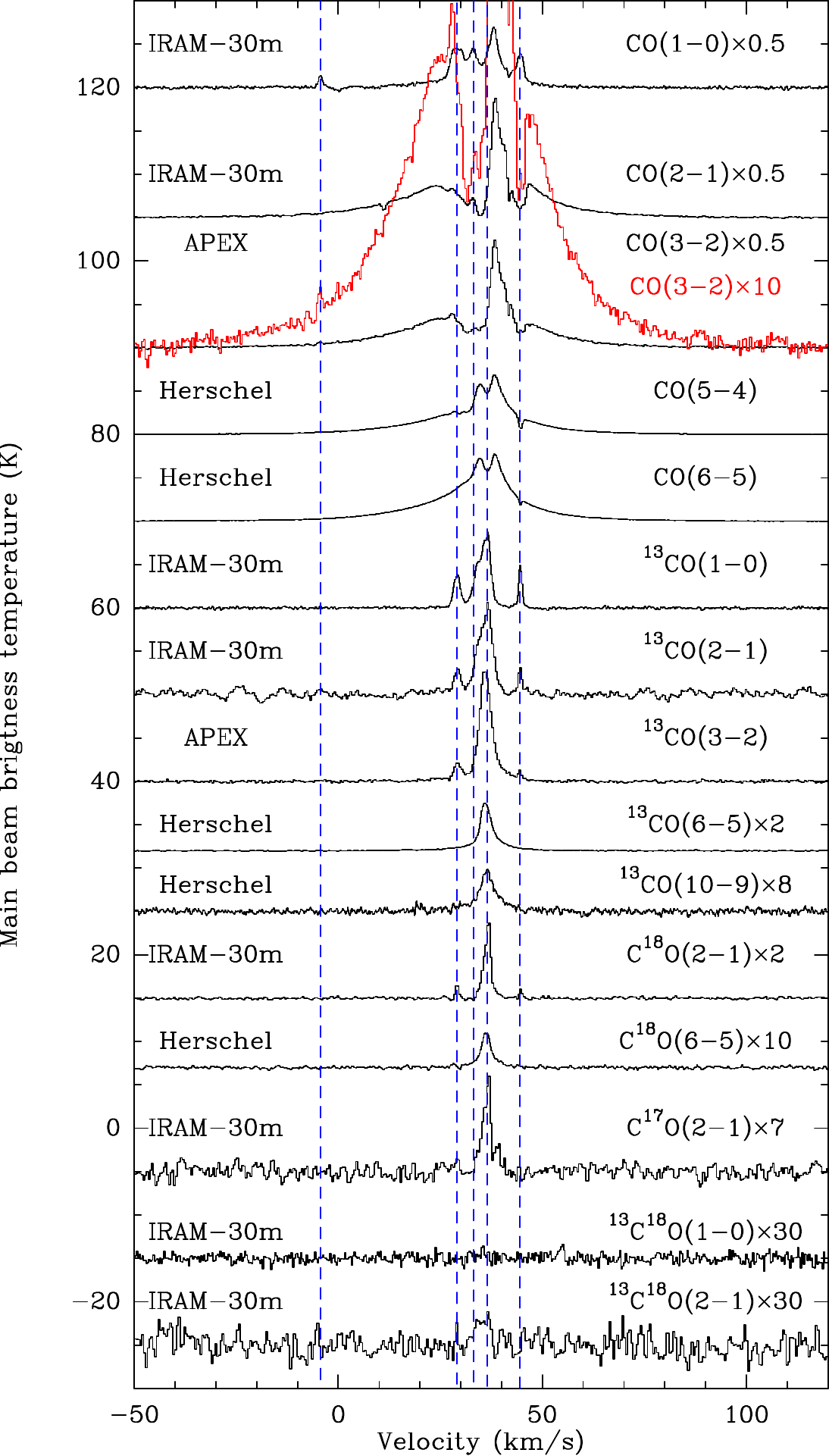}
\caption{{Observed spectra of CO and its isotopologues toward NDO. The transition is given in the upper right of each spectrum. The five dashed vertical lines mark the velocity components at $-$4.4~\kms, 29.1~\kms, 33.1~\kms, 36.5~\kms, and 44.5~\kms, respectively.}\label{Fig:co}}
\end{figure}

\begin{figure}[!htbp]
\centering
\includegraphics[width = 0.45 \textwidth]{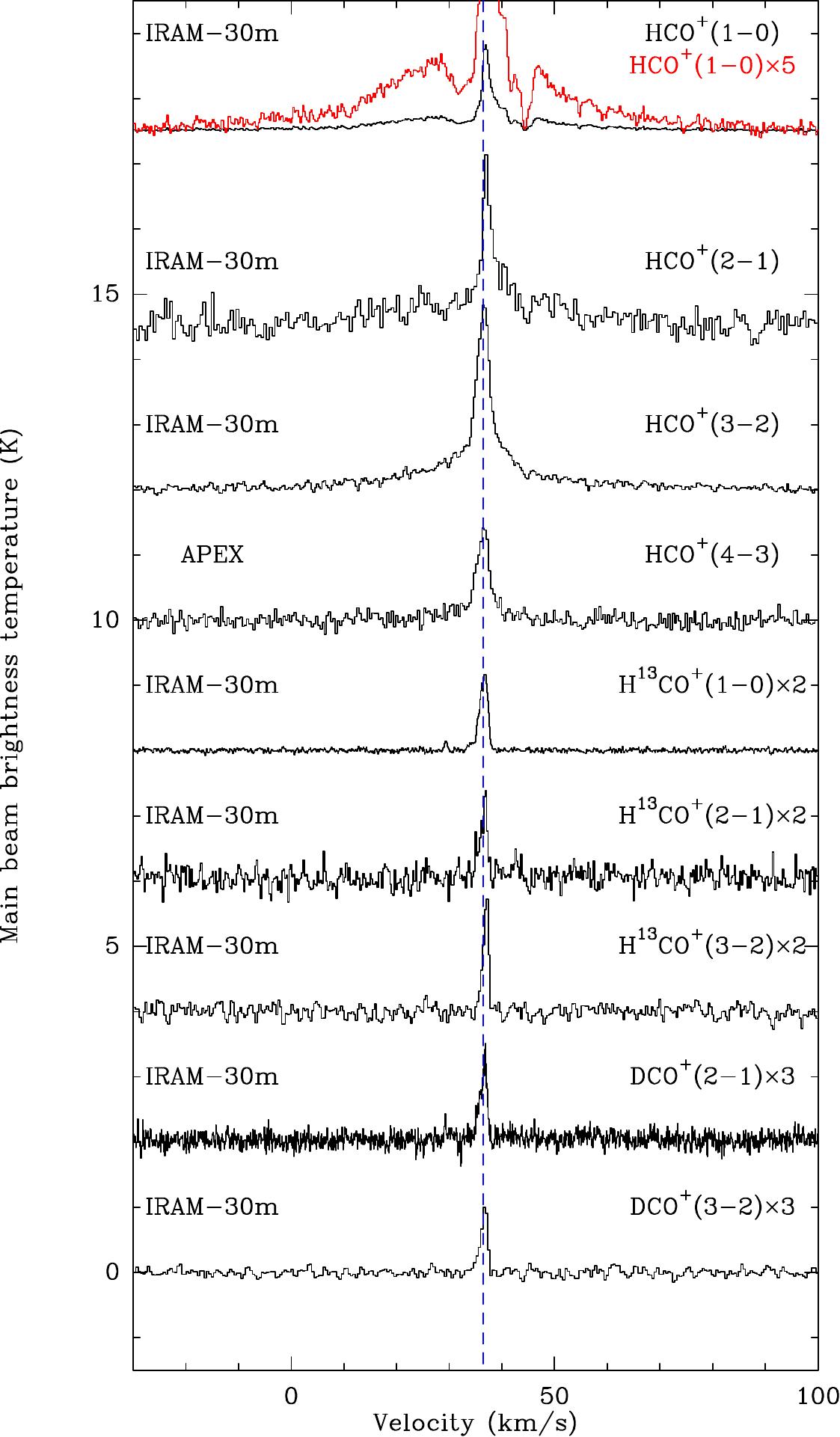}
\caption{{Observed spectra of HCO$^{+}$ and its isotopologues toward NDO. The corresponding transition is indicated in the upper-right of each spectrum. The dashed vertical line marks the systemic velocity of 36.5~\kms. The factors given after each transition name denote the scaling applied for visualization.}\label{Fig:hcop}}
\end{figure}

\begin{figure}[!htbp]
\centering
\includegraphics[width = 0.45 \textwidth]{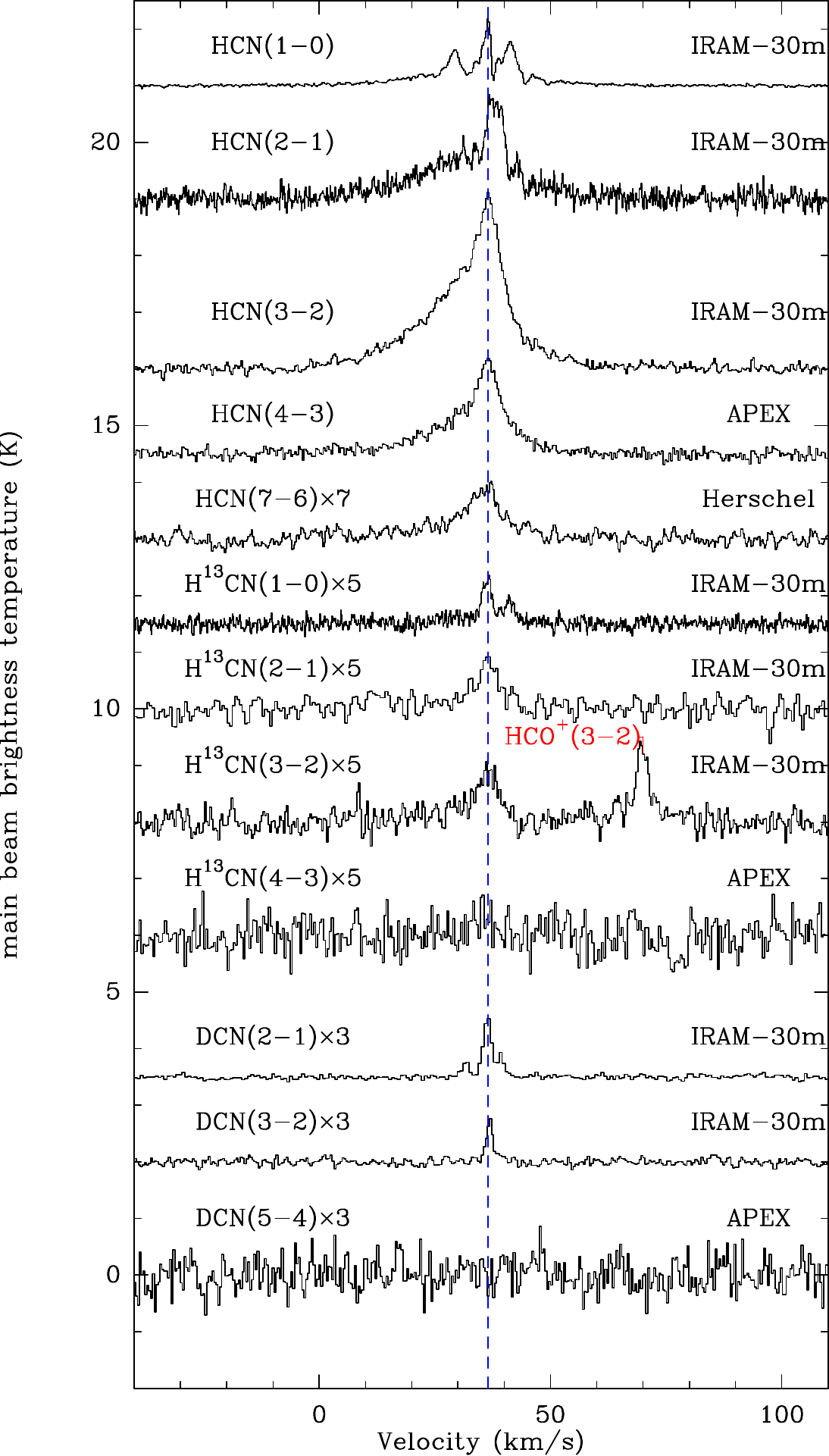}
\caption{{Observed spectra of HCN and its isotopologues toward NDO. The corresponding transition is indicated in the upper-left of each spectrum. The dashed vertical line marks the systemic velocity of 36.5~\kms. The line next to H$^{13}$CN (3--2) arises from HCO$^{+}$ (3--2) from its image band.}\label{Fig:hcn}}
\end{figure}

\begin{figure}[!htbp]
\centering
\includegraphics[width = 0.45 \textwidth]{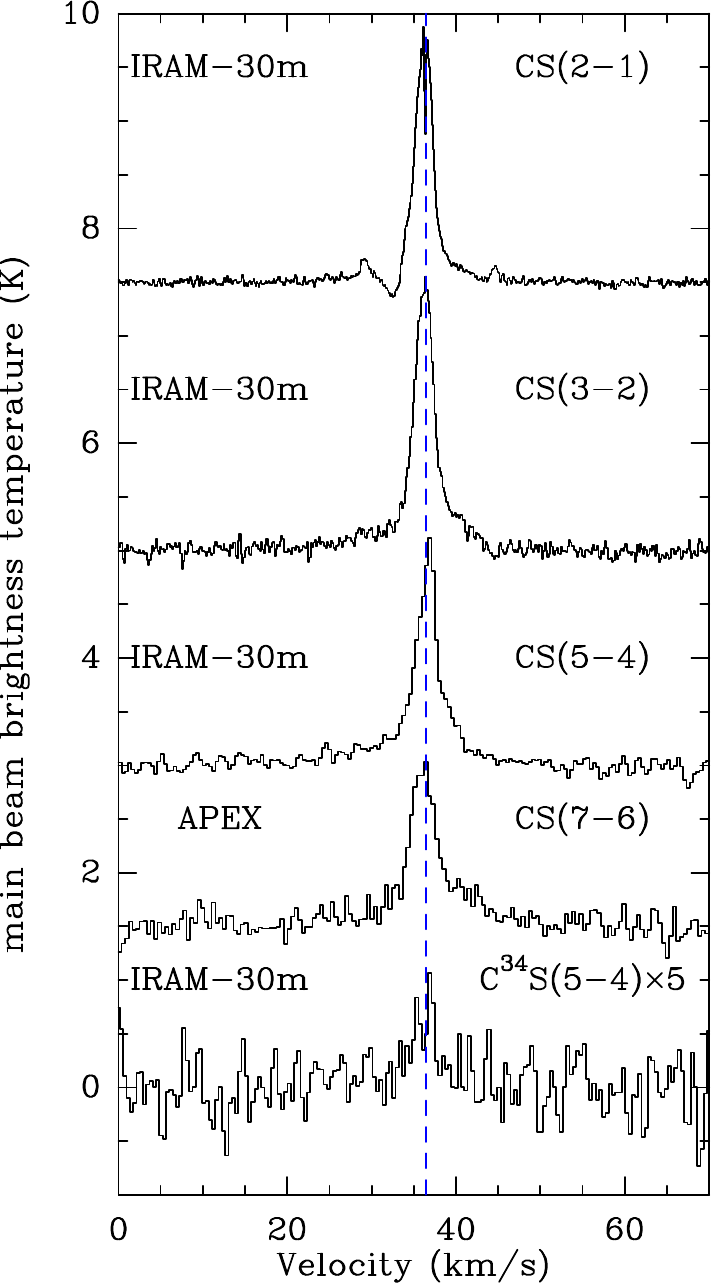}
\caption{{Observed spectra of CS and its isotopologues toward NDO. The corresponding transition is indicated in the upper-right of each spectrum. The dashed vertical line marks the systemic velocity of 36.5~\kms.}\label{Fig:cs}}
\end{figure}

\begin{figure}[!htbp]
\centering
\includegraphics[width = 0.45 \textwidth]{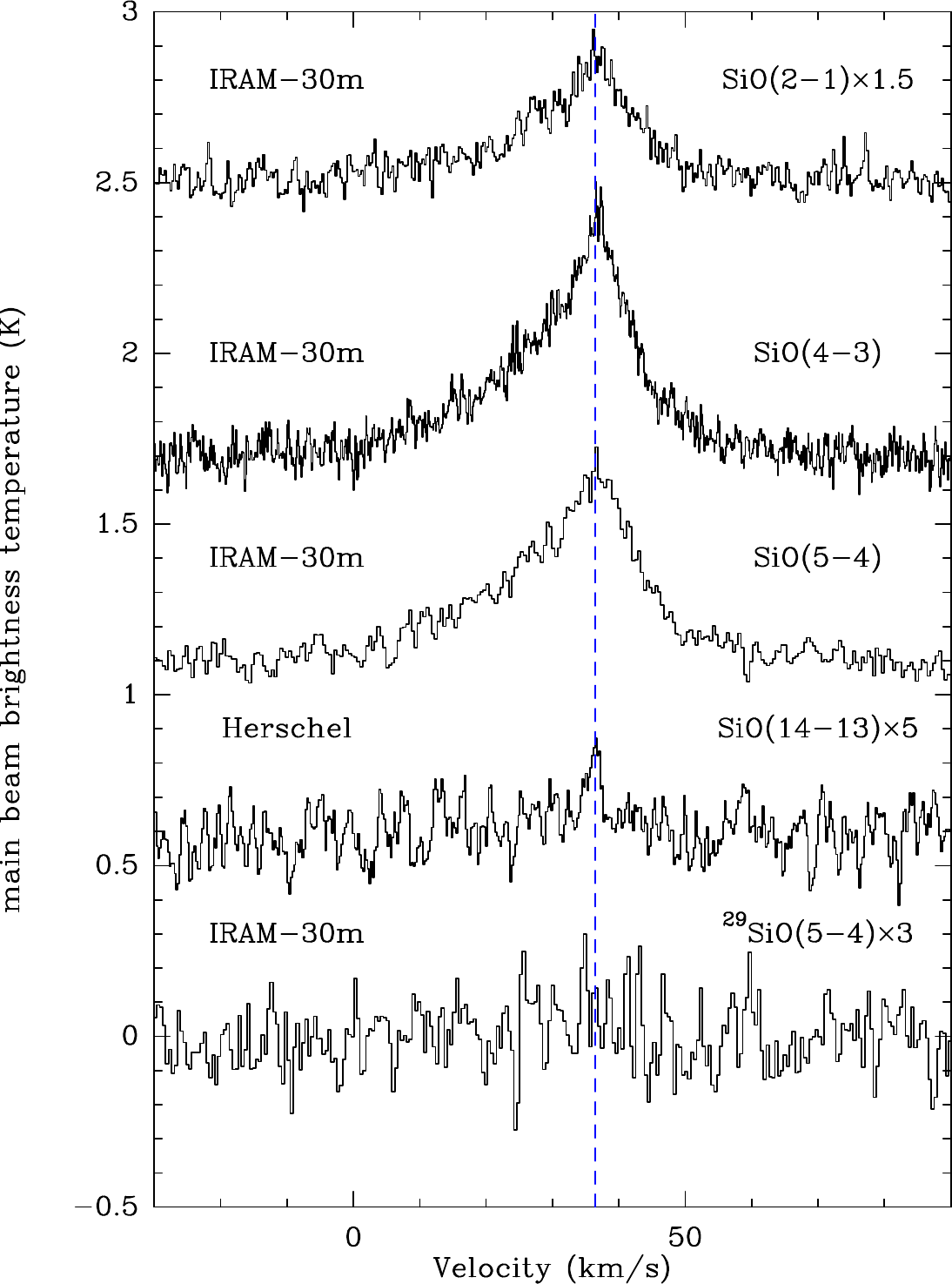}
\caption{{Observed spectra of SiO and its isotopologues toward NDO. The corresponding transition is indicated in the upper-right of each spectrum. The dashed vertical line marks the systemic velocity of 36.5~\kms.}\label{Fig:sio}}
\end{figure}

\begin{figure}[!htbp]
\centering
\includegraphics[width = 0.45 \textwidth]{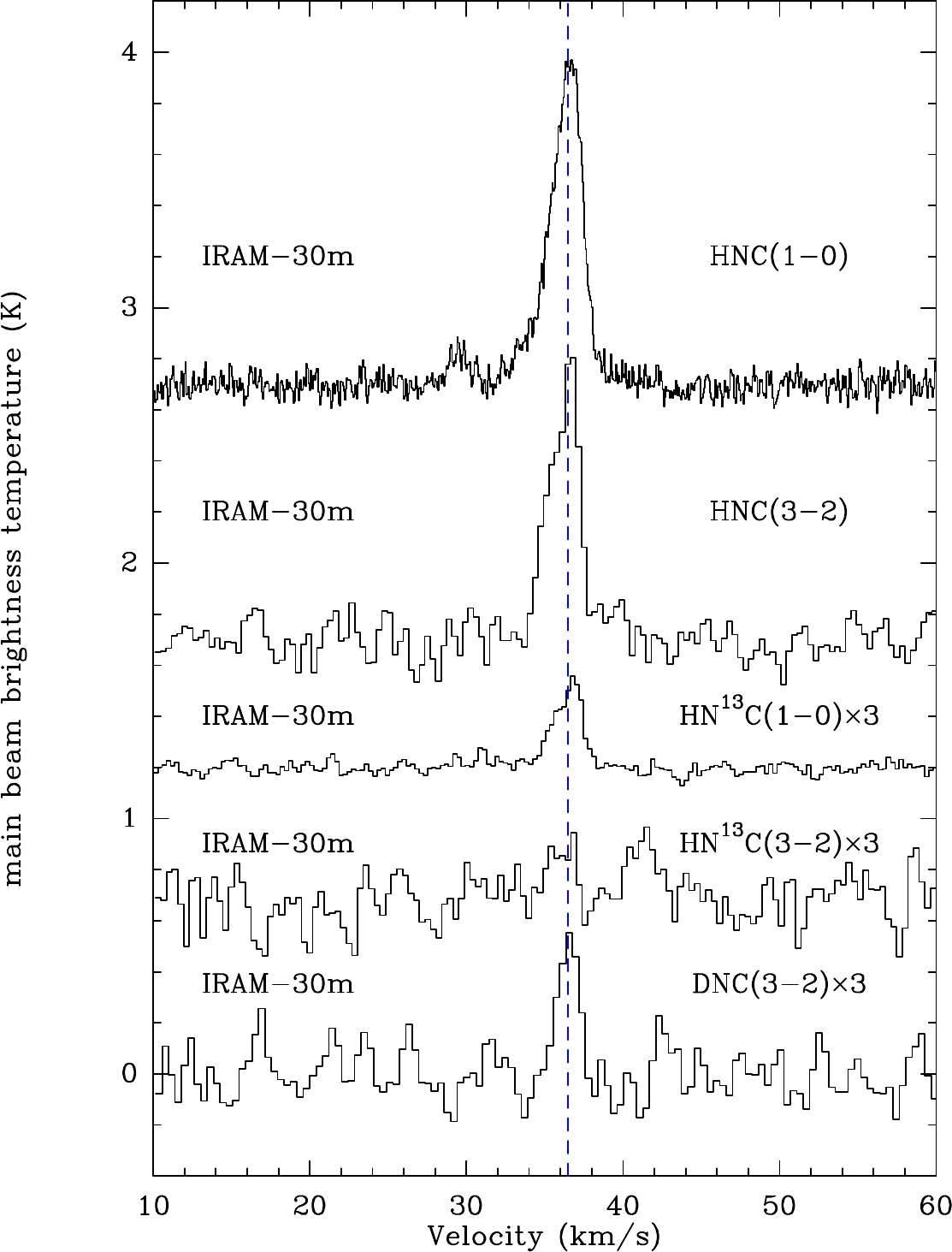}
\caption{{Observed spectra of HNC and its isotopologues toward NDO. The corresponding transition is indicated in the upper-right of each spectrum. The dashed vertical line marks the systemic velocity of 36.5~\kms.}\label{Fig:hnc}}
\end{figure}

\begin{figure}[!htbp]
\centering
\includegraphics[width = 0.45 \textwidth]{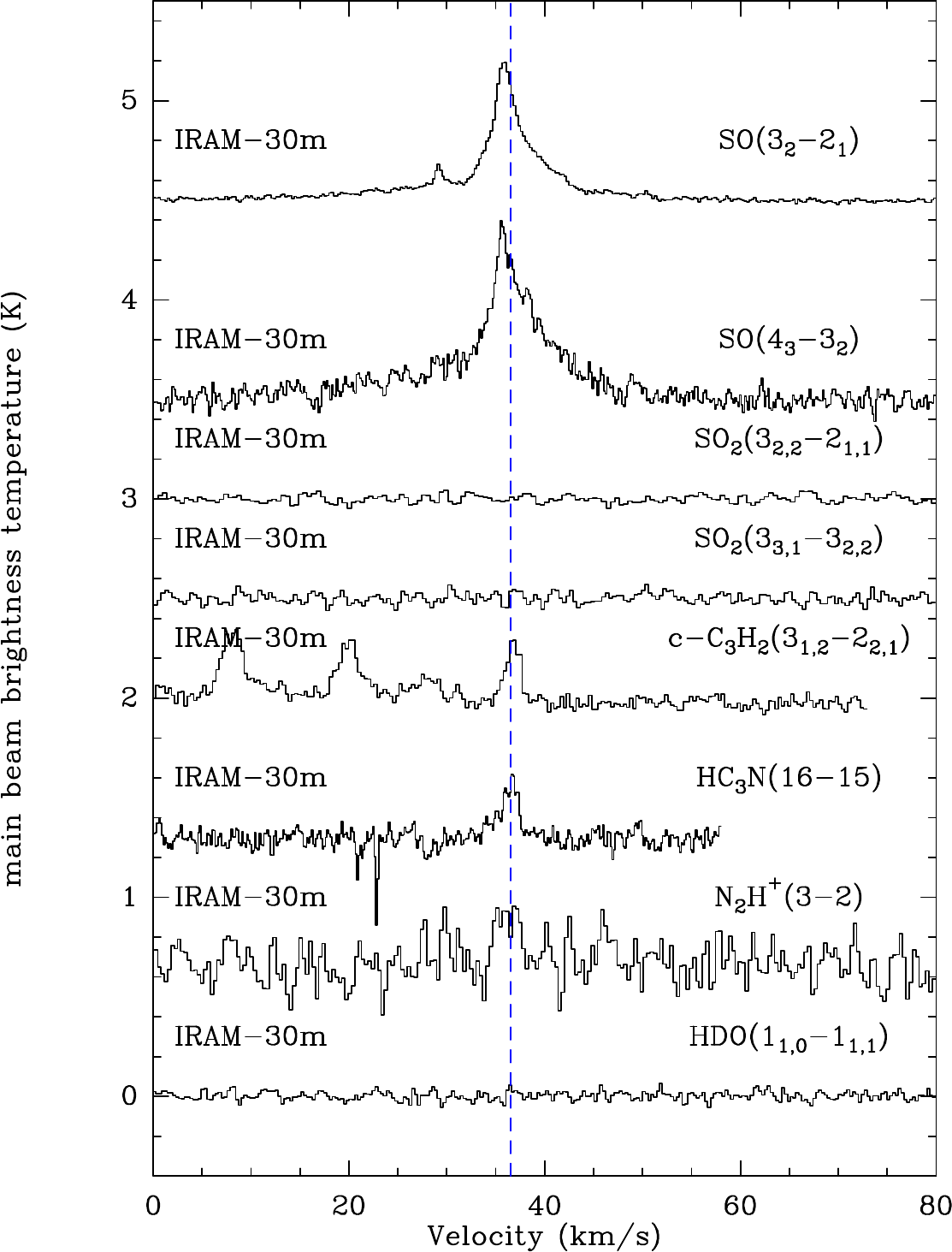}
\caption{{Observed spectra of SO, SO$_{2}$, c-C$_{3}$H$_{2}$, HC$_{3}$N, N$_{2}$H$^{+}$, and HDO toward NDO. The corresponding transition is indicated in the upper-right of each spectrum. The dashed vertical line marks the systemic velocity of 36.5~\kms.}\label{Fig:others}}
\end{figure}

\begin{figure}[!htbp]
\centering
\includegraphics[width = 0.45 \textwidth]{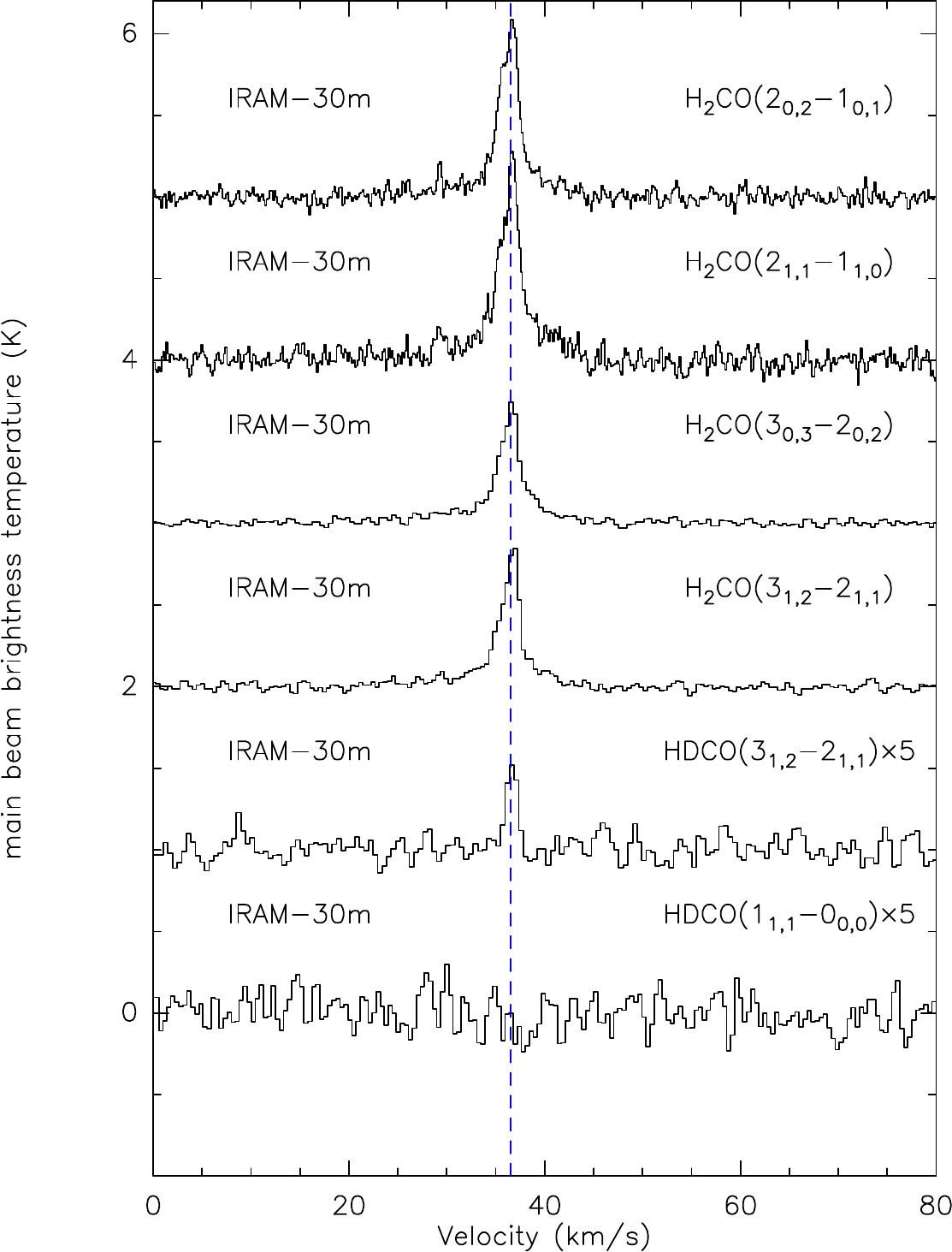}
\caption{{Spectra of H$_{2}$CO and its isotopologues toward NDO. The corresponding transition is indicated in the upper-right of each spectrum. The dashed vertical line marks the ystemic velocity of 36.5~\kms.}\label{Fig:h2co}}
\end{figure}

\begin{figure}[!htbp]
\centering
\includegraphics[width = 0.45 \textwidth]{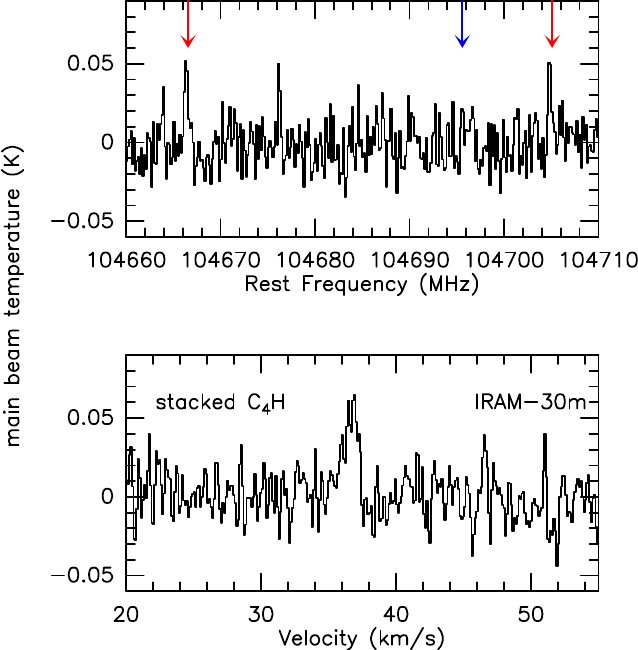}
\caption{{\textit{Top:} the observed C$_{4}$H spectrum with three hyperfine structure lines indicated by the arrows. \textit{Bottom:} the stacked C$_{4}$H spectrum averaged over the two hyperfine structure lines indicated by the two red arrows in the top panel.}\label{Fig:c4h}}
\end{figure}

\begin{figure}[!htbp]
\centering
\includegraphics[width = 0.45 \textwidth]{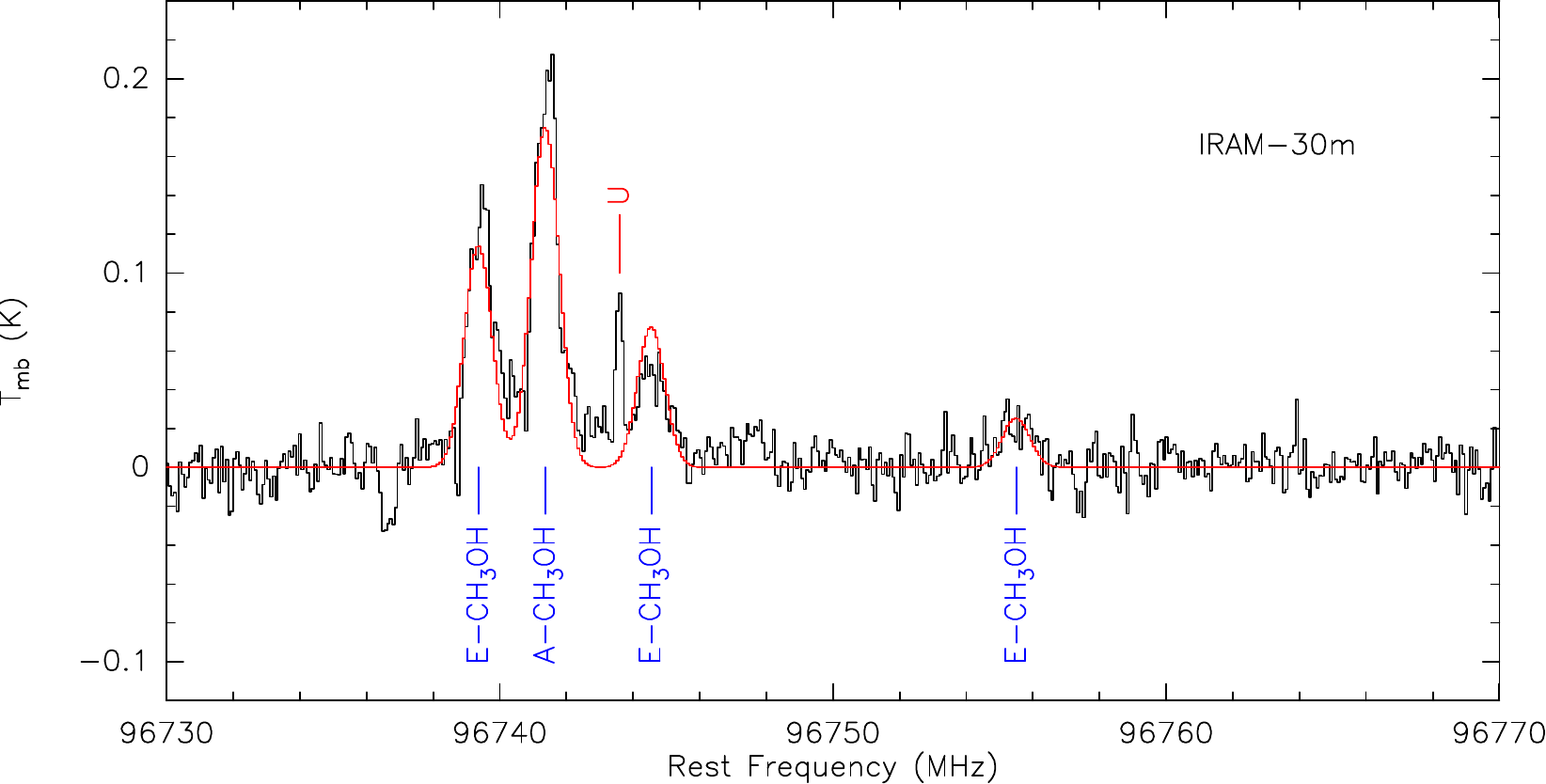}
\includegraphics[width = 0.45 \textwidth]{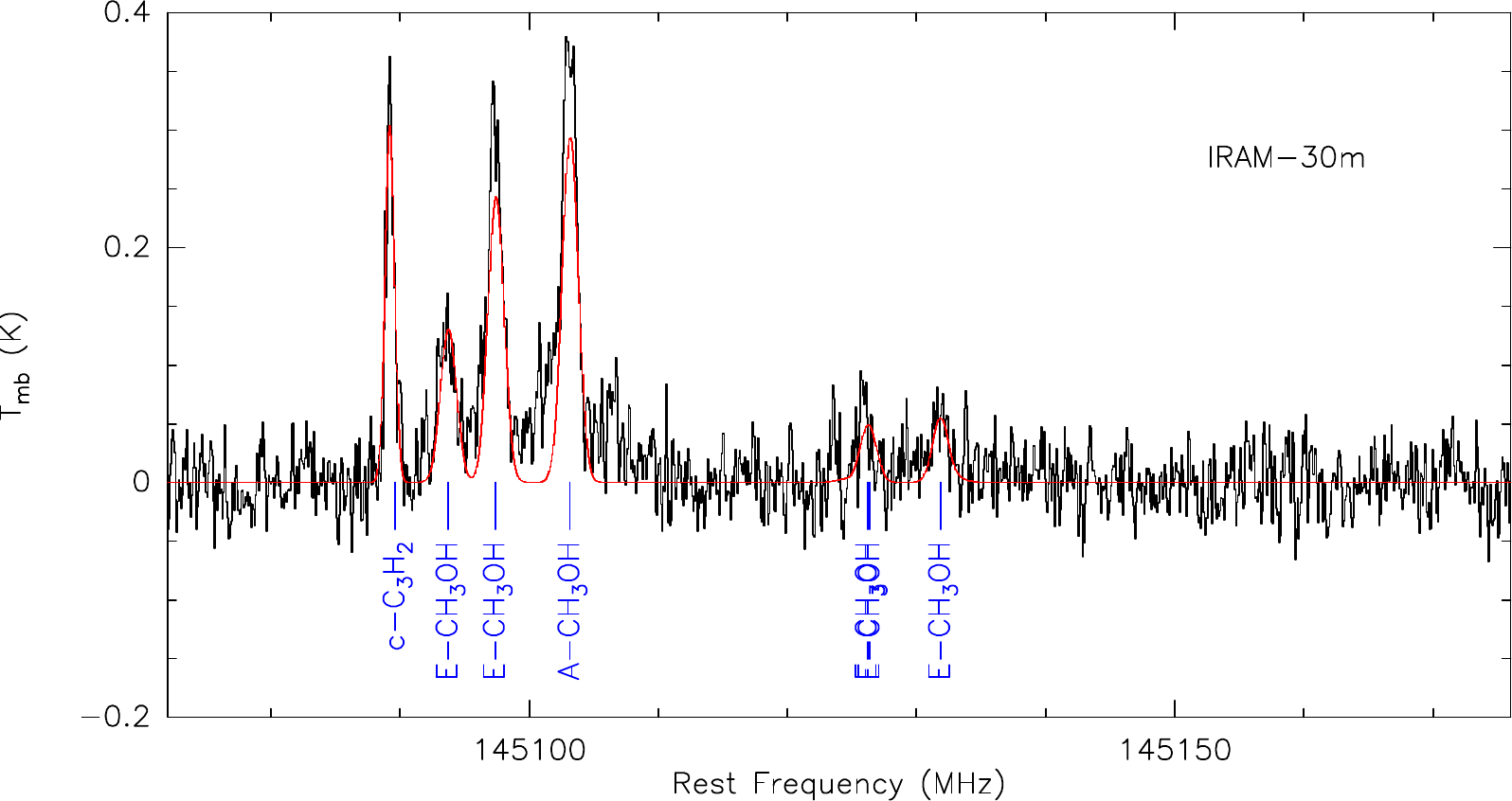}
\includegraphics[width = 0.45 \textwidth]{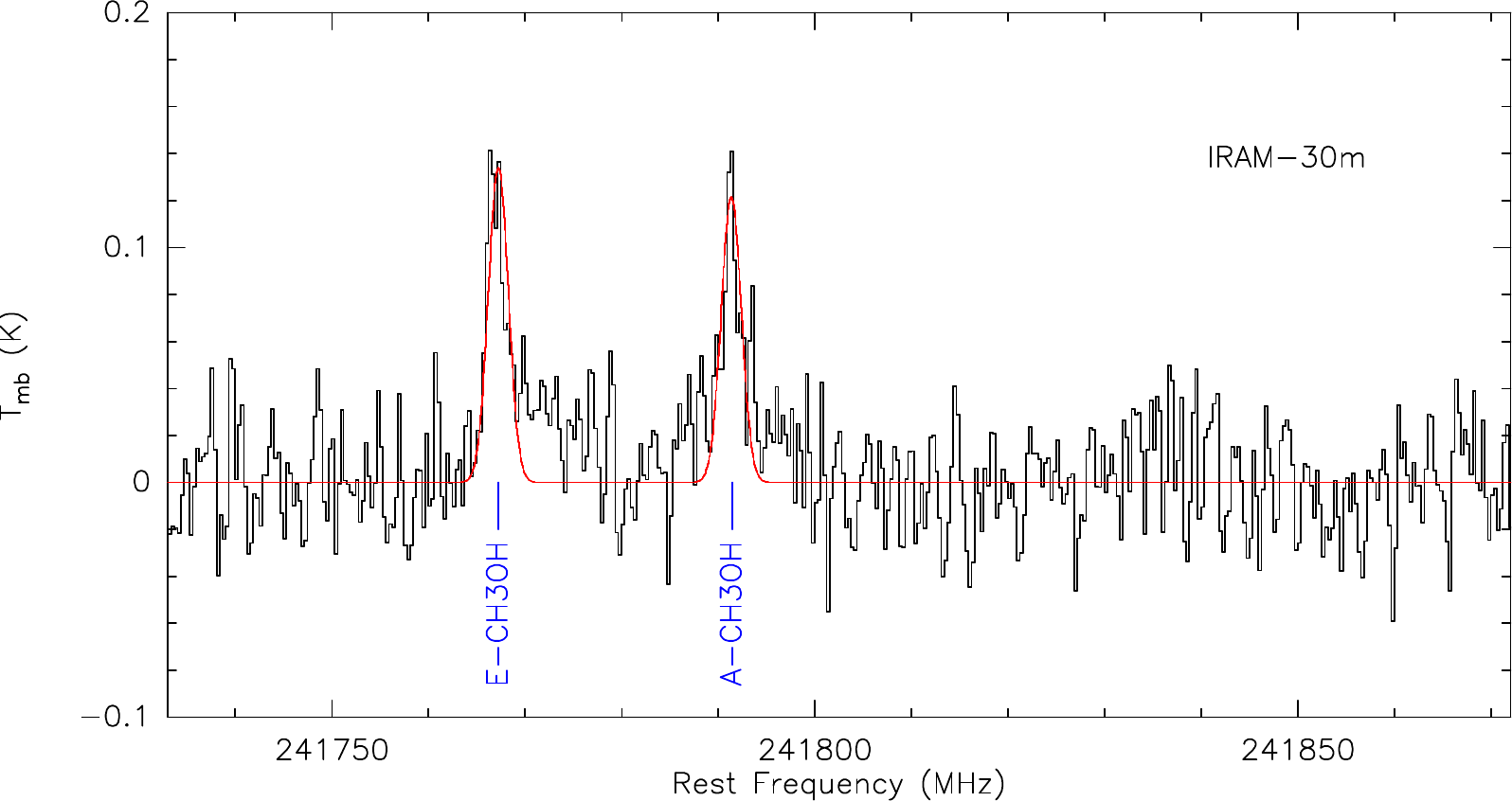}
\caption{{Observed CH$_{3}$OH spectra (black) overlaid with the LTE synthetic spectra (red), generated using the Weeds software \citep{2011A&A...526A..47M} with the fitted parameters listed in Table~\ref{Tab:colum}. The blue labels indicate the molecules assigned to the detected lines. The red label indicates an unidentified line.}\label{Fig:ch3oh}}
\end{figure}

\begin{figure*}[!htbp]
\centering
\includegraphics[width = 0.33 \textwidth]{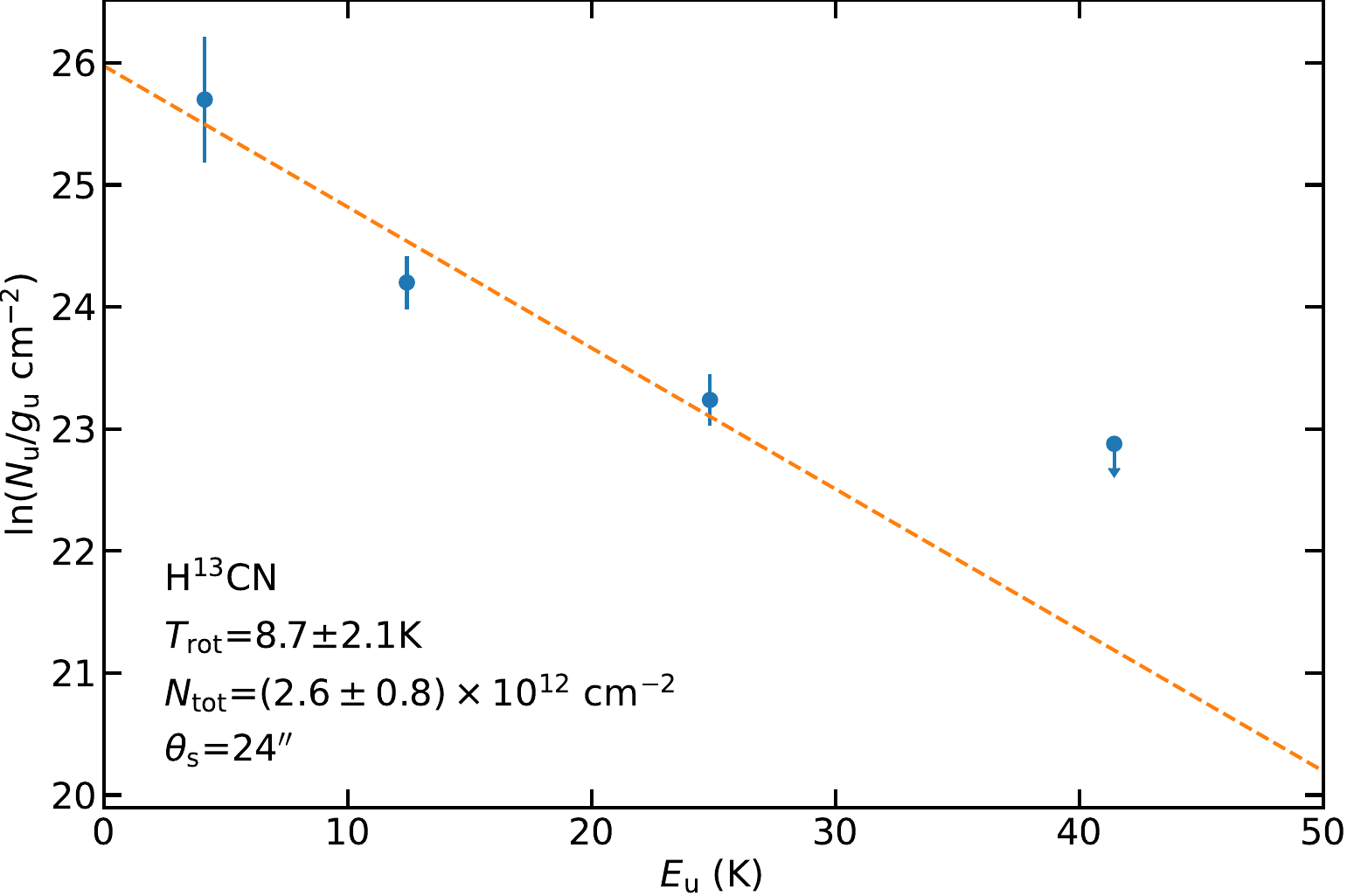}
\includegraphics[width = 0.33 \textwidth]{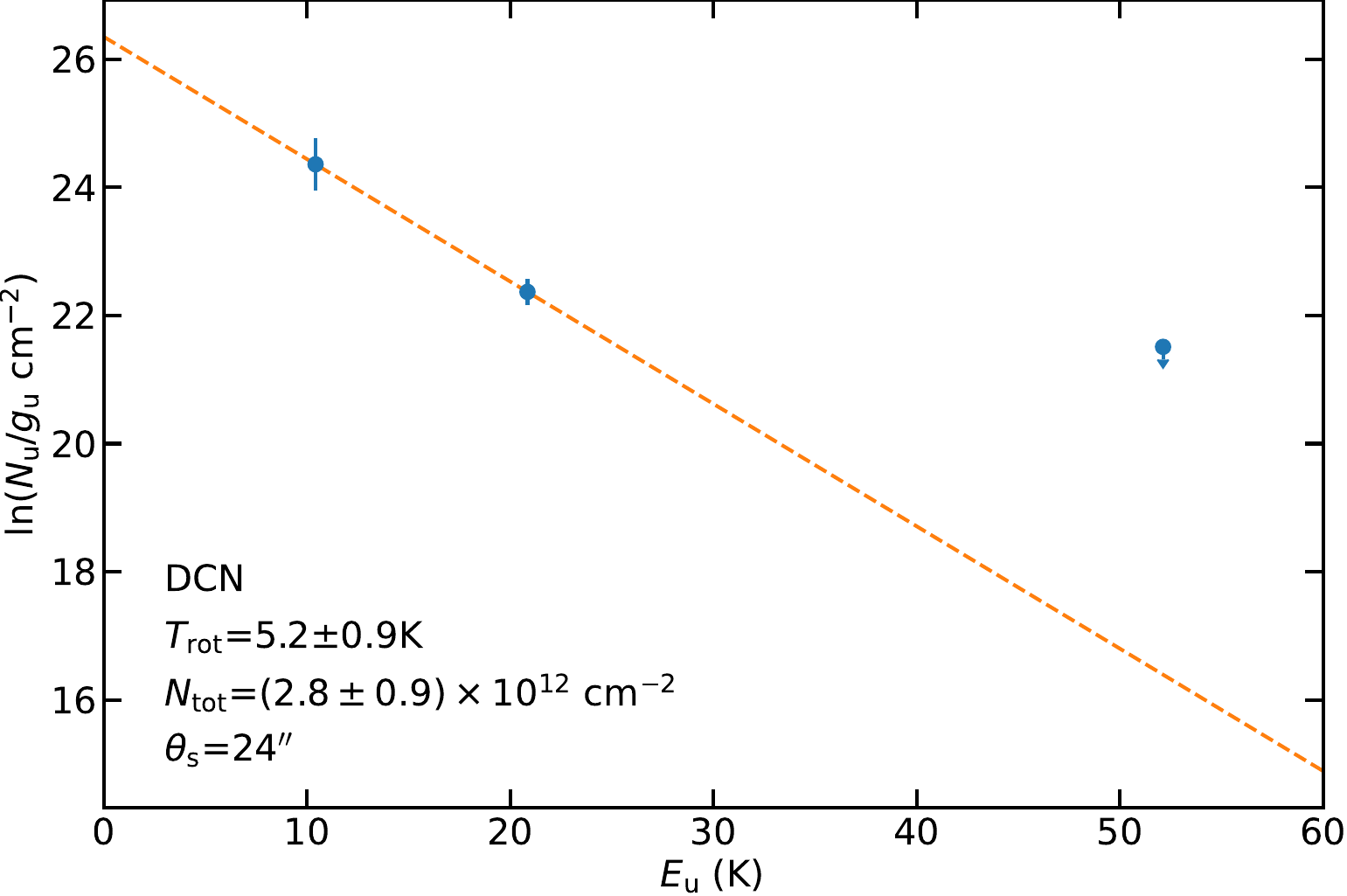}
\includegraphics[width = 0.33 \textwidth]{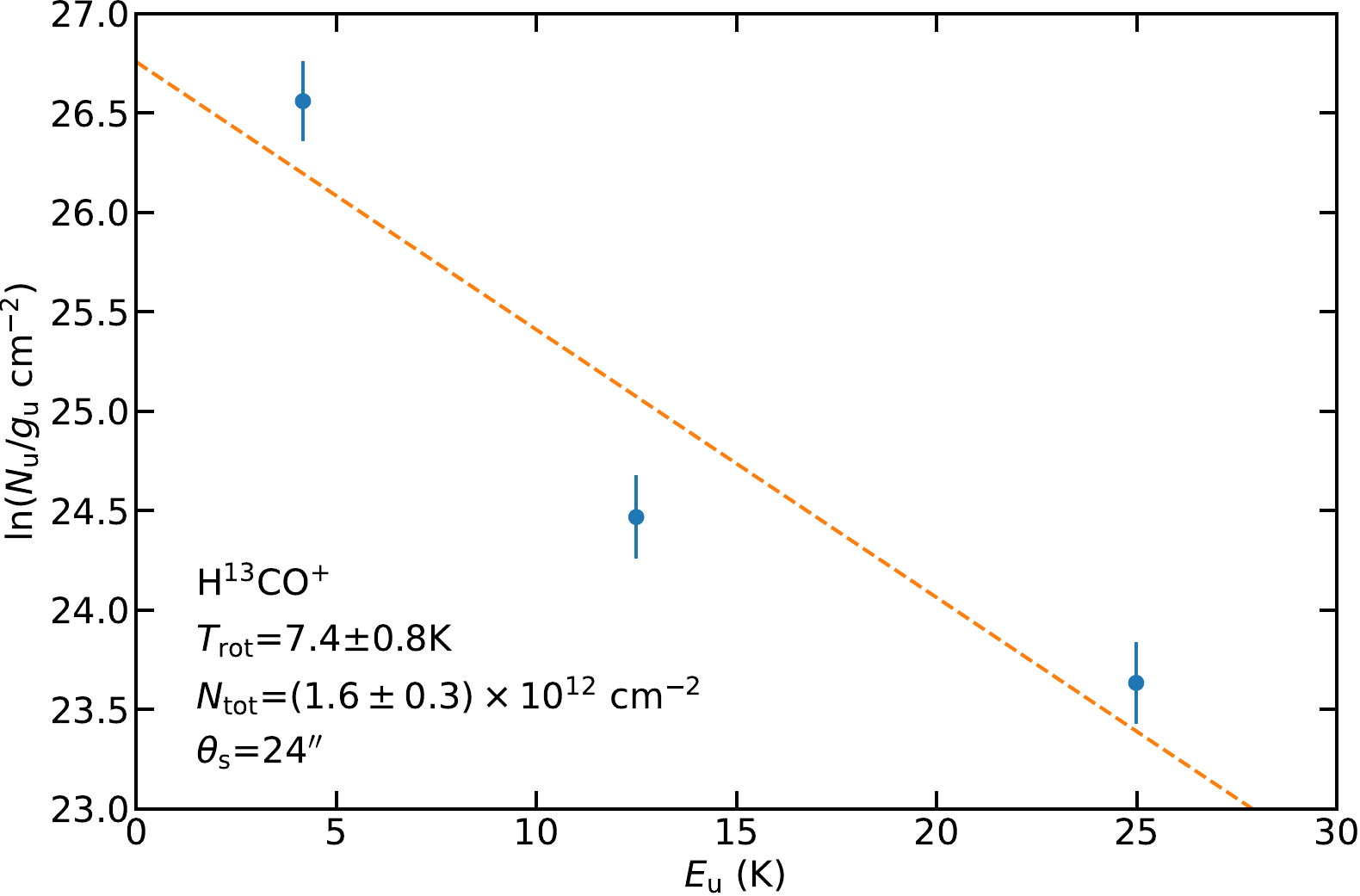}
\includegraphics[width = 0.33 \textwidth]{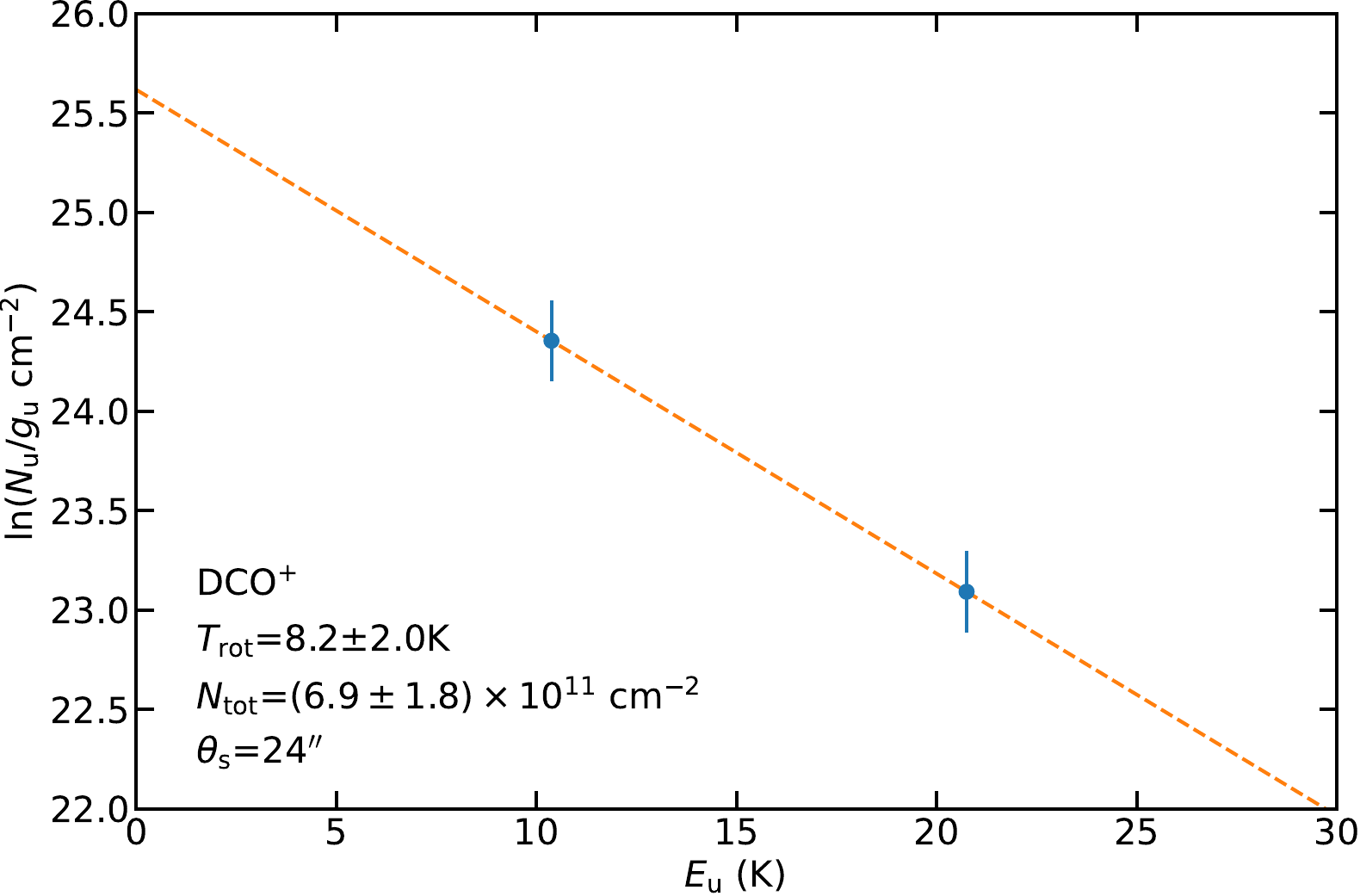}
\includegraphics[width = 0.33 \textwidth]{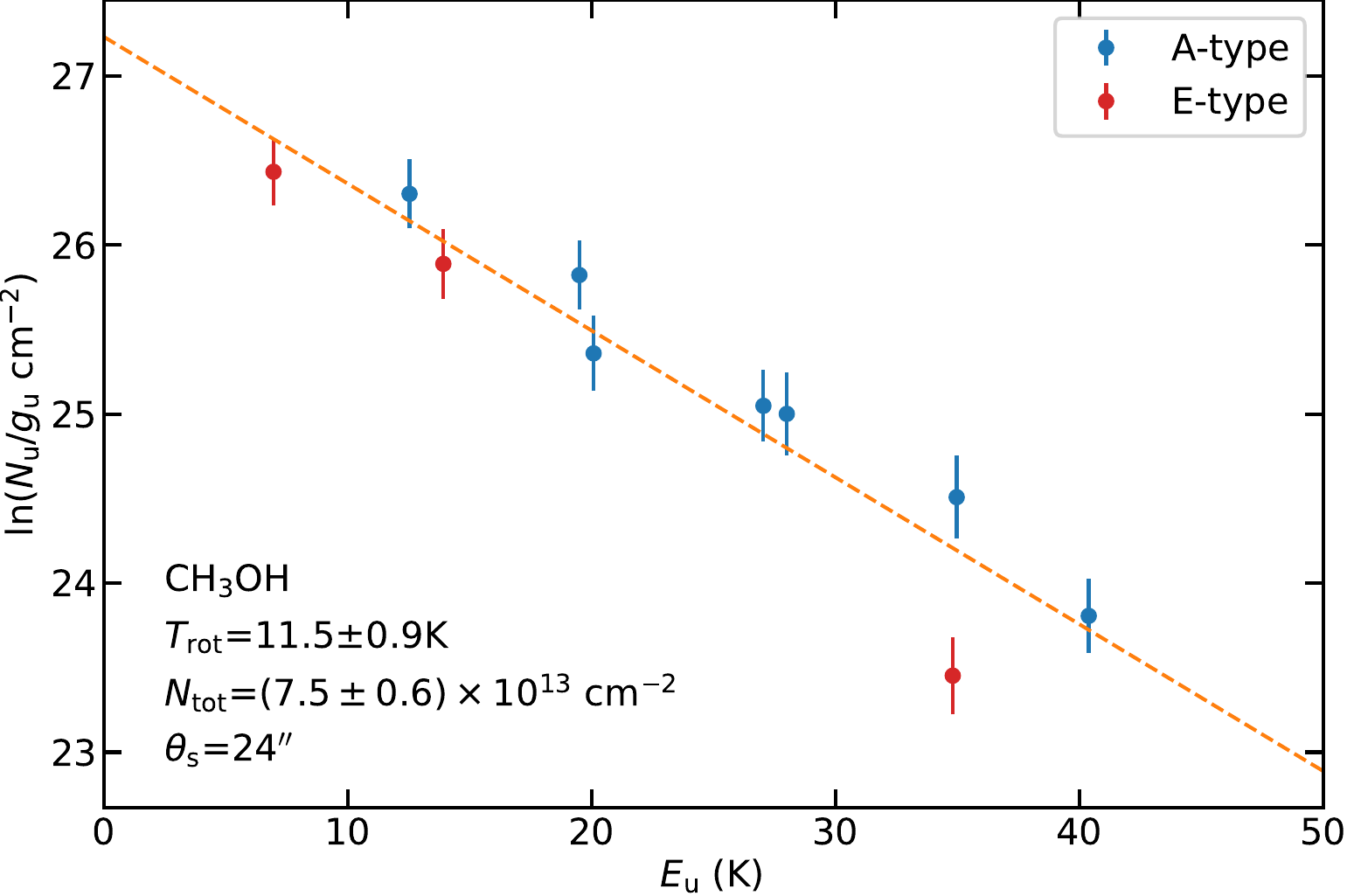}
\includegraphics[width = 0.33 \textwidth]{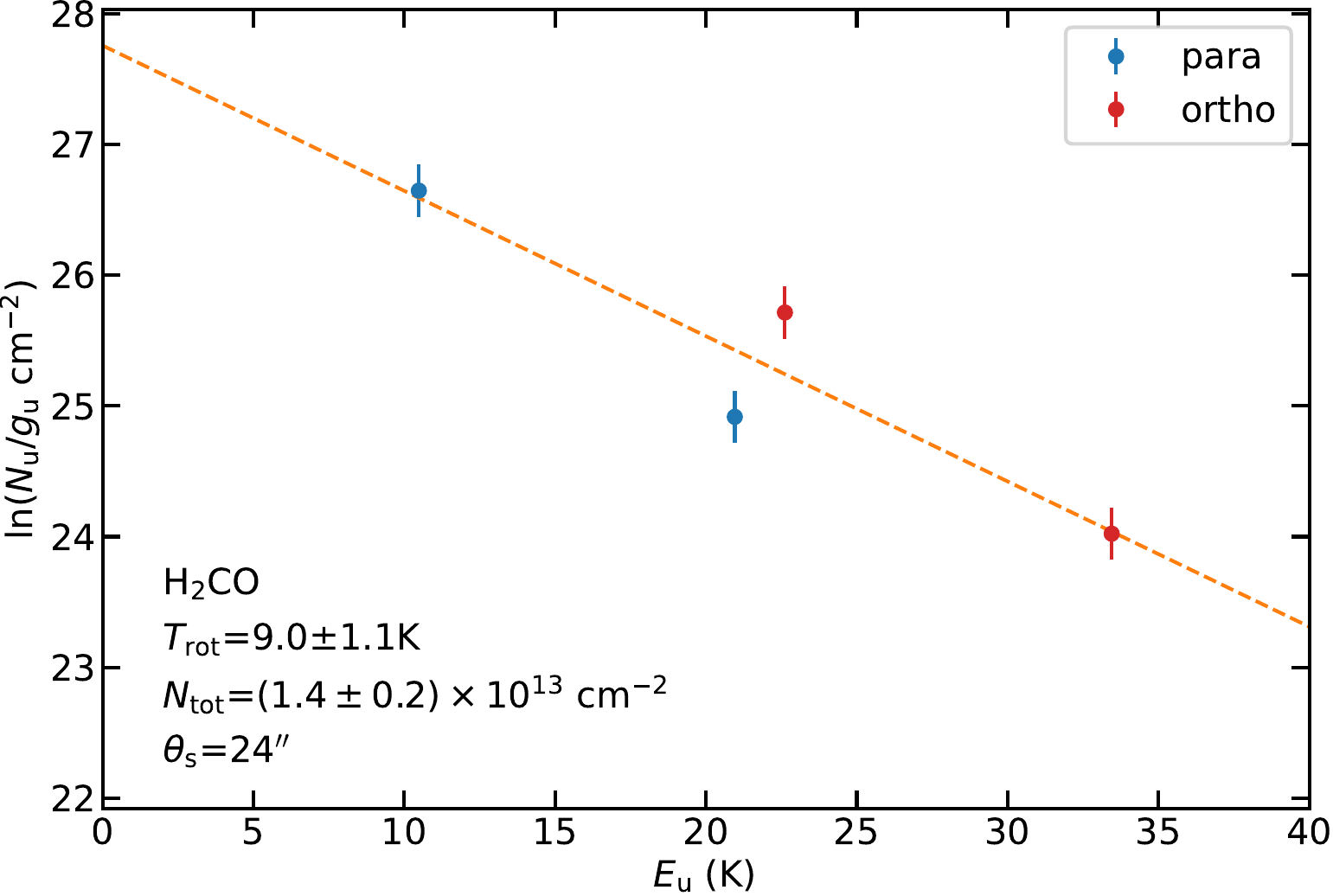}
\includegraphics[width = 0.33 \textwidth]{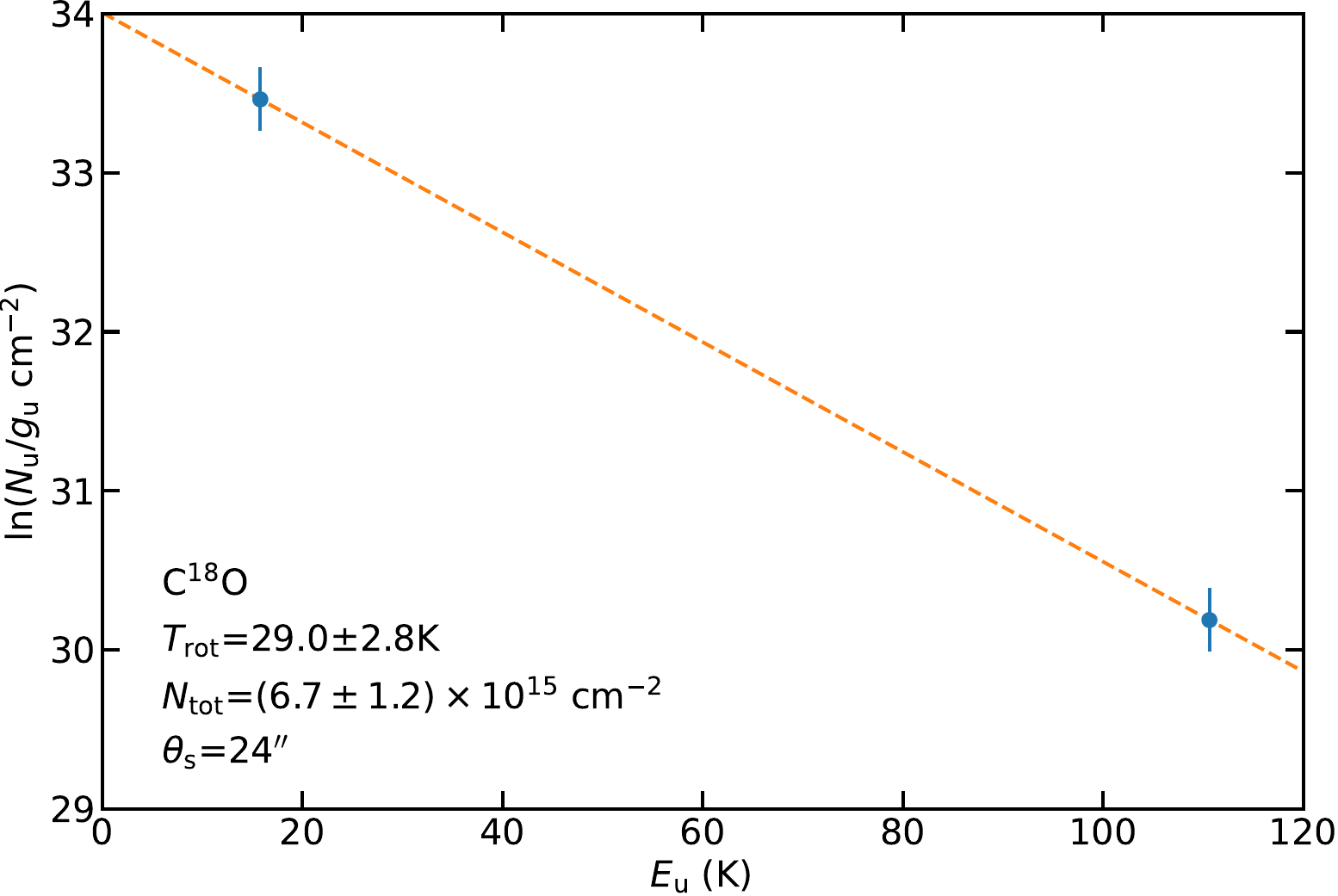}
\hspace{0.66 \textwidth} 
\caption{{Rotational diagram of H$^{13}$CN, DCN, H$^{13}$CO$^{+}$, DCO$^{+}$, CH$_{3}$OH, H$_{2}$CO, and C$^{18}$O detected by our single-dish observations. The species, rotational temperature, and molecular column density are indicated in the lower left corner of each panel. The dashed lines represent the linear least-square fits to the observed data.}\label{Fig:RD-SD}}
\end{figure*}

\begin{figure*}[!htbp]
\centering
\includegraphics[width = 0.33 \textwidth]{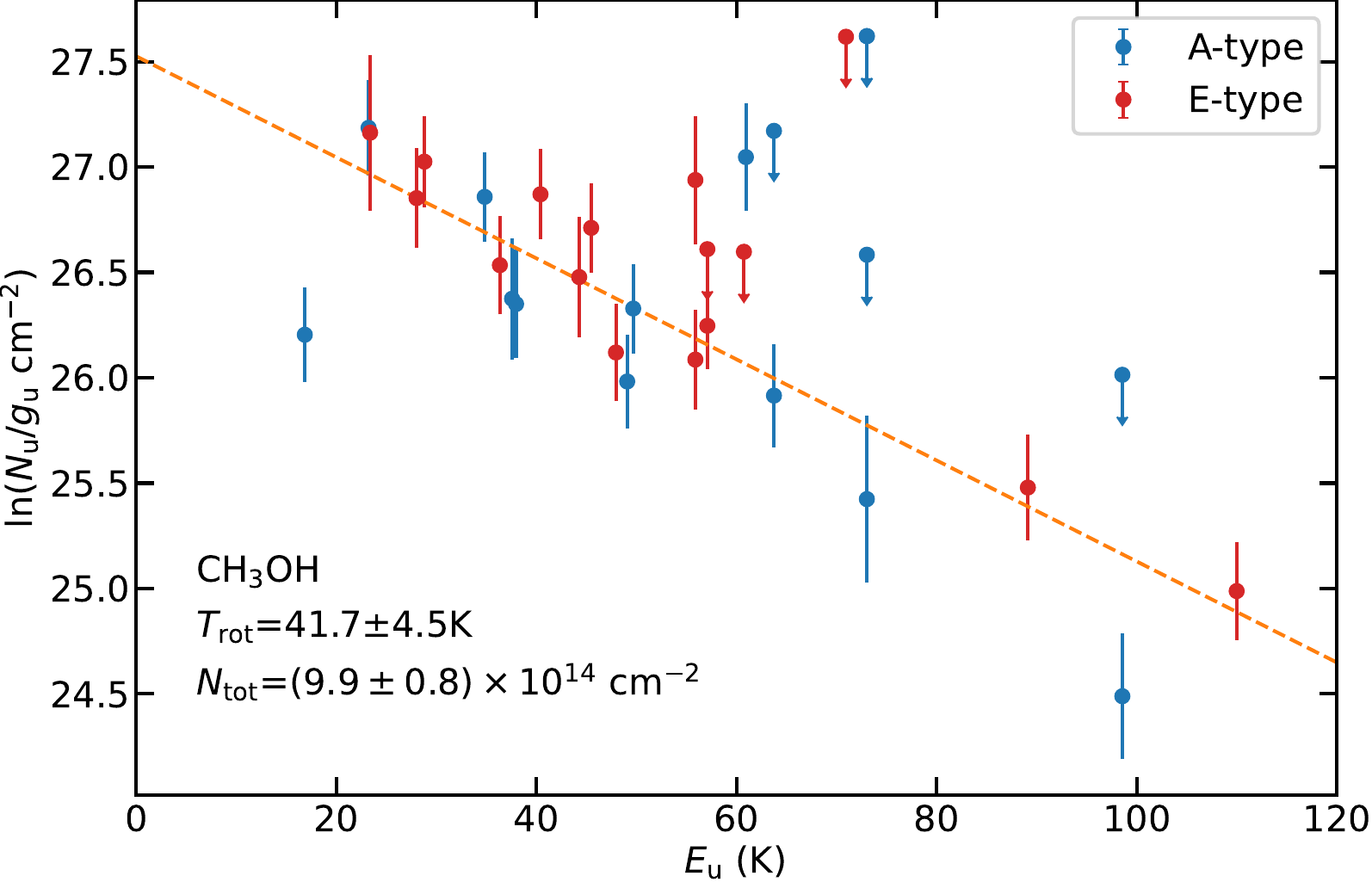}
\includegraphics[width = 0.33 \textwidth]{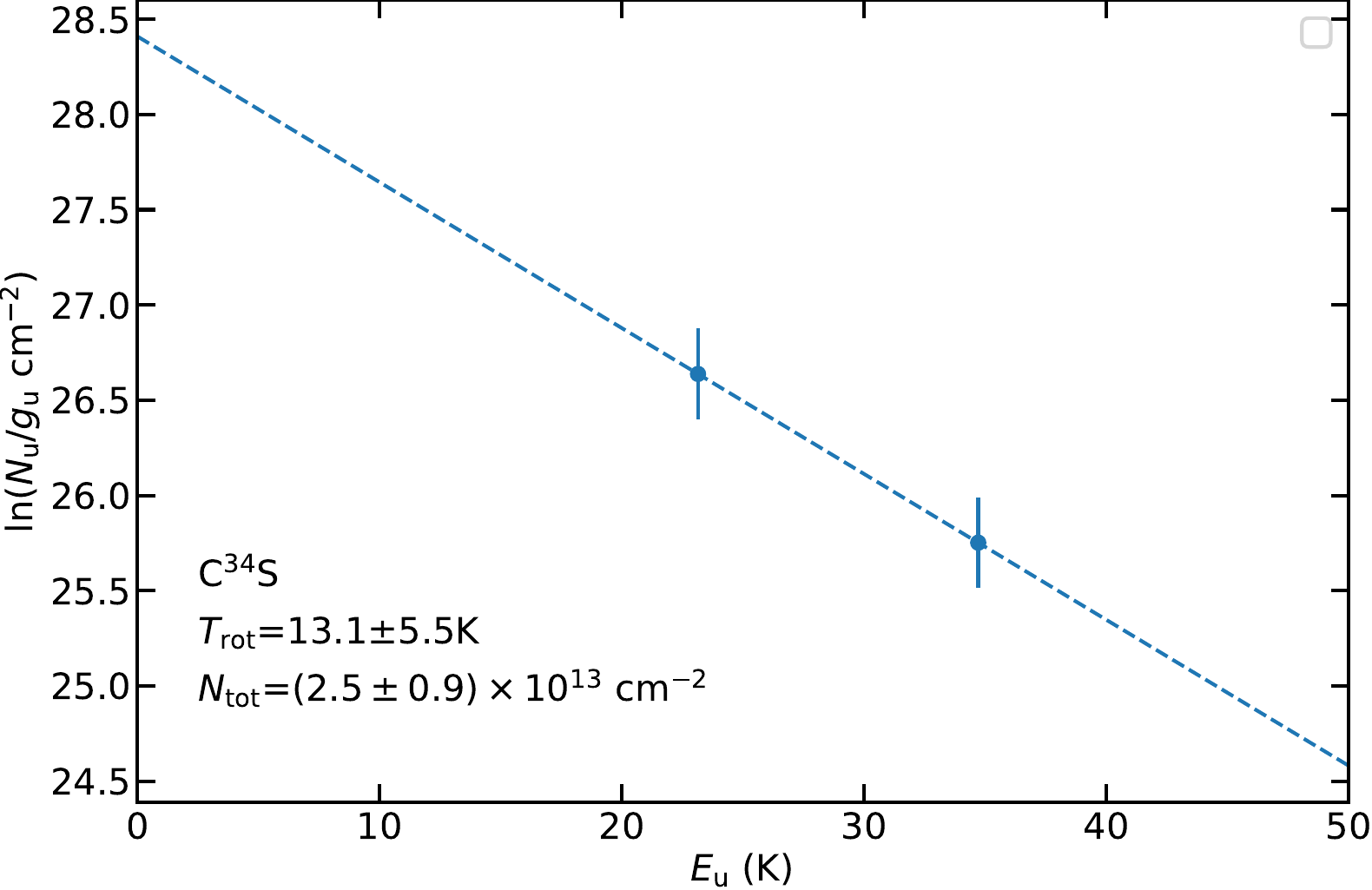}
\includegraphics[width = 0.33 \textwidth]{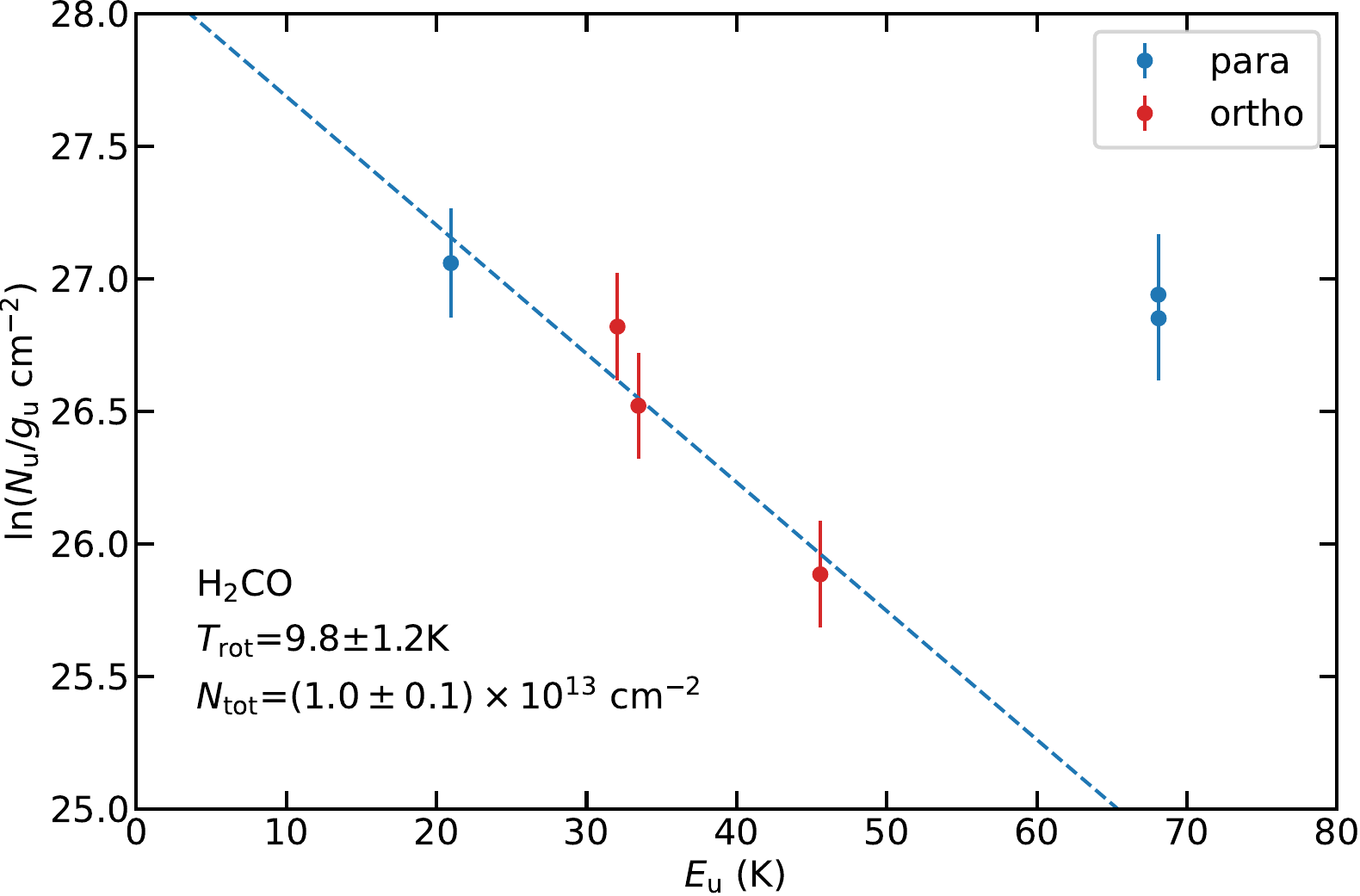}
\includegraphics[width = 0.33 \textwidth]{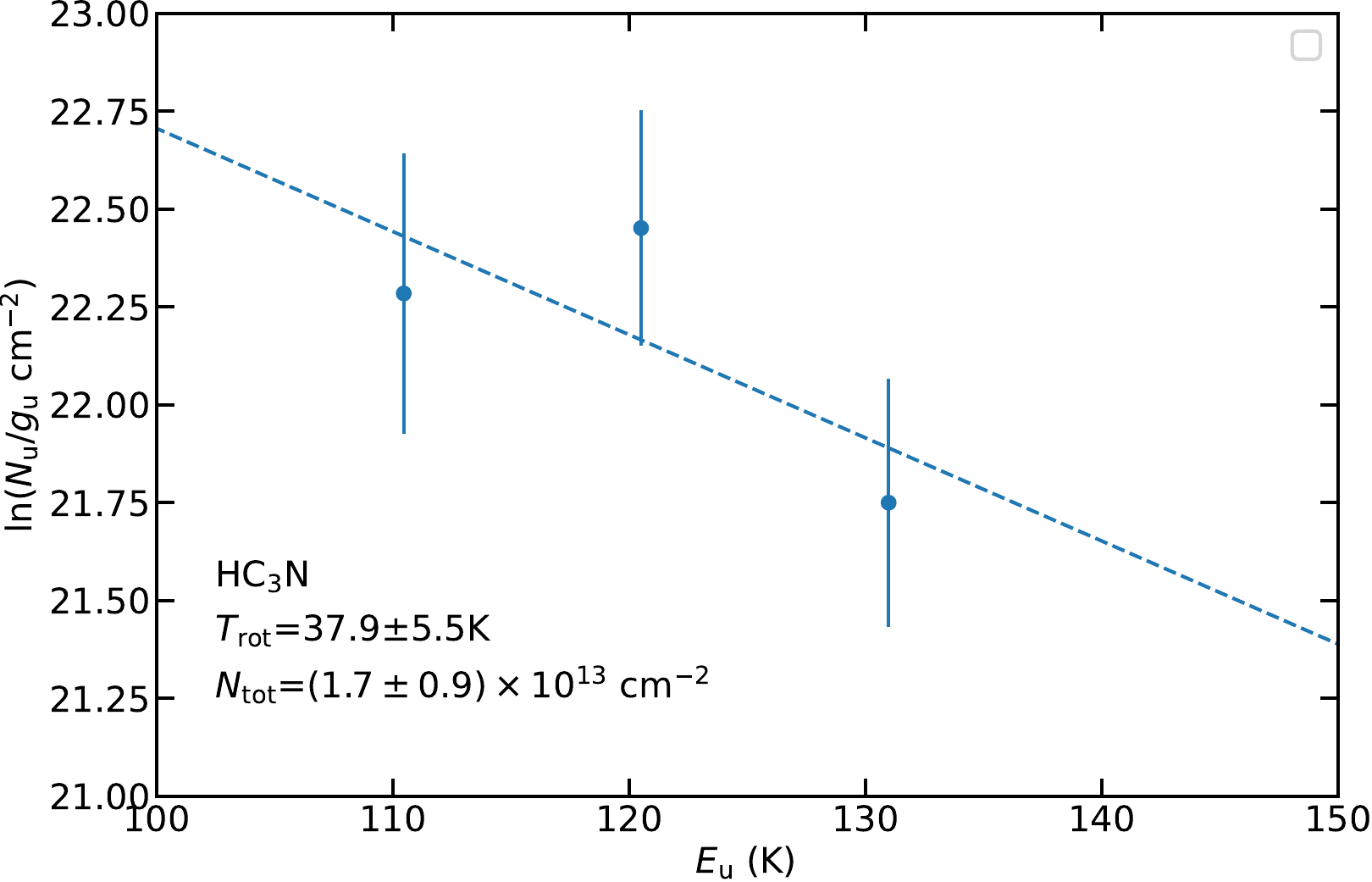}
\includegraphics[width = 0.33 \textwidth]{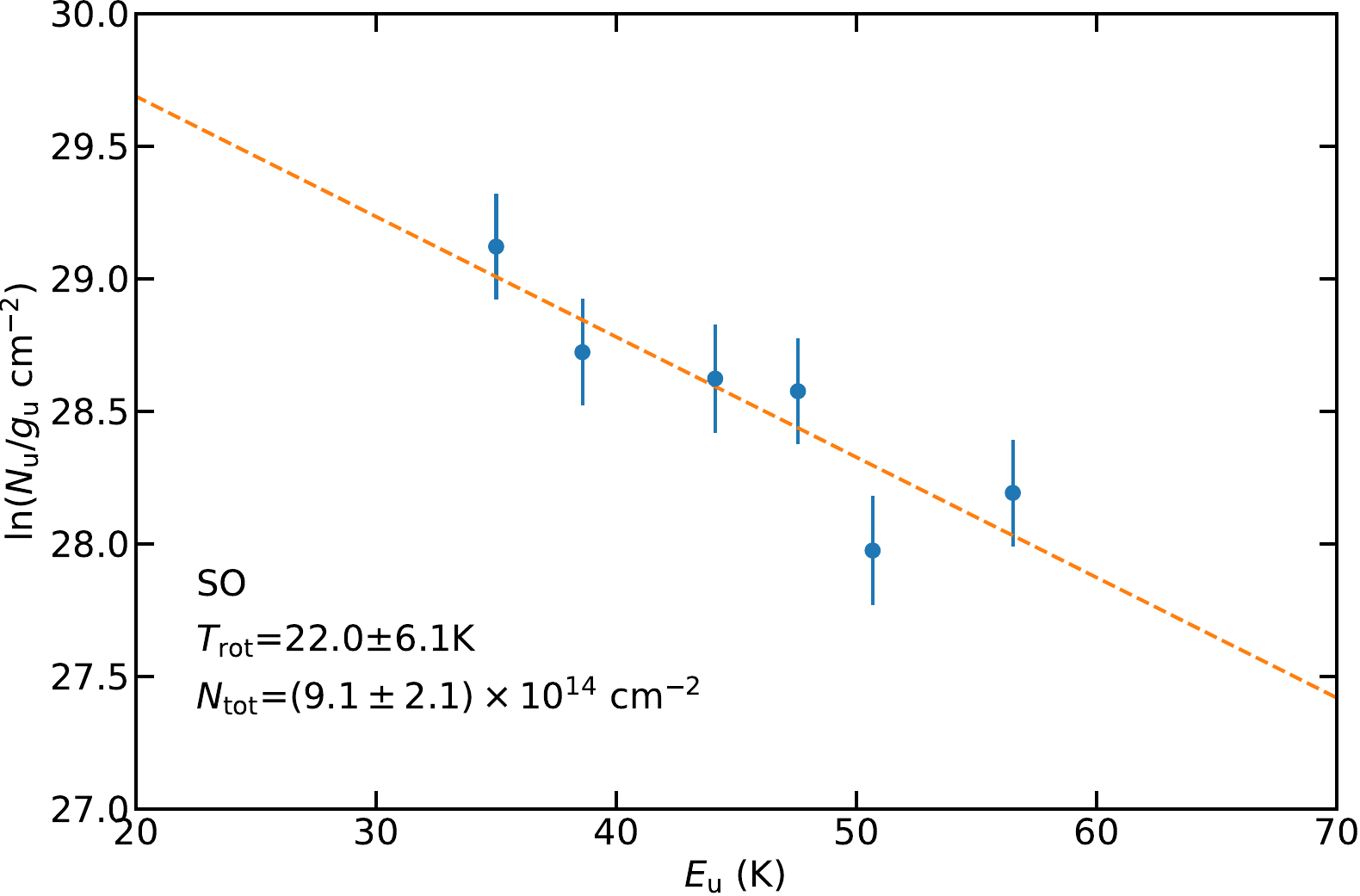}
\includegraphics[width = 0.33 \textwidth]{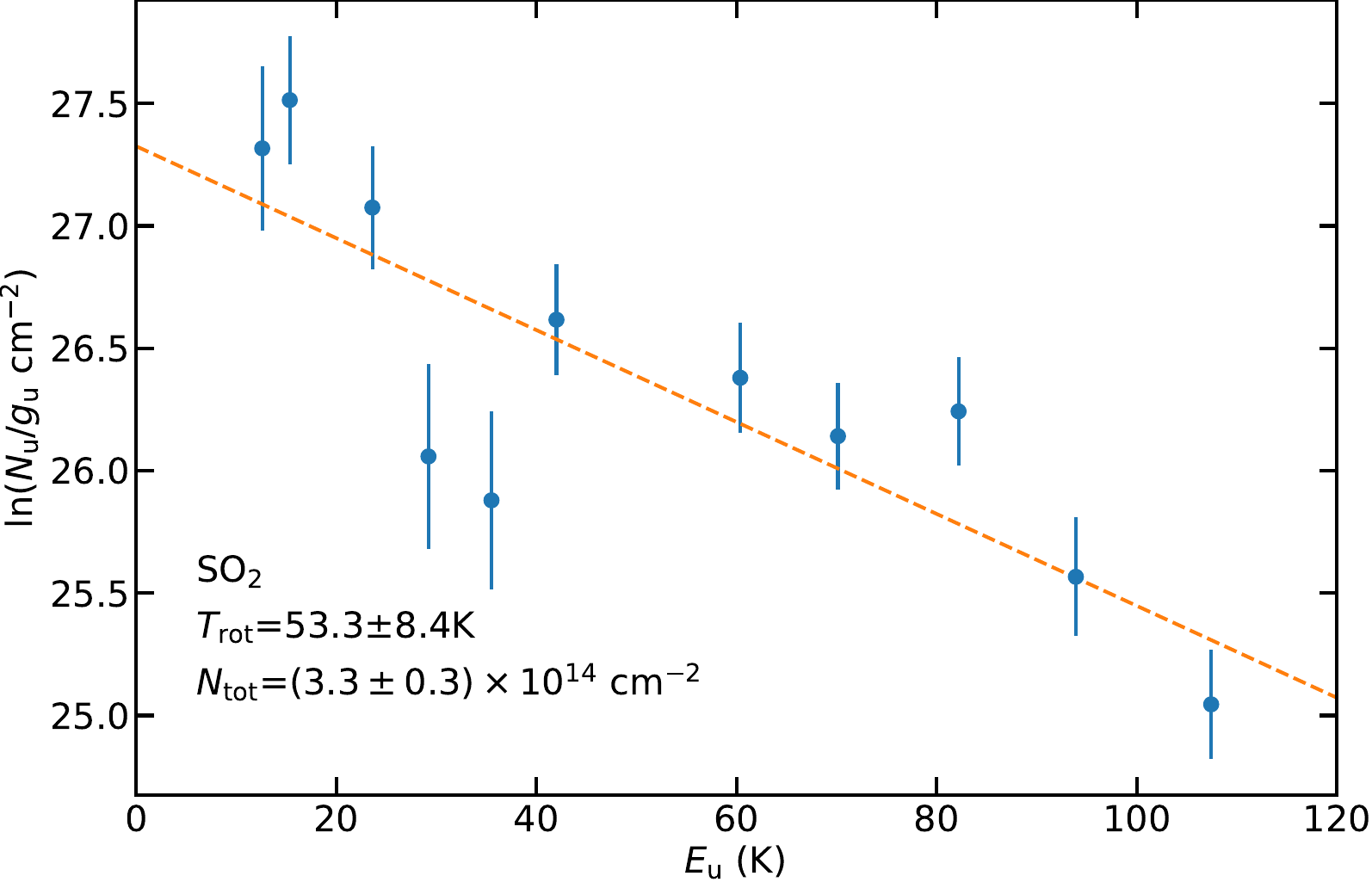}
\caption{{Rotational diagram of CH$_{3}$OH, C$^{34}$S, H$_{2}$CO, HC$_{3}$N, SO and SO$_{2}$ detected by our SMA observations. The species, fitted rotational temperature, fitted molecular column density, and assumed emission size are indicated in the lower left corner of each panel. The dashed lines represent the linear least-square fits to the observed data. In the H$_{2}$CO panel, the two high-energy lines appear to deviate from the overall trend, and are thus excluded from the fit.}\label{Fig:RD-SMA}}
\end{figure*}

\section{Zoom-in plots of the observed SMA spectrum of MM1}\label{app.c}
\begin{figure*}[!htbp]
\centering
\includegraphics[width = 0.9 \textwidth]{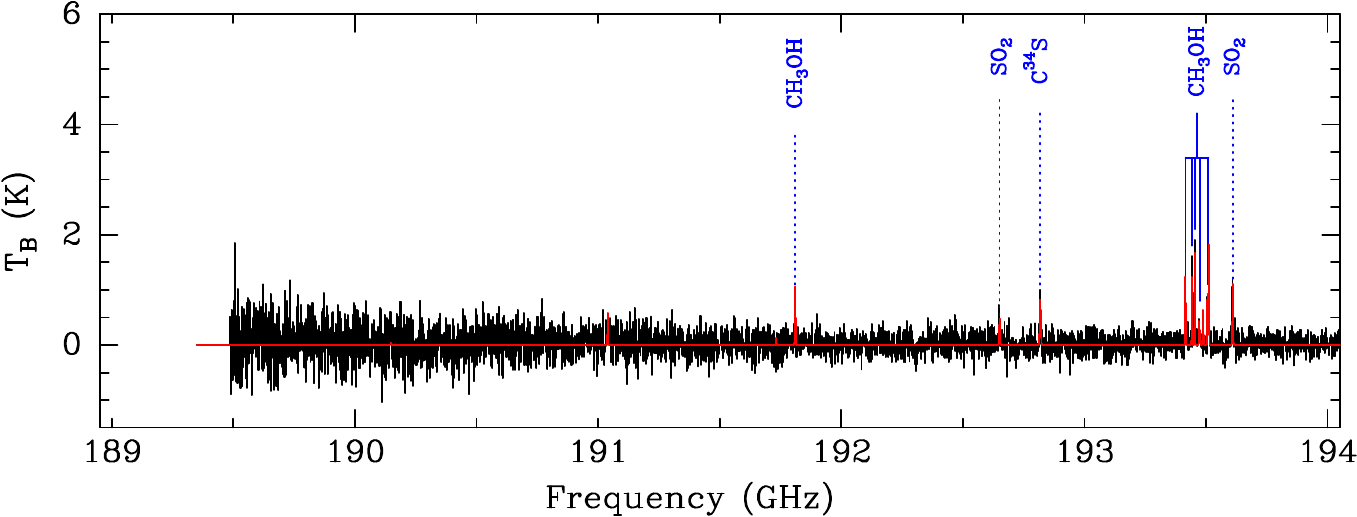}
\includegraphics[width = 0.9 \textwidth]{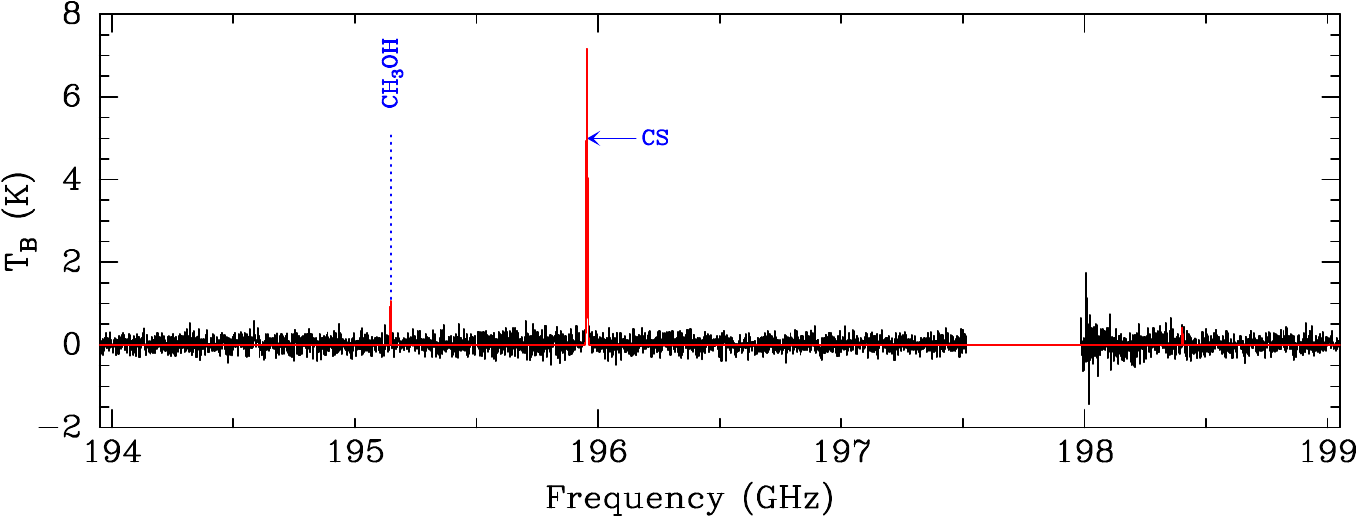}
\includegraphics[width = 0.9 \textwidth]{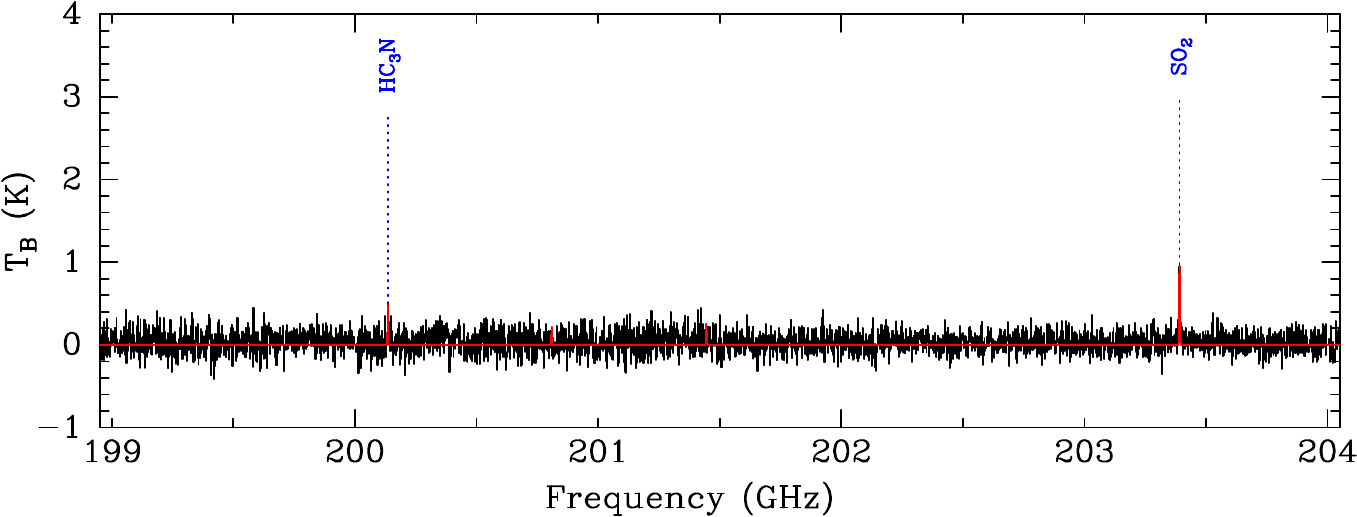}
\includegraphics[width = 0.9 \textwidth]{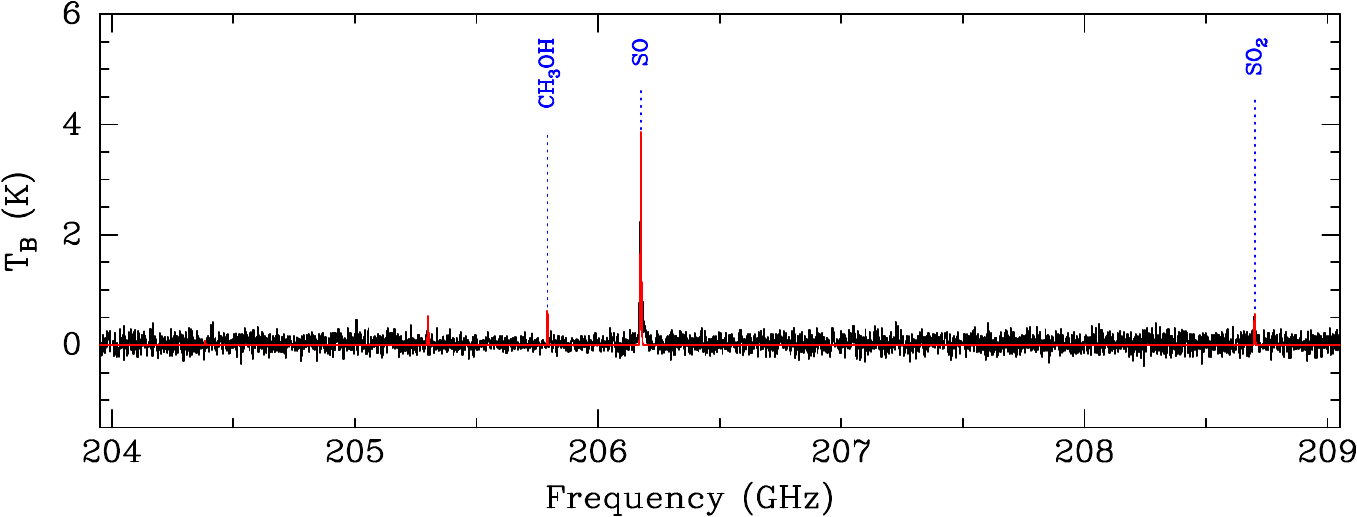}
\caption{{Observed SMA spectrum of MM1. The best-fit LTE synthetic spectrum is displayed in red and overlaid on the observed spectrum shown in black. The identified spectral features are labeled with the names of their corresponding molecules, while those of unknown origin are designated as "U".}\label{Fig:sma-line-survey}}
\end{figure*}

Figure~\ref{Fig:sma-line-survey} present the observed SMA spectrum of MM1 in detail (see also \url{https://gongyan2444.github.io/NDO.html}). To model the observed spectrum, we utilized the Weeds software \citep{2011A&A...526A..47M} which can generate synthetic LTE spectra by accounting for the radiative transfer along the line of sight and for the finite angular resolution of the telescope. The spectroscopic information was taken from the CDMS and JPL databases \citep{1998JQSRT..60..883P,2016JMoSp.327...95E}. We performed the modeling for each species individually and aggregated the contributions of all molecules to compose the final synthetic spectrum. 

For the synthetic LTE model, we assume a single Gaussian component for all molecules for simplicity. The LTE model is defined by a set of five parameters: molecular column density, excitation temperature, source size, systemic velocity, and FWHM line width. The excitation temperatures and column densities of different molecules are derived from the rotational diagram analysis (see Table~\ref{Tab:colum}). Since the SMA spectrum was convolved to a common beam across the whole frequency coverage, we simply take a sufficiently large source size of 24\arcsec\,to create the synthetic model with Weeds. The systemic velocity and FWHM line width are determined by the fitted values listed in Table~\ref{Tab:sma-lines}. The final synthetic spectrum is shown in red in \url{https://gongyan2444.github.io/NDO.html}.

\begin{table*}[!hbt]
\caption{Observed properties of the molecular lines detected by our SMA observations toward MM1.}\label{Tab:sma-lines}
\scriptsize
\centering
\begin{tabular}{cccccccc}
\hline \hline
Line                & Frequency     & $E_{\rm u}/k$ & $\int T_{\rm mb}{\rm d}\varv$ & $\varv$ & $\Delta \varv$ & $T_{\rm mb}$ & note\\ 
                    & (MHz)         &  (K)       & (K~\kms)                     & (\kms)  & (\kms)         & (K)    &    \\ 
   (1)                 & (2)         &  (3)       & (4)                     & (5)  & (6)         & (7)     & (8)   \\                     
\hline              
CH$_{3}$OH ($4_{1}-3_{1}$ A$^{+}$) & 191810.503(4) & 37.6  &   3.97$\pm$0.82 &   38.2$\pm$0.7   &  6.0$\pm$1.4 & 0.62$\pm$0.13 & single \\
SO$_{2}$ ($2_{2,0}-1_{1,1}$)       & 192651.02(30) & 12.6  &   3.32$\pm$0.90 &   39.1$\pm$1.0   &  6.5$\pm$2.0 & 0.48$\pm$0.13 & single \\ 
C$^{34}$S ($4-3$)                  & 192818.457(1) & 23.1  &   6.47$\pm$0.84 &   37.4$\pm$0.4   &  6.3$\pm$0.9 & 0.96$\pm$0.06 & single \\
CH$_{3}$OH ($4_{0}-3_{0}$ E)       & 193415.324(4) & 36.3  &   5.00$\pm$0.60 &   36.9$\pm$0.3   &  4.9$\pm$0.7 & 0.96$\pm$0.08 & single \\
CH$_{3}$OH ($4_{-1}-3_{-1}$ E)     & 193441.600(4) & 28.8  &   7.67$\pm$0.65 &   37.1$\pm$0.2   &  4.5$\pm$0.4 & 1.61$\pm$0.04 & single \\
CH$_{3}$OH ($4_{0}-3_{0}$ A$^{+}$) & 193454.358(4) & 23.2  &   9.60$\pm$1.00 &   37.6$\pm$0.3   &  5.6$\pm$0.8 & 1.6$\pm$0.31  & single \\
CH$_{3}$OH ($4_{3}-3_{3}$ A$^{+}$) & 193471.434(3) & 73.0  & $<$6.48 & \nodata & \nodata & \nodata & blended\\
CH$_{3}$OH ($4_{3}-3_{3}$ A$^{-}$) & 193471.545(3) & 73.0  & $<$6.48 & \nodata & \nodata & \nodata & blended\\
CH$_{3}$OH ($4_{-3}-3_{-3}$ E)     & 193474.414(3) & 73.0  & $<$6.48 & \nodata & \nodata & \nodata & blended \\
CH$_{3}$OH ($4_{-1}-3_{-1}$ E)     & 193506.559(4) & 44.3  &  4.55$\pm$0.93 &   36.8$\pm$0.4    &  4.1$\pm$1.0 & 1.04$\pm$0.05 & single \\
CH$_{3}$OH ($4_{2}-3_{2}$ A$^{+}$) & 193510.750(3) & 60.9  &  6.31$\pm$1.00 &   36.8$\pm$0.5    &  6.7$\pm$1.4 & 0.88$\pm$0.12 & single \\
SO$_{2}$ ($9_{1,9}-8_{0,8}$)       & 193609.49(10) & 42.0  &  6.68$\pm$0.7  &   38.2$\pm$0.3    &  4.8$\pm$0.6 & 1.31$\pm$0.06 & single \\
CH$_{3}$OH ($4_{1}-3_{1}$ A$^{-}$) & 195146.790(4) & 38.0  &  3.94$\pm$0.64 &   36.9$\pm$0.3    &  4.3$\pm$1.0 & 0.87$\pm$0.11 & single \\
CS ($4-3$)                         & 195954.211(2) & 23.5  & 42.96$\pm$1.11 &   37.8$\pm$0.1    &  7.3$\pm$0.2 & 5.53$\pm$0.31 & single \\
HC$_{3}$N ($22-21$)                & 200135.392(22)& 110.5 & 1.75$\pm$0.52  &   36.6$\pm$0.5    &  3.3$\pm$1.5 & 0.49$\pm$0.03 & single \\
SO$_{2}$ ($12_{0,12}-11_{1,11}$)   & 203391.55(10) & 70.1  & 6.17$\pm$0.54  &   38.6$\pm$0.3    &  8.1$\pm$0.9 & 0.72$\pm$0.11 & single \\
CH$_{3}$OH ($1_{1}-2_{0}$ A$^{+}$) & 205791.270(11)& 16.8  & 2.06$\pm$0.21  &   37.5$\pm$0.3    &  5.0$\pm$0.6 & 0.39$\pm$0.03 & single \\
SO ($4_{5}-3_{4}$)                 & 206176.005(40)& 38.6  & 32.78$\pm$0.57 &   38.1$\pm$0.1    &  9.8$\pm$0.2 & 3.1$\pm$0.30  & single \\
SO$_{2}$ ($3_{2,2}-2_{1,1}$)       & 208700.336(1) & 15.3  & 4.94$\pm$0.83  &   37.0$\pm$0.8    &  10.9$\pm$2.3& 0.43$\pm$0.08 & single \\
HC$_{3}$N ($23-22$)                & 209230.234(30)& 120.5 & 2.26$\pm$0.51  &   36.7$\pm$0.7    &  5.6$\pm$1.6 & 0.39$\pm$0.10 & single  \\
H$_2$CO ($3_{1,3}-2_{1,2}$)        & 211211.468(10)& 32.1  & 24.43$\pm$0.81 &   37.6$\pm$0.1    &  8.6$\pm$0.4 & 2.68$\pm$0.17 & single  \\
CH$_{3}$OH ($1_{-1}-0_{0}$ E)      & 213427.061(6) & 23.4  & 2.86$\pm$0.89  &   38.9$\pm$0.7    &  5.2$\pm$2.3 & 0.52$\pm$0.09 & single \\
SO ($5_{5}-4_{4}$)                 & 215220.653(40)& 44.1  & 39.31$\pm$1.74 &   37.8$\pm$0.2    &  10.9$\pm$0.7 & 3.38$\pm$0.45 & single \\
CH$_{3}$OH ($5_{-1}-4_{-2}$ E)     & 216945.521(12)& 55.9  &  2.92$\pm$0.67 &   34.3$\pm$0.7    &  8.1$\pm$2.8 & 0.34$\pm$0.05 & single \\
SiO ($5-4$)                        & 217104.98(5)  & 31.3  & 87.54$\pm$1.26 &   \nodata         &   \nodata    &   \nodata   & multiple   \\
DCN ($3-2$)                        & 217238.538(4) & 20.9  & 4.80$\pm$0.46  &   36.5$\pm$0.2    &  5.15$\pm$0.58 & 0.88$\pm$0.04 & single \\
H$_2$CO ($3_{0,3}-2_{0,2}$)        & 218222.192(10)& 21.0  & 12.02$\pm$0.63 &   37.5$\pm$0.1    &  5.21$\pm$0.33 & 2.17$\pm$0.14 & single \\
HC$_{3}$N ($24-23$)                & 218324.723(10)& 131.0 & 1.22$\pm$0.30  &   37.4$\pm$0.36   &  3.0$\pm$0.9 & 0.38$\pm$0.01 & single \\
CH$_{3}$OH ($4_{-2}-3_{-1}$ E)     & 218440.063(13)&  45.5 & 7.25$\pm$0.51  &   37.3$\pm$0.2    &  4.8$\pm$0.5 & 1.41$\pm$0.13 & single \\
H$_2$CO ($3_{2,2}-2_{2,1}$)        & 218475.632(10)&  68.1 & 5.94$\pm$0.64  &   36.9$\pm$0.3    &  7.7$\pm$1.2 & 0.72$\pm$0.16 & single \\
H$_2$CO ($3_{2,1}-2_{2,0}$)        & 218760.066(10)&  68.1 & 5.44$\pm$0.66  &   36.9$\pm$0.3    &  6.1$\pm$1.1 & 0.84$\pm$0.10 & single \\
C$^{18}$O ($2-1$)                  & 219560.354(2) &  15.8 & 8.19$\pm$0.77  &   37.6$\pm$0.3    &  6.9$\pm$1.0 & 1.11$\pm$0.20 & single \\
SO ($6_{5}-5_{4}$)                 & 219949.442(40)&  35.0 & 81.92$\pm$1.10 &   \nodata         &   \nodata    &   \nodata    & multiple  \\
$^{13}$CO ($2-1$)                  & 220398.684(1) &  15.9 & 87.04$\pm$1.24 &   \nodata         &   \nodata    &   \nodata    & multiple \\
SO$_{2}$ ($11_{1,11}-10_{0,10}$)   & 221965.220(1) &  60.4 &  7.82$\pm$0.83 &   38.0$\pm$0.4    &  7.3$\pm$1.0 & 1.00$\pm$0.11 & single \\
C$^{17}$O (2--1)                   & 224714.187(80) &  16.2 & 1.37$\pm$0.41  &   37.7$\pm$0.4    &  3.3$\pm$1.5 & 0.39$\pm$0.10 & single \\
H$_2$CO ($3_{1,2}-2_{1,1}$)        & 225697.775(10)&  33.4 & 19.37$\pm$0.39 &   37.5$\pm$0.1    &  7.7$\pm$0.2 & 2.35$\pm$0.21 & single \\
CH$_{3}$OH ($8_{1}-7_{0}$ E)       & 229758.756(12)&  89.1 & 3.23$\pm$0.49  &   36.3$\pm$0.3    &  4.0$\pm$0.9 & 0.76$\pm$0.09 & single \\
CO ($2-1$)                         & 230538.000(1) &  16.6 & 434.25$\pm$1.41&   \nodata         &   \nodata    &   \nodata    & multiple  \\
CH$_{3}$OH ($5_{1}-4_{1}$ A$^{+}$) & 239746.219(4) &  49.1 &  4.29$\pm$0.43 &   37.4$\pm$0.3    &  6.2$\pm$0.7 & 0.65$\pm$0.08 & single \\
C$^{34}$S ($5-4$)                  & 241016.089(1) &  34.7 &  4.13$\pm$0.51 &   37.5$\pm$0.4    &  6.9$\pm$1.0 & 0.57$\pm$0.14 & single \\
SO$_{2}$ ($5_{2,4}-4_{1,3}$)       & 241615.797(2) &  23.6 &  4.70$\pm$0.71 &   37.9$\pm$0.8    & 10.7$\pm$1.8 & 0.41$\pm$0.12 & single \\
CH$_{3}$OH ($5_{0}-4_{0}$ E)       & 241700.159(4) &  47.9 &  5.16$\pm$0.59 &   37.3$\pm$0.3    &  5.1$\pm$0.8 & 0.96$\pm$0.13 & single \\
CH$_{3}$OH ($5_{1}-4_{1}$ E)       & 241767.234(4) &  40.4 & 10.51$\pm$0.79 &   36.9$\pm$0.2    &  5.3$\pm$0.5 & 1.88$\pm$0.17 & single \\
CH$_{3}$OH ($5_{0}-4_{0}$ A$^{+}$) & 241791.352(4) &  34.8 & 10.81$\pm$0.79 &   37.1$\pm$0.2    &  5.3$\pm$0.5 & 1.90$\pm$0.21 & single \\
CH$_{3}$OH ($5_{-1}-4_{-1}$ E)     & 241879.025(4) &  55.9 &  4.92$\pm$0.61 &   37.0$\pm$0.3    &  5.5$\pm$0.8 & 0.84$\pm$0.11 & single \\
CH$_{3}$OH ($5_{2}-4_{2}$ E)       & 241904.147(4) &  60.7 &  $<$7.01       &   \nodata         &  \nodata     & \nodata      & blended \\
CH$_{3}$OH ($5_{-2}-4_{-2}$ E)     & 241904.643(4) &  57.1 &  $<$7.01       &   \nodata         &  \nodata     & \nodata      & blended \\
CH$_{3}$OH ($5_{1}-4_{1}$ A$^{-}$) & 243915.788(4) &  49.7 &  6.17$\pm$0.44 &   37.6$\pm$0.2    &  5.6$\pm$0.5 & 1.03$\pm$0.11 & single \\
SO$_{2}$ ($14_{0,14}-13_{1,13}$)   & 244254.218(1) &  93.9 & 5.20$\pm$0.71  &   \nodata         &  \nodata     & \nodata      & multiple \\
CS ($5-4$)                         & 244935.557(3) &  35.3 & 44.34$\pm$1.07 &   37.6$\pm$0.1    &  8.0$\pm$0.3 & 5.18$\pm$0.40 & single \\
SO$_{2}$ ($13_{1,13}-12_{0,12}$)   & 251199.675(1) &  82.2 &  9.64$\pm$0.89 &   \nodata         &  \nodata     & \nodata      & multiple \\
CH$_{3}$OH ($6_{3}-6_{2}$ A$^{+}$) & 251738.437(13)&  98.5 &  1.36$\pm$0.30 &   36.8$\pm$0.3    & 2.9$\pm$0.8  & 0.44$\pm$0.01 & single \\
SO ($5_{6}-4_{5}$)                 & 251825.77(5)  &  50.7 & 24.23$\pm$1.13 &   38.4$\pm$0.2    & 7.6$\pm$0.5  & 3.00$\pm$0.26 & single \\
CH$_{3}$OH ($4_{3}-4_{2}$ A$^{+}$) & 251866.524(14)&  73.0 & 1.96$\pm$0.67  &   37.0$\pm$0.7    & 4.3$\pm$1.9  & 0.43$\pm$0.07 & single \\
CH$_{3}$OH ($6_{3}-6_{2}$ A$^{-}$) & 251895.728(13)&  98.5 &  $<$6.25       &   \nodata         &  \nodata     & \nodata     & blended \\
CH$_{3}$OH ($4_{3}-4_{2}$ A$^{-}$) & 251900.452(14)&  73.0 &  $<$6.25       &   \nodata         &  \nodata     & \nodata     & blended \\
CH$_{3}$OH ($3_{3}-3_{2}$ A$^{+}$) & 251905.729(14)&  63.7 &  $<$6.25       &   \nodata         &  \nodata     & \nodata     & blended \\
CH$_{3}$OH ($3_{3}-3_{2}$ A$^{-}$) & 251917.065(14)&  63.7 &  1.78$\pm$0.25 &   37.7$\pm$0.3    & 3.9$\pm$0.6  & 0.43$\pm$0.07 & single \\
$^{29}$SiO ($6-5$)                 & 257255.216(17)&  43.2 &  2.72$\pm$0.38 &   38.0$\pm$0.4    & 5.5$\pm$0.9  & 0.46$\pm$0.03 & single \\
SO ($6_{5}-5_{5}$)                 & 258255.826(5) &  56.5 & 37.26$\pm$0.88 &   \nodata         &  \nodata     & \nodata     & multiple \\
H$^{13}$CN ($3-2$)                 & 259011.798(1) &  24.9 & 5.36$\pm$0.56  &   36.9$\pm$0.3    & 5.4$\pm$0.7  & 0.93$\pm$0.05 & single \\
SiO ($6-5$)                        & 260518.02(5)  &  43.8 & 59.57$\pm$1.01 &   \nodata         &  \nodata     & \nodata     & multiple \\
CH$_{3}$OH ($2_{-1}-1_{0}$ E)      & 261805.675(6) &  28.0 &  3.84$\pm$0.48 &   37.8$\pm$0.4    & 6.4$\pm$0.9  & 0.56$\pm$0.07 & single \\
SO ($7_{6}-6_{5}$)                 & 261843.72(3)  &  47.6 & 66.07$\pm$1.01 &   \nodata         &  \nodata     & \nodata     & multiple \\
C$_{2}$H ($3-2$)                   & 262004.26(5)  &  25.1 &  1.64$\pm$0.49 &   36.4$\pm$0.8    & 5.8$\pm$2.2  & 0.27$\pm$0.04 & single \\
HCN ($3-2$)                        & 265886.434(1) &  25.5 &114.18$\pm$0.97 &   \nodata         &  \nodata     & \nodata     & multiple \\
CH$_{3}$OH ($5_{-2}-4_{-2}$ E)     & 266838.148(13)&  57.1 & 6.16$\pm$0.27  &   37.1$\pm$0.1    & 3.6$\pm$0.2  & 1.6$\pm$0.08 & single \\
HCO$^{+}$ ($3-2$)                  & 267557.626(1) &  25.7 & 31.80$\pm$1.02 &   37.3$\pm$0.1    & 9.8$\pm$0.5  & 3.06$\pm$0.41 & single \\
SO$_{2}$ ($7_{2,6}-6_{1,5}$)       & 271529.013(1) &  35.5 &  2.01$\pm$0.61 &   38.5$\pm$0.7    & 4.9$\pm$1.7  & 0.39$\pm$0.17 & single \\
HNC ($3-2$)                        & 271981.142(20)&  26.1 &  6.38$\pm$0.61 &   36.2$\pm$0.2    & 4.6$\pm$0.5  & 1.30$\pm$0.14 & single \\
CH$_{3}$OH ($9_{-1}-8_{0}$ E)      & 278304.512(12)& 110.0 &  2.76$\pm$0.33 &   36.7$\pm$0.2    & 3.7$\pm$0.6  & 0.70$\pm$0.03 & single \\
H$_{2}$CO ($4_{1,4}-3_{1,3}$)      & 281526.929(10)& 45.6  & 17.99$\pm$0.40 &   37.7$\pm$0.1    & 7.2$\pm$0.2  & 2.36$\pm$0.22 & single \\
SO$_{2}$ ($15_{1,15}-14_{0,14}$)   & 281762.600(1) & 107.4 &  3.95$\pm$0.39 &   37.6$\pm$0.3    & 5.6$\pm$0.8  & 0.66$\pm$0.11 & single \\
SO$_{2}$ ($6_{2,4}-5_{1,5}$)       & 282036.566(1) & 29.2  &  1.75$\pm$0.56 &   38.2$\pm$1.0    & 6.1$\pm$3.4  & 0.27$\pm$0.08  & single \\
\hline
\end{tabular}
\tablefoot{(1) Transition. (2) Rest frequency taken from the CDMS or JPL databases \citep{1998JQSRT..60..883P,2016JMoSp.327...95E}. Uncertainties in the last decimal digits are given in parentheses. (3) Upper level energy. (4) Integrated intensity. (5) Velocity centroid. (6) FWHM line width. (7) Peak main beam brightness temperature. (8) Notes on the observed line profile.}
\normalsize   
\end{table*}

\begin{figure*}[!htbp]
\centering
\includegraphics[width = 0.9 \textwidth]{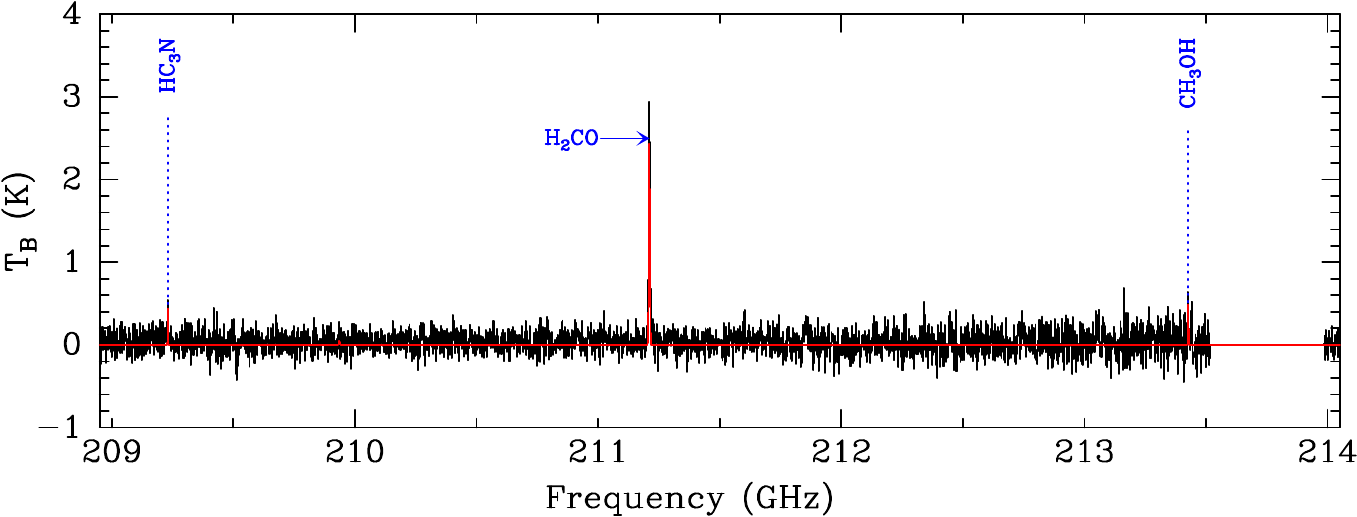}
\includegraphics[width = 0.9 \textwidth]{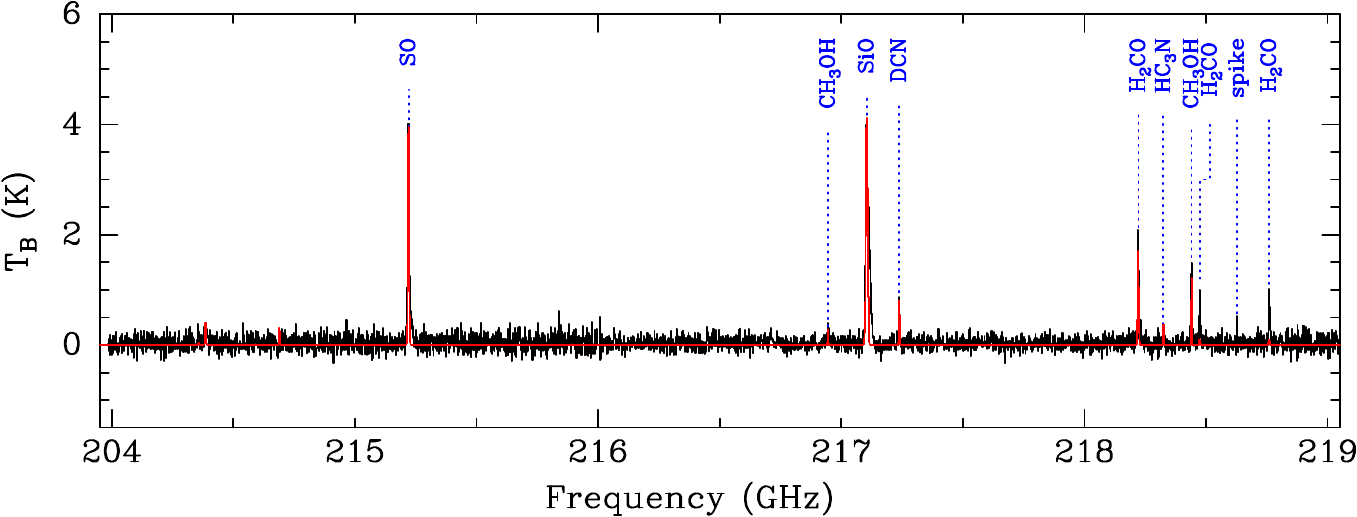}
\includegraphics[width = 0.9 \textwidth]{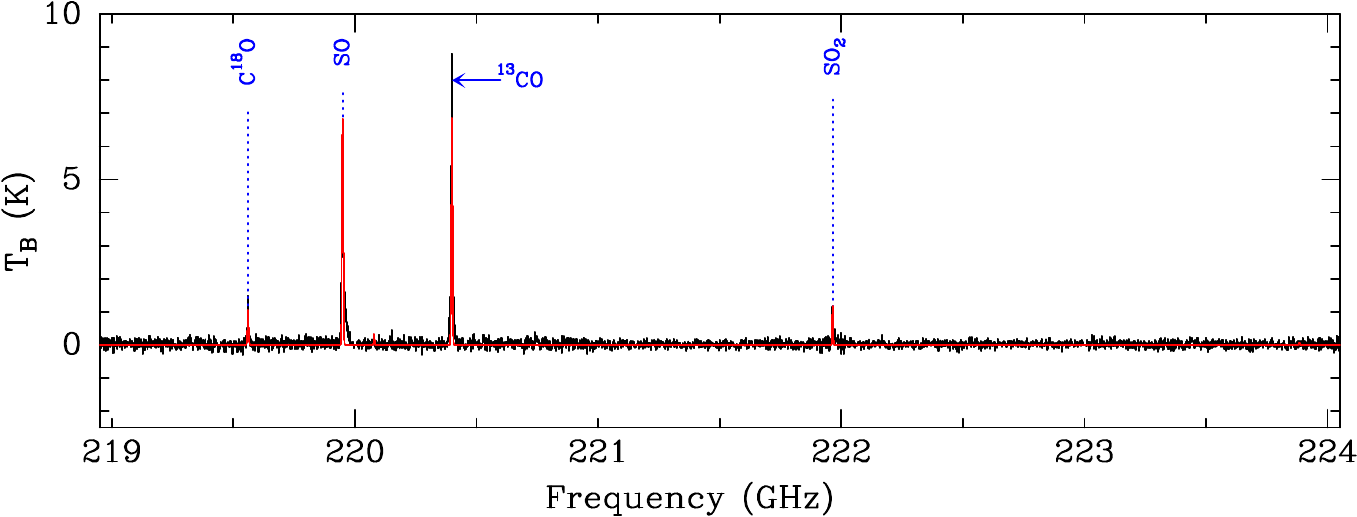}
\includegraphics[width = 0.9 \textwidth]{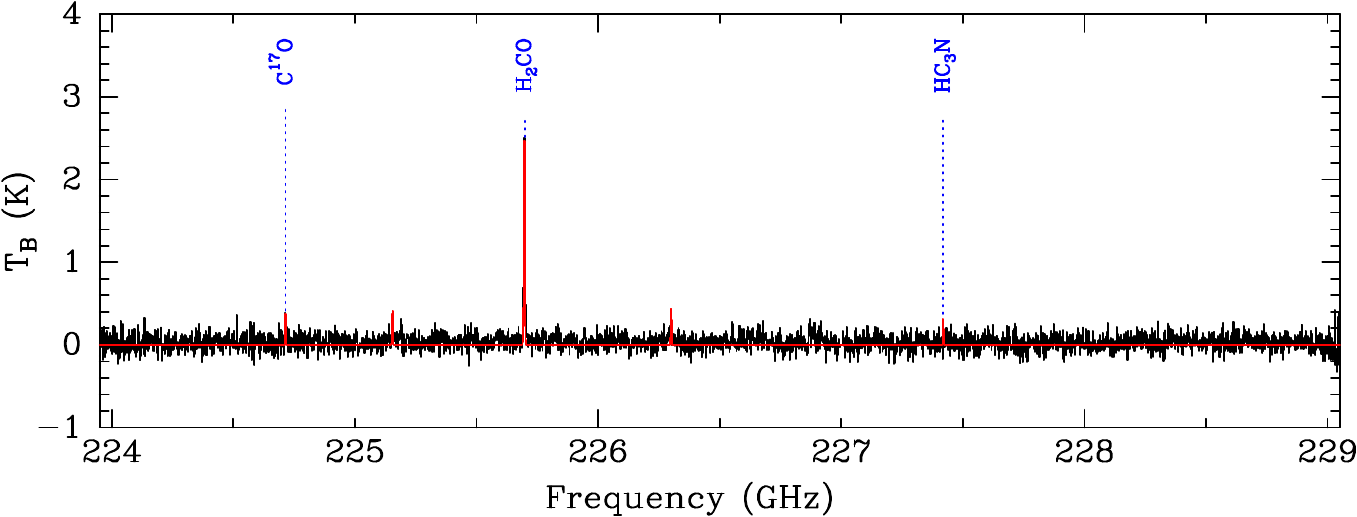}
\centerline{Fig.~\ref{Fig:sma-line-survey} --- Continued.}
\end{figure*}
\begin{figure*}[!htbp]
\centering
\includegraphics[width = 0.9 \textwidth]{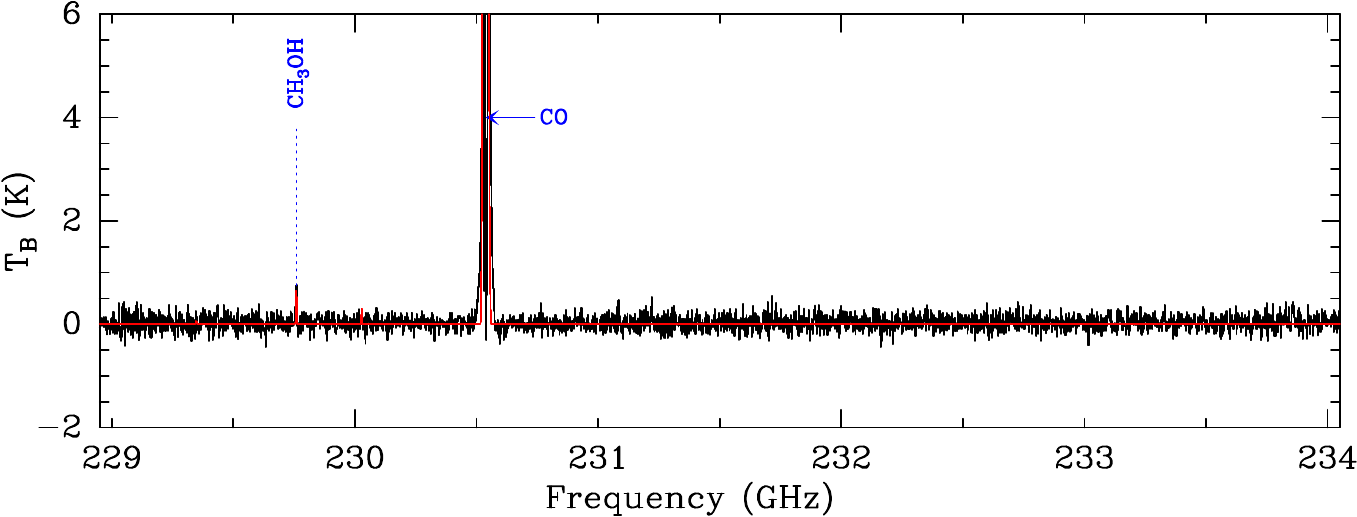}
\includegraphics[width = 0.9 \textwidth]{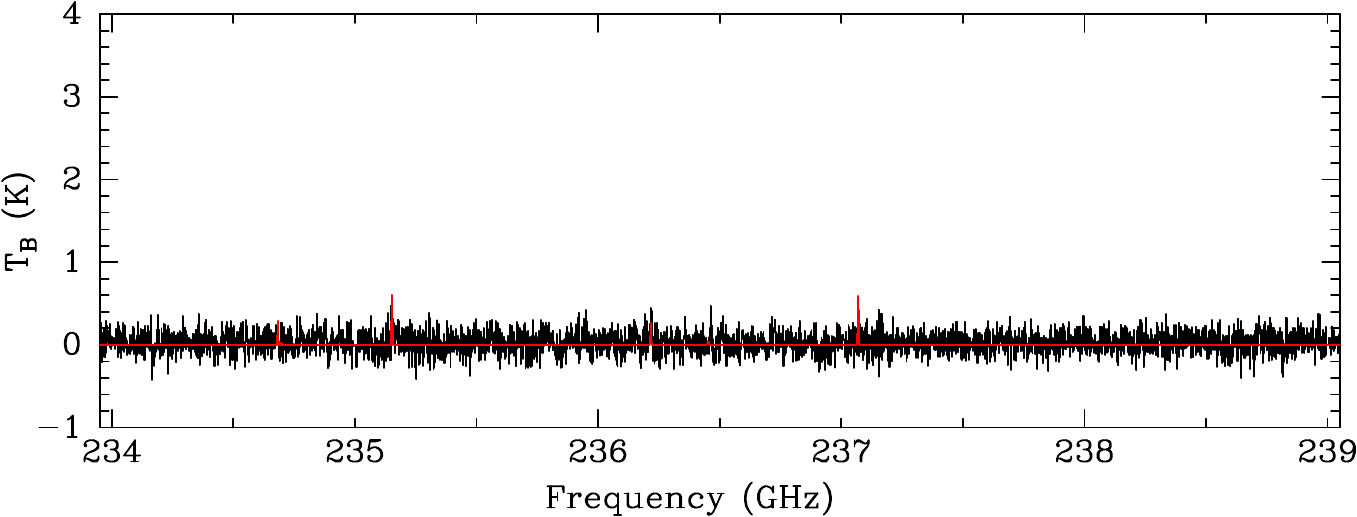}
\includegraphics[width = 0.9 \textwidth]{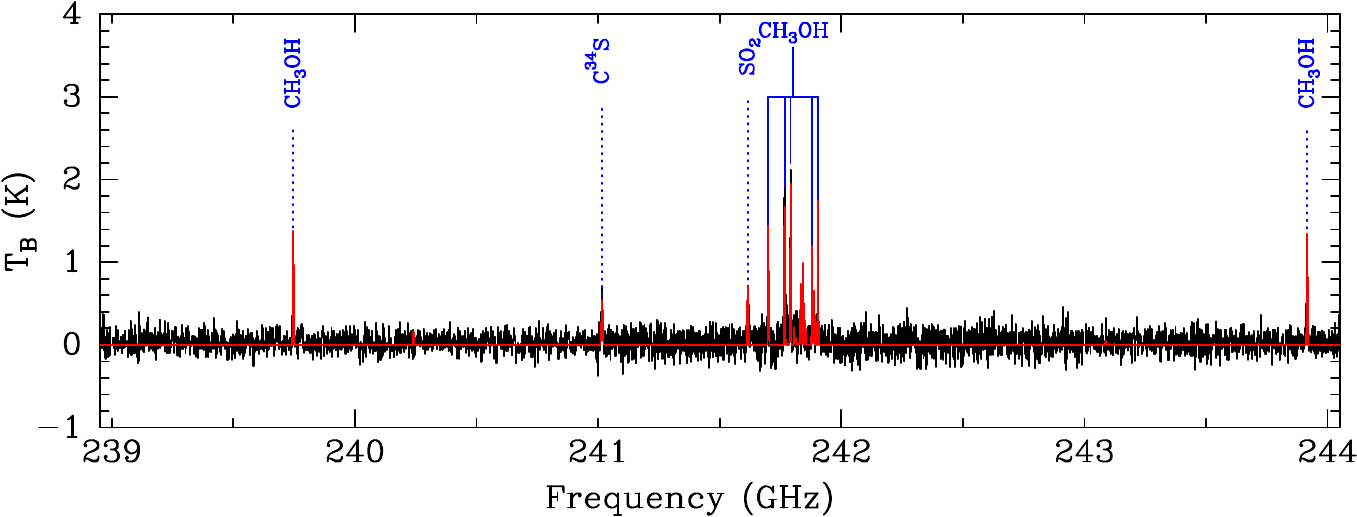}
\includegraphics[width = 0.9 \textwidth]{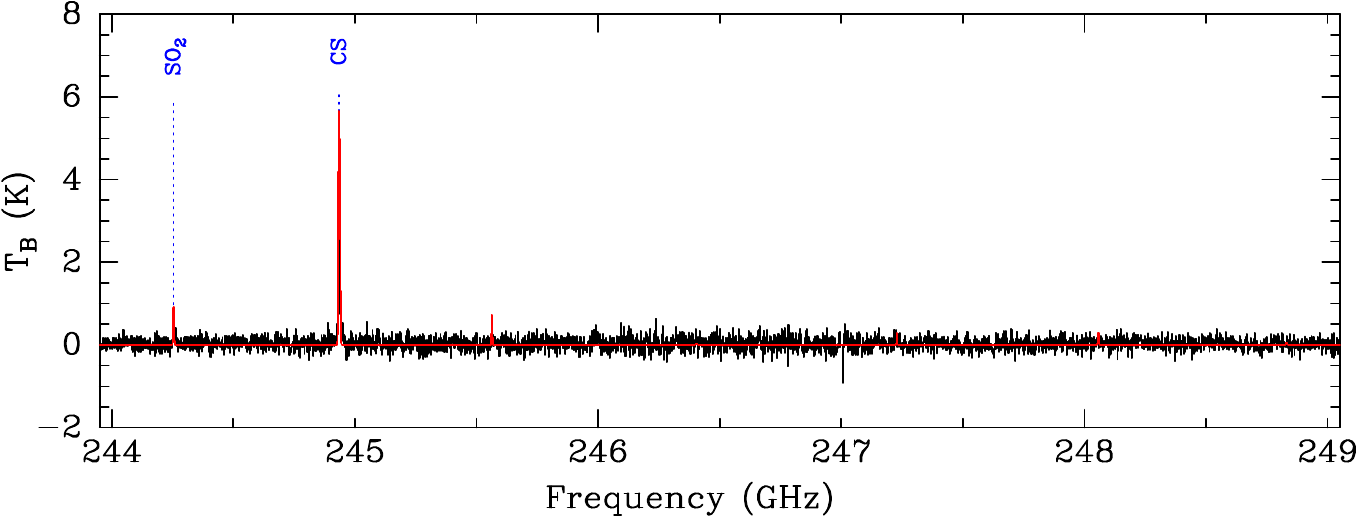}
\centerline{Fig.~\ref{Fig:sma-line-survey} --- Continued.}
\end{figure*}
\begin{figure*}[!htbp]
\centering
\includegraphics[width = 0.9 \textwidth]{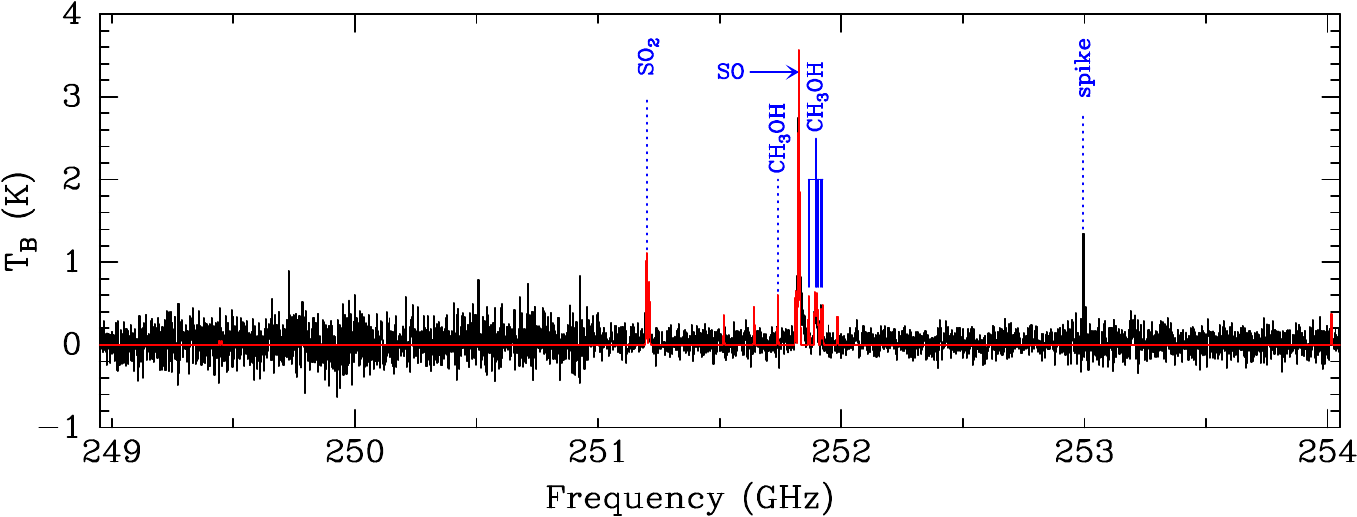}
\includegraphics[width = 0.9 \textwidth]{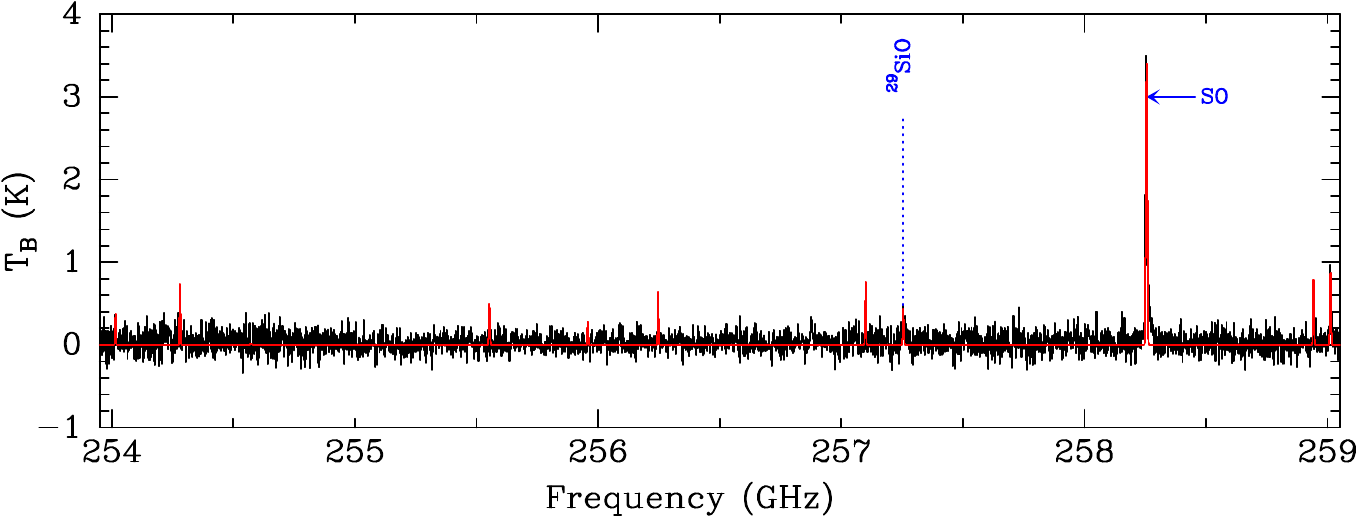}
\includegraphics[width = 0.9 \textwidth]{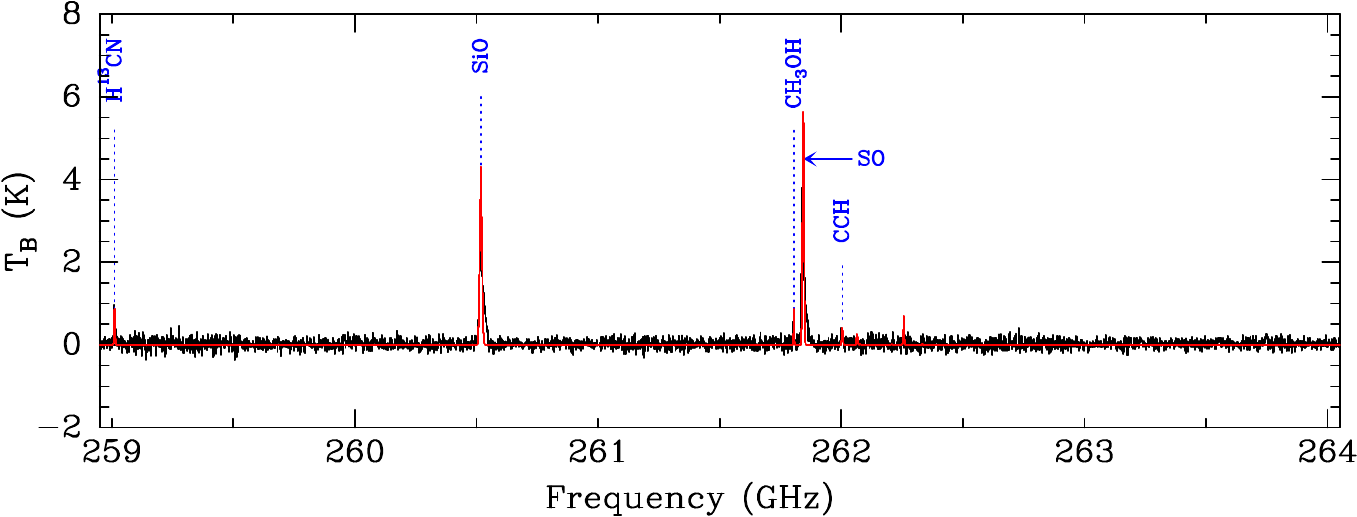}
\includegraphics[width = 0.9 \textwidth]{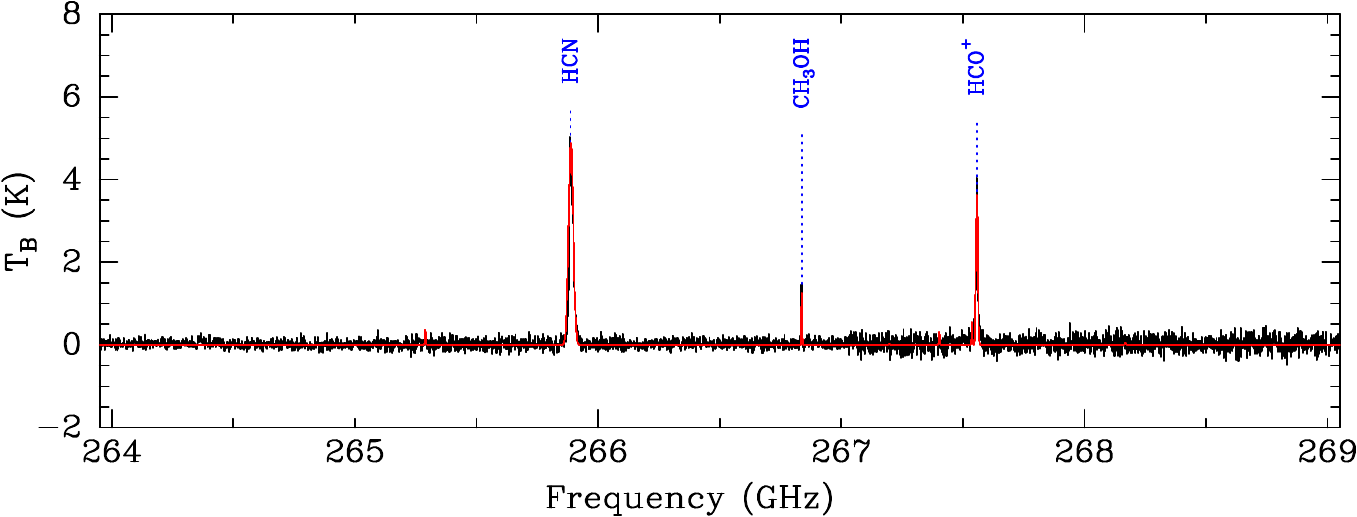}
\centerline{Fig.~\ref{Fig:sma-line-survey} --- Continued.}
\end{figure*}
\begin{figure*}[!htbp]
\centering
\includegraphics[width = 0.9 \textwidth]{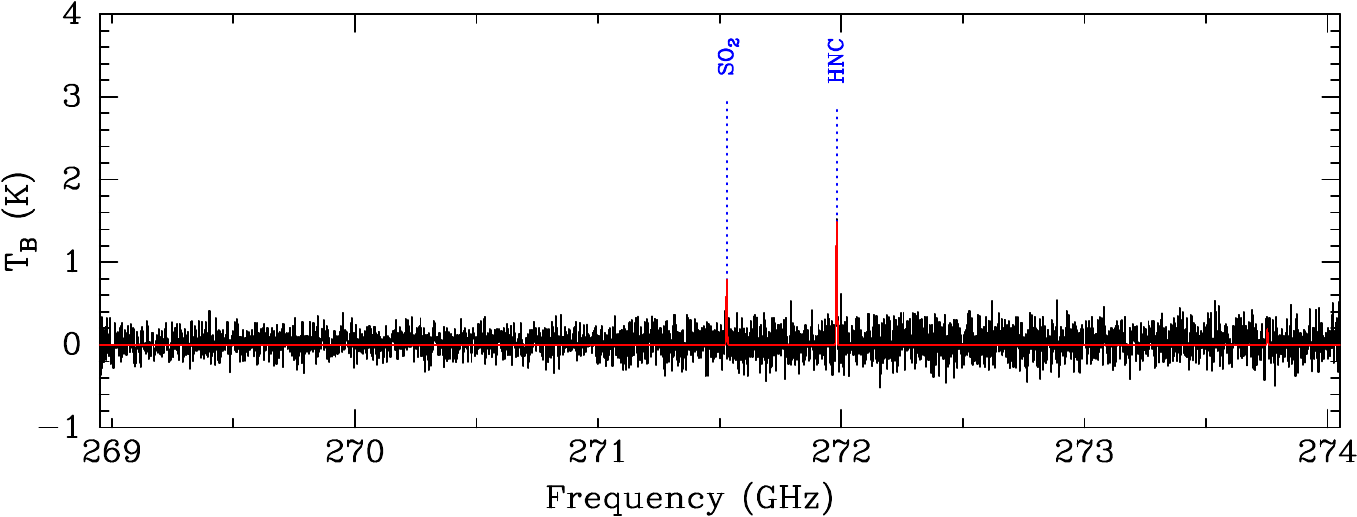}
\includegraphics[width = 0.9 \textwidth]{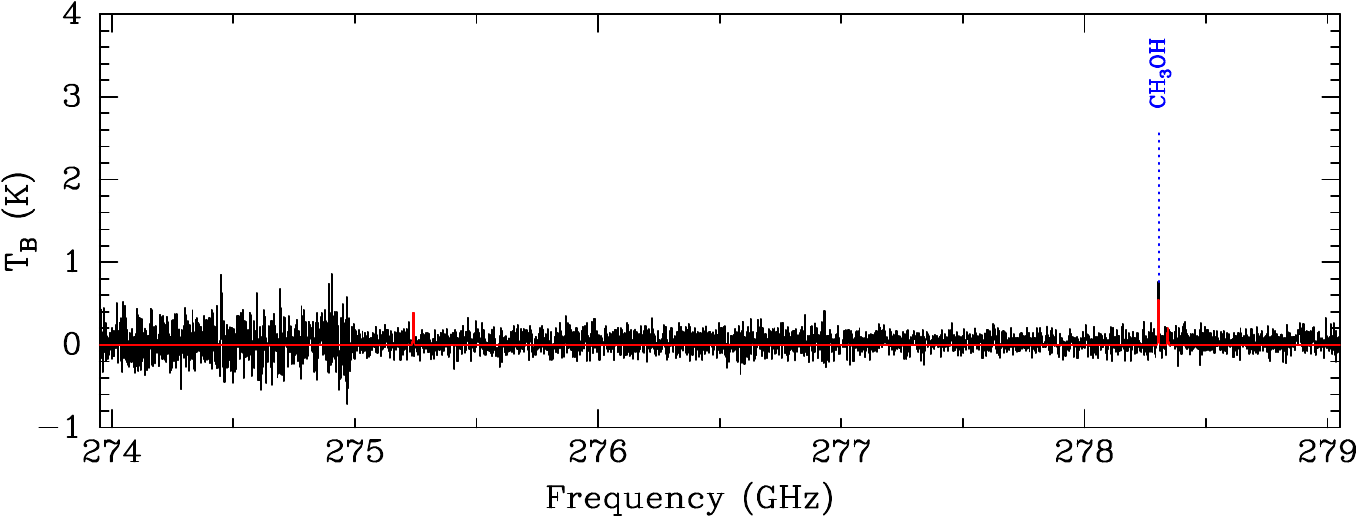}
\includegraphics[width = 0.9 \textwidth]{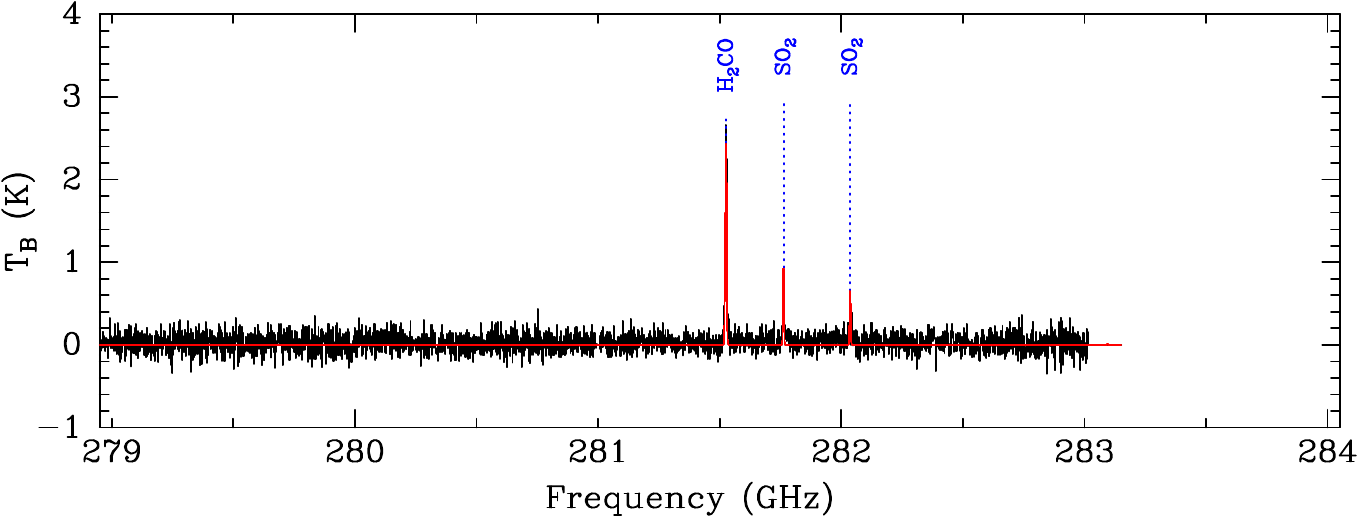}
\centerline{Fig.~\ref{Fig:sma-line-survey} --- Continued.}
\end{figure*}

\begin{figure*}[!htbp]
\centering
\includegraphics[width = 0.9 \textwidth]{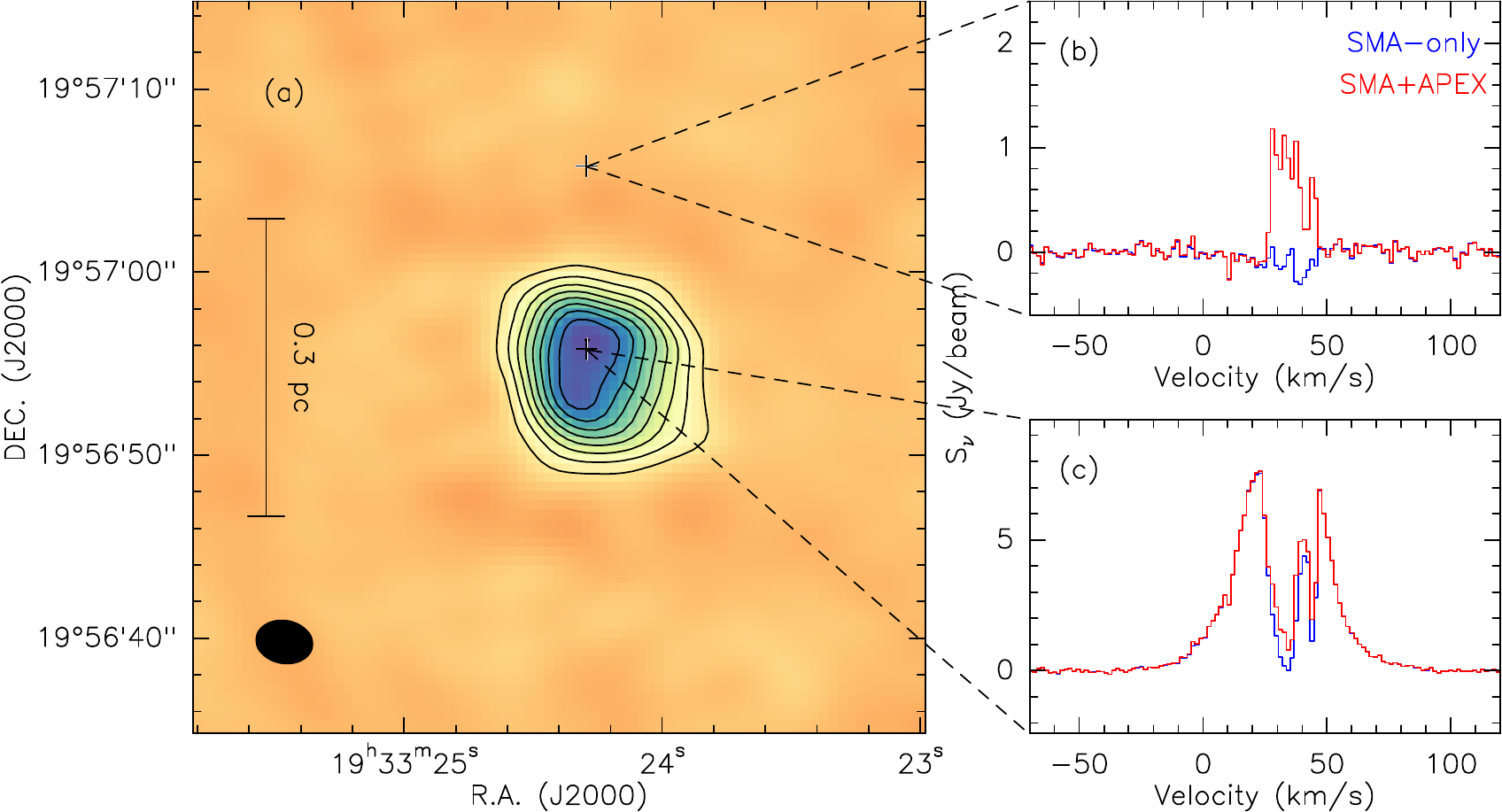}
\caption{{Comparison between the SMA-only and SMA+APEX CO (2--1) data. (a) SMA+APEX CO (2--1) intensity map integrated from $-$50~\kms\,to 100~\kms. The contours start at 20\% of the peak (i.e., 50.2~Jy~beam$^{-1}$~\kms) and increase in steps of 25.1~Jy~beam$^{-1}$~\kms. The restored beam is shown in the lower-left corner of the panel. The two plus symbols mark the positions where the spectra of both SMA-only and SMA+APEX CO (2--1) data are compared in panels (b) and (c).}\label{Fig:co-zero}}
\end{figure*}

\end{appendix}

\end{document}